\documentclass[10pt]{article}

\usepackage[T1]{fontenc}
\usepackage{lmodern}

\usepackage{amsmath,amssymb,amsfonts}
\usepackage{graphicx}
\usepackage{indentfirst,csquotes}
\usepackage{paralist,titlesec,fancyhdr,etoolbox}
\usepackage{colortbl}
\usepackage{xcolor}
\usepackage{pifont}
\usepackage{booktabs}
\usepackage{multirow}
\usepackage{array}
\usepackage{geometry}
\usepackage{setspace}
\usepackage{caption}
\usepackage{subcaption}
\usepackage[numbers,sort&compress]{natbib}
\makeatletter
\usepackage[
    colorlinks=true,
    linkcolor=blue,
    citecolor=blue,
    urlcolor=blue
]{hyperref}
\usepackage{authblk}
\usepackage{float}

\providecommand{\keywords}[1]{\textbf{Keywords:} #1}

\begin{document}

\title{PK/PD-integrated Bayesian platform design for phase II dose regimen optimization}

\author[1]{Axel Vuorinen}
\author[2,3,4]{Antoine Guillon}
\author[5,6]{Emmanuelle Comets}
\author[1,*]{Moreno Ursino}

\affil[1]{Inserm, Université Paris Cité, Inria, HeKA, F-75015 Paris, France}

\affil[2]{Inserm, Centre d'\'Etude des Pathologies Respiratoires (CEPR), UMR 1100, Tours, France}

\affil[3]{Université de Tours, Tours, France}

\affil[4]{Service de Médecine Intensive Réanimation, CHRU de Tours, Tours, France}

\affil[5]{Université Paris Cité et Université Sorbonne Paris Nord, Inserm, IAME, Paris, France}

\affil[6]{Univ Rennes, Inserm, EHESP, Irset, UMRS 1085, Rennes, France}

\date{}

\maketitle

\let\thefootnote\relax
\footnotemark
\footnotetext{$^{*}$Corresponding author: \href{mailto:moreno.ursino@inserm.fr}{moreno.ursino@inserm.fr}.} 

\begin{abstract}
Early-phase dose-finding methods increasingly assess toxicity and efficacy jointly, but comparisons based only on administered dose may inadequately characterize regimens differing in schedule. We developed a Bayesian phase II adaptive platform design for regimen optimization that integrates pharmacokinetic/pharmacodynamic (PK/PD) modelling into toxicity, efficacy, regimen selection and adaptation decisions.
The proposed PK/PD-informed Regimen Optimization Platform (PROP) design uses a population PK/PD model to generate patient- and population-level predictions of exposure and biological activity. Acute and cumulative toxicities are analysed using a discrete-time time-to-event model informed by PK exposure. Efficacy is evaluated through Bayesian model averaging of exposure-driven and biomarker-driven time-to-event models. The design supports regimen graduation, discontinuation for futility or safety, and addition of unexplored regimens. Performance was evaluated through simulations motivated by an influenza intensive-care setting.
Across six scenarios, PROP generally improved graduation and futility decisions, reduced inappropriate graduation, and supported the addition of promising regimens compared with dose-based alternatives. It also more accurately estimated regimen-specific toxicity and arm-specific efficacy, while the model-averaging framework favored the efficacy model consistent with the data-generating mechanism. Dose-based approaches performed better for safety stopping in some scenarios, despite less accurate characterization of the regimen--toxicity relationship.
PK/PD-informed platform designs can improve adaptive regimen selection and knowledge generation when dose alone cannot adequately characterize treatment regimens.

\noindent \keywords{Adaptive platform trials; Bayesian methods; Dose regimen optimization; Population pharmacokinetics/pharmacodynamics; Time-to-event models; Toxicity; Efficacy}
\end{abstract} 

\bigskip

\section{Introduction}\label{sec1}
The high attrition rate during clinical development leads to substantial loss of time and resources, especially when compounds fail in large and expensive late-stage trials \cite{hay2014clinical}. Failures in phase III trials are mainly due to insufficient efficacy or unacceptable safety for the dosing regimen being tested \citep{arrowsmith2011phase} and highlight the need for more informative early-phase studies that support biomarker-guided patient selection and dose optimization \citep{wong2019estimation, sun202290}. If dose optimization is postponed until the confirmatory stage, suboptimal regimens may enter into large trials, potentially exposing many patients to an unfavorable benefit-risk balance \citep{korn2023dose}.
 
In early-stage development, phase I studies, performed in healthy volunteers or in patients for treatments with high expected toxicity, primarily aim to characterize the relationship between dose and safety, while generating pharmacokinetic (PK) and, when feasible, pharmacodynamic (PD) information, to identify one or more acceptable doses or dosing regimens for further clinical evaluation. Clinical development progresses towards phase II trials that evaluate preliminary efficacy in the target patient population, while continuing to monitor safety and providing an opportunity to refine dose and regimen selection. Dose-ranging trials can use a variety of designs depending on the development objective, the knowledge context, and the number of candidate regimens under evaluation.

When several doses, dosing regimens, or treatments remain under consideration at this stage, randomized controlled trials (RCTs) provide direct concurrent comparisons and are thus better suited to support comparative dose-selection decisions. In this context, multi-arm designs can naturally  accommodate the concurrent evaluation of multiple experimental arms within a single randomized trial, often using a shared control to enable direct comparisons \citep{royston2003novel, wason2016some}. For the optimization of multiple treatments or dosing regimens, adaptive designs provide additional flexibility by allowing prespecified modifications based on accumulating data, including early stopping for safety or futility, sample-size adaptation, or preferential allocation to promising arms. Seamless designs integrate different development phases that are traditionally separated, such as phase I and II or phase II and III, within a single protocol. Adaptive platform trials (APTs) extend this framework by using a master protocol to evaluate multiple interventions over time, with arms entering or leaving the platform according to predefined decision rules~\citep{adaptive2019adaptive}. Although platform trials may incorporate features such as response-adaptive randomization, information sharing, subgroup selection, and Bayesian decision-making, fixed randomization and frequentist approaches can also be used. Most platform trials have been developed for later-stage evaluation, 
including phase II proof-of-concept\citep{barker2009spy, ritchie2016development}, phase III \citep{bateman2017dian}, seamless phase II/III \citep{alexander2018adaptive, kaplan2015focus4}, or phase IV settings \citep{angus2020remap, bongard2018antivirals}. Their application to early-phase dose optimization based on the joint assessment of toxicity and efficacy remains comparatively limited, although methodological proposals have emerged in immunotherapy~\citep{mu2021bayesian, mu2022bayesian, mu2024bayesian, shi2026bayesian}, oncology~\citep{yuan2016midas, rossoni2018phase}, and COVID-19~\citep{griffiths2020agile}.

As already implied above, for a given drug, dose optimization extends beyond the selection of a dose level. A dosing regimen is defined by the amount of drug administered at each dosing occasion, together with the route, frequency, and schedule of administration. Two regimens delivering the same total dose can produce distinct exposure profiles and pharmacological effects. Comparing dosing regimens thus implies evaluating how differences in dosing history can translate into exposure, biological activity, toxicity, and efficacy. Dose optimization has received particular attention in oncology, where the US Food and Drug Administration (FDA)'s Project Optimus \cite{shah2021drug, projectoptimus} encouraged earlier characterization of dose--exposure--response relationships using PK/PD information alongside safety, tolerability, and early evidence of efficacy, with the aim of supporting randomized comparisons of candidate doses and identify regimens that achieve a favorable benefit-risk balance.
Extending these principles beyond oncology could help bridge the evidence gap between exploratory dose-finding and confirmatory trials. Several methodological studies have shown that integrating pharmacokinetic (PK, describing the time course of drug concentration) and pharmacodynamic (PD, describing the biological effect) information can improve the characterization of dose--toxicity, exposure--response, and biomarker--response relationships, thereby supporting safer and more informative dose or regimen selection \citep{piantadosi1996improved, ursino2017dose, vuorinen2026comparative}. Early approaches incorporated exposure summaries, such as the area under the concentration--time curve (AUC), into Bayesian models for binary toxicity \citep{piantadosi1996improved, ursino2017dose}. More recent designs use longitudinal individual PK data to fit a mechanistic population PK model, whose structure can be updated as data accrue, allowing the model to better describe the observed concentration--time profiles \citep{micallef2022exposure}. Semi-mechanistic approaches further model the dose/schedule--concentration relationship, the link between concentration and pharmacological activity, and the association between cumulative activity and toxicity within a unified framework \citep{su2022semi, yang2024extended}. The PEEDOOP design applies this framework of incorporating PK measurements and latent PD information into toxicity and efficacy models, respectively, to support optimal dose identification in a seamless oncology phase I/II trial, although without the flexibility of a platform framework \citep{yuan2026pharmacometrics}. TITE-PK extended PK-informed modelling to multiple schedules and time-to-event toxicity outcomes \citep{gunhan2020bayesian}, illustrating the feasibility of incorporating the timing and magnitude of longitudinal PK/PD information into toxicity and efficacy assessments. These approaches can handle nonlinear dose--exposure relationships, inter-patient variability, and updates to the PK model structure as data accumulates. They can also allow information to be shared across regimens, including regimens not yet evaluated, by linking outcomes to predicted exposure or pharmacological activity rather than to dose alone \citep{gerard2022bayesian}. Their added value is particularly relevant when exposure and/or effect is not approximately dose proportional \citep{ursino2017dose, vuorinen2026comparative}. 

Motivating by new prospective trials in anti-influenza therapies for critically ill patients, we propose a phase II adaptive platform trial design to compare multiple dosing regimens of the same investigational drug recommended from a first-in-human phase I study in healthy volunteers.
Influenza is an acute respiratory viral infection caused by influenza viruses. Worldwide, annual seasonal influenza epidemics are estimated to result in approximately 1 billion annual clinical cases, 3 to 5 million cases of severe illness and approximately 290 000 to 650 000 deaths \citep{iuliano2018estimates, WHO2025seasonalinfluenza}. In addition to seasonal epidemics, influenza pandemics can result in high illness attack rates responsive of high number of hospitalizations, and death because most people lack immunity to the novel virus. The clinical management of patients with severe influenza virus infection requires not only optimal supportive care but also effective specific antiviral therapy. However, current therapeutic options for influenza remain limited, and their effectiveness can be reduced by factors such as the timing of treatment initiation, antiviral resistance, and the emergence of influenza viruses with reduced susceptibility to available agents. To date, 2024 WHO clinical practice guidelines for the treatment of patients with severe influenza virus infection recommend the administration of oseltamivir \citep{WHO2024influenza}. This is a conditional recommendation based on very low-quality evidence \citep{WHO2024influenza}. Indeed, skepticism persists regarding the efficacy of approved anti-influenza drugs, especially when administered later in the course of infection \citep{bartlett1995community, jefferson2014neuraminidase}. In June 2026, the global Bayesian adaptive platform trial REMAP-CAP, which evaluates multiple interventions in hospitalized patients with respiratory tract infections, reported that oseltamivir was ineffective and was associated with a high probability of increased 90-day mortality in critically ill patients with influenza \citep{remapcap2026oseltamivir}. This important finding, from the first RCT comparing oseltamivir with no antiviral treatment in critically ill patients with influenza, may prompt a reassessment of the current 2024 WHO guidelines. Thus, severe influenza represents a significant area of unmet medical need, highlighting the need for innovative therapeutic strategies.
Innovative drug candidates for severe influenza virus infection are likely to require short treatment courses (3--5 days) with once- or twice-daily administration, preferably via intravenous or nebulized routes, particularly given the potential for impaired enteric absorption in critically ill patients with severe influenza \citep{may2019paracetamol}. Aconitate is one example of anti-influenza innovative molecule currently undergoing drug development \citep{cezard2026cis}.

Within the proposed design, efficacy remains the primary focus, as in conventional phase II trials, while tolerability continues to be monitored throughout the trial. Drug exposure and pharmacological activity, estimated using a mechanistic population PK/PD model, are incorporated into a Bayesian framework for the joint evaluation of efficacy and tolerability. In the context of the motivating trial, the terms toxicity and safety are occasionally used in a broad sense to refer, respectively, to mild adverse events and tolerability.
In the remainder of the manuscript, we first introduce the PK/PD, toxicity, and efficacy models underlying the proposed design and describe how pharmacological information is incorporated into the Bayesian decision framework. We then assess the operating characteristics of the design through an extensive simulation study and conclude with a discussion of its practical implementation, limitations, and potential extensions.

\section{Platform design and methods for dose optimization}

Multiple dosing regimens of a single investigational drug can be compared with a common control in a homogeneous patient population with the same disease. The initial regimens are selected from the recommended phase II dosing regimens identified in the preceding phase I study. Regimens showing insufficient efficacy or unacceptable safety in patients may be discontinued early. Additional candidates may be identified from accumulating patient PK data during phase II. Rather than stopping the trial or continuing to allocate patients to suboptimal regimens, newly identified candidates can replace discontinued arms and be evaluated within the same platform. This flexibility can increase the likelihood of identifying a regimen suitable for confirmatory evaluation while avoiding the time and resources required to initiate a new phase II trial.
\begin{figure}
    \centering
    \includegraphics[
        width=1\textwidth,
        keepaspectratio
    ]{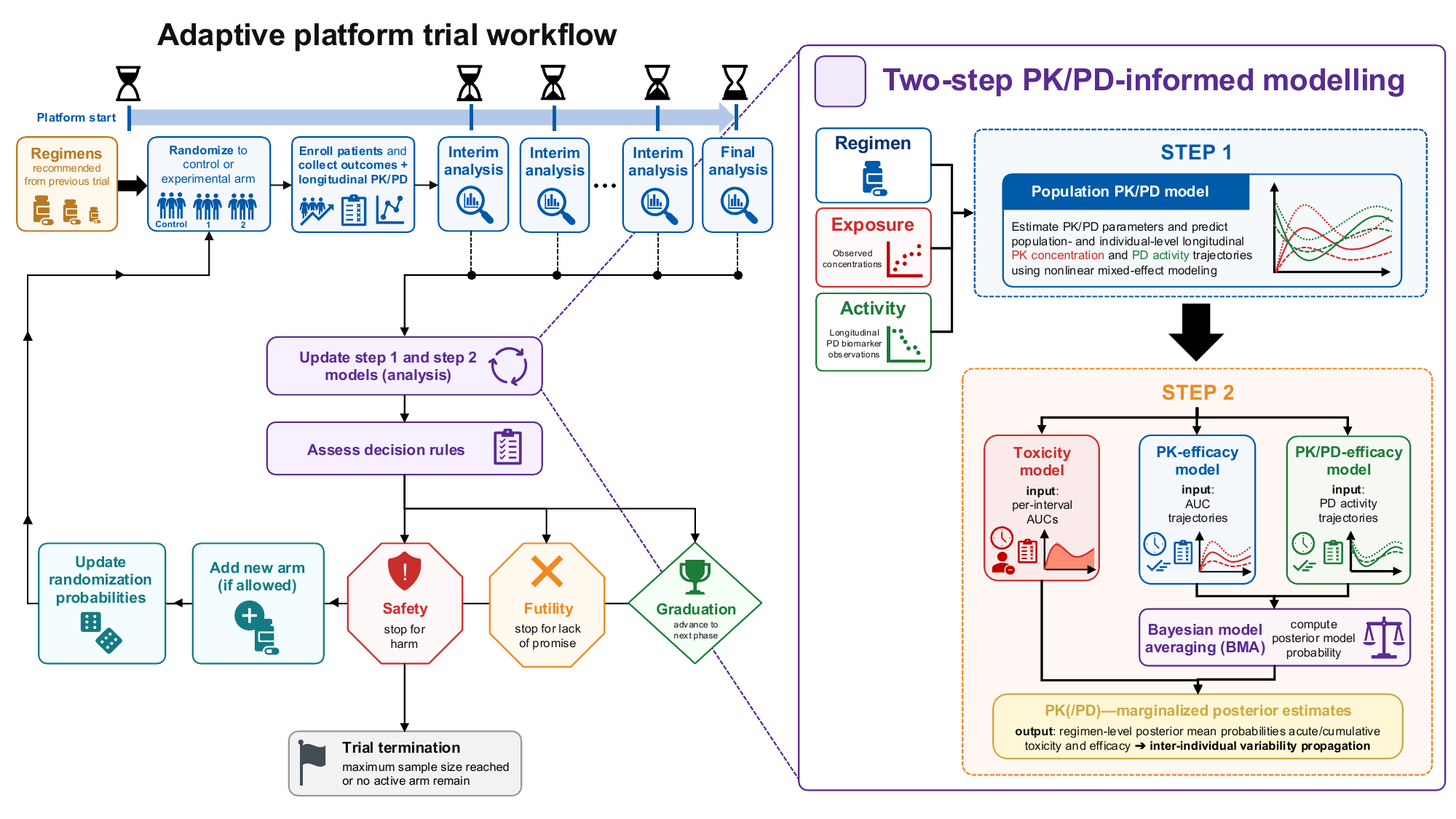}
    \caption{Overview of the PROP framework: adaptive platform trial workflow and two-step PK/PD-informed modelling updated at each analysis.}
    \label{PROP_design}
\end{figure}
The PK/PD-informed Regimen Optimisation Platform (PROP) design, illustrated in Figure \ref{PROP_design}, relies on a two-step modelling framework. In the first step, accumulating PK and PD data are analysed using a frequentist nonlinear mixed-effects population PK/PD (popPK/PD) model with the Stochastic Approximation Expectation Maximization (SAEM) algorithm. In the second step, the resulting individual PK and PD estimates are linked to toxicity and efficacy outcomes in different Bayesian models to inform trial decisions.

In this work, longitudinal PK and PD measurements are collected only from patients randomized to the experimental arms, as only the PK of the investigational treatment is explicitly modelled. No natural-history or alternative PD model is specified for patients assigned to the control arm, precluding the estimation or simulation of longitudinal biomarker measurements in this group. Here, PK exposure is quantified from the estimated concentration--time profile, for example using the area under the curve (AUC) as a summary measure, whereas PD activity refers to the drug-induced pharmacological effect on the longitudinal biomarker response estimated by the PD model. The aim is to obtain patient- and population-level predictions of PK exposure and PD activity trajectories to inform subsequent toxicity and efficacy assessments. 
Unlike conventional approaches based on nominal or cumulative dose, PK/PD-informed regimen selection uses quantified exposure--toxicity and exposure--activity--efficacy relationships to account for inter-individual variability and support the identification of regimens with an optimal benefit--risk profile. In the second modelling step, safety is monitored using a Bayesian time-to-event (TTE) PK--toxicity model that relates the interval-specific exposure metric to toxicity risk within each dosing interval, allowing unsafe regimens to be identified and discontinued early. Efficacy is assessed using a Bayesian model-averaged TTE PK/PD approach combining two competing models: one driven solely by predicted drug exposure and another incorporating predicted PD activity, which is itself partly determined by PK exposure. By weighting the competing TTE models according to their posterior probabilities, the approach accounts for structural uncertainty in the exposure-activity--efficacy relationship, estimates the probability of efficacy for each intervention arm, and identifies the most plausible pathway linking exposure, activity, and efficacy.

To define the PROP framework, we introduce the following notation. Let $N$ denote the maximum number of patients who may be enrolled. The trial may terminate before reaching $N$ if all experimental arms are discontinued before the final analysis. Accordingly, let $n$ denote the number of patients enrolled at trial completion, with $0 < n \leq N$ and index patients by $i \in\{1,\ldots,n\}$. Let $d_k$ denote dose level $k$, with $k\in \{1,\ldots,K\}$, from the ordered set of candidate doses $D={d_1,\ldots,d_K}$, and let $S={\mathbf{s}_1,\ldots,\mathbf{s}_L}$ denote the set of $L$ candidate dosing schedules. For clarity, the schedules may be indexed according to a nested structure, for example by increasing numbers of administrations or increasing treatment duration, although other definitions and orderings may also be considered. Each schedule $\mathbf{s}_l$, for $l\in{1,\ldots,L}$, is represented by the vector of administration times $\mathbf{s}_l=(\tau_{l,1},\ldots,\tau_{l,U_l})$, where $\tau_{l,u}$ is the time of the $u$-th administration and $U_l$ is the total number of administrations under schedule $l$.
A dosing regimen is defined by pairing a dose level $d_k$ with a schedule $\mathbf{s}_l$. Let $\mathcal R={r_1,\ldots,r_M}$ denote the set of all $M=K\times L$ possible dose-schedule combinations, with $r_m=(d_k,\mathbf{s}_l)$, where $m$ uniquely indexes each combination. The reference regimen is denoted by $r_*=(d_*,\mathbf{s}_*)$, where $d_*\in D$ and $\mathbf{s}_*\in S$ are the reference dose and schedule, respectively.
Only a subset $\mathcal R_0\subseteq\mathcal R$, with $|\mathcal R_0|=O$, is selected for inclusion at the start of the phase II dose-optimization trial. The platform initially comprises one control arm $A_0$ and $O$ experimental arms, with $\mathcal A^{\mathrm{exp}}={A_1,\ldots,A_O}$, each corresponding to one regimen in $\mathcal R_0$. The full set of trial arms is therefore $\mathcal A={A_0}\cup\mathcal A^{\mathrm{exp}}={A_0,\ldots,A_O}$. For patient $i$, let $A_{(i)}\in\mathcal A$ denote the assigned trial arm. If $A_{(i)}\in\mathcal A^{\mathrm{exp}}$, let $r_{(i)}\in\mathcal R$ denote the corresponding assigned dosing regimen.

In the remainder of this section, we describe the models used for the population PK/PD, toxicity, and efficacy components of the two-step PROP framework, followed by the adaptive phase II decision rules governing arm graduation (i.e., meeting the prespecified success criterion), arm discontinuation, arm addition, and response-adaptive randomization within the platform design.

\subsection{Population PK/PD modelling: longitudinal concentration and activity data}

We assume that a previously completed first-in-human phase I study in healthy volunteers provided sufficient data to develop the population PK model used during phase II, whereas the PD model is based on preclinical data or evidence from similar compounds. Patients enrolled in the phase II trial may have different PK parameter distributions from healthy volunteers, but the underlying structural models are assumed to remain applicable. Although structural PK/PD model development and updating are not explicitly considered, they could also be incorporated into the proposed framework.

Let \(\mathbf{t}=(t_1,\ldots,t_J)\) be the vector of PK/PD sampling times, where \(t_j\) is the $j$-th sampling time relative to treatment initiation, with $j\in\{1,\ldots,J\}$. In the proposed framework, all patients follow the same prespecified sampling schedule throughout the trial; therefore, the notation is independent of the patient index. Although blood sampling times could be individualized for each newly enrolled patient using optimal-design methods to maximize the informativeness of the collected PK/PD observations, a common fixed sampling schedule is assumed here for simplicity. The population PK/PD model is specified as nonlinear mixed-effects models, with structural models \(f^{\mathrm{PK}}(\cdot)\) and \(f^{\mathrm{PD}}(\cdot)\) and residual-error models \(\phi^{\mathrm{PK}}(\cdot)\) and \(\phi^{\mathrm{PD}}(\cdot)\). For patient \(i\), receiving regimen \(r_{(i)}\), let \(y^{\mathrm{PK}}_{ij}=y_i^{\mathrm{PK}}(t_j)\) and \(y^{\mathrm{PD}}_{ij}=y_i^{\mathrm{PD}}(t_j)\) denote the observed PK and PD responses at time \(t_j\), respectively:
\begin{align}
\begin{split}
y^{\mathrm{PK}}_{ij} &= f^{\mathrm{PK}}\left(t_j,r_{(i)},\boldsymbol{\psi}^{\mathrm{PK}}_i\right)+\phi^{\mathrm{PK}}\left(t_j,r_{(i)},\boldsymbol{\psi}^{\mathrm{PK}}_i,\boldsymbol{\sigma}^{\mathrm{PK}}\right)\varepsilon^{\mathrm{PK}}_{ij},\\
y^{\mathrm{PD}}_{ij} &= f^{\mathrm{PD}}\left(t_j,C_{i}\left(t_j\right),\boldsymbol{\psi}^{\mathrm{PK/PD}}_i\right)+\phi^{\mathrm{PD}}\left(t_j,C_{i}\left(t_j\right), \boldsymbol{\psi}^{\mathrm{PK/PD}}_i,\boldsymbol{\sigma}^{\mathrm{PD}}\right)\varepsilon^{\mathrm{PD}}_{ij},
\end{split}
\label{PKPDmodel}
\end{align}
where \(C_{i}(t) =f^{\mathrm{PK}}\left(t,r_{(i)},\boldsymbol{\psi}^{\mathrm{PK}}_i\right)\) denotes the individual drug concentration predicted by the structural PK model at time \(t\), which is used as the exposure driving the PD response. Individual PK and PD parameters are expressed as functions of the corresponding population fixed effects, \(\boldsymbol{\mu}^{\mathrm{PK}}\) and \(\boldsymbol{\mu}^{\mathrm{PD}}\), and individual random effects, \(\boldsymbol{\eta}^{\mathrm{PK}}_i\) and \(\boldsymbol{\eta}^{\mathrm{PD}}_i\), such that \(\boldsymbol{\psi}^{\mathrm{PK(/PD)}}_i=\boldsymbol{\mu}^{\mathrm{PK(/PD)}}\odot\exp\left(\boldsymbol{\eta}^{\mathrm{PK(/PD)}}_i\right)\), where \(\odot\) denotes element-wise multiplication and \(\boldsymbol{\eta}^{\mathrm{PK(/PD)}}_i\sim\mathcal{N}\left(\mathbf{0},\boldsymbol{\Omega}^{\mathrm{PK(/PD)}}\right)\). Here, \(\boldsymbol{\Omega}^{\mathrm{PK(/PD)}}\) denotes the variance-covariance matrix of the random effects and captures between-patient variability in the PK and PD parameters. The residual-error parameters \(\boldsymbol{\sigma}^{\mathrm{PK}}\) and \(\boldsymbol{\sigma}^{\mathrm{PD}}\) define the observation-error models for the PK and PD responses, respectively. For example, \(\boldsymbol{\sigma}={a}\), \({b}\), or \({a,b}\) may be used for constant, proportional, or combined residual-error models, respectively. The residual errors \(\varepsilon^{\mathrm{PK}}_{ij}\) and \(\varepsilon^{\mathrm{PD}}_{ij}\) are assumed to be mutually independent and independently distributed as \(\mathcal{N}(0,1)\), and to be independent of the individual random effects. To maintain model simplicity, the proposed population PK/PD model does not incorporate covariates, although these could be included to support regimen optimization in specific patient subpopulations at different stages of drug development.

\subsection{Toxicity: Time-to-event model for Acute and Cumulative Toxicities using PharmacoKinetics information (TACT-PK)}

Continuous Bayesian safety monitoring of the experimental arms is used to verify that regimens identified as acceptable in phase I, as well as newly introduced regimens as experimental arms, remain appropriate in the patient population and randomized setting. Since regimens combining different doses and schedules may not admit a natural ordering, toxicity is modelled using predicted PK exposure rather than administered dose. In particular, regimens with the same total dose may produce different exposure profiles both across and within regimens. Individual exposure measures are therefore derived from the population PK/PD model and incorporated into the toxicity model. Building on the DICE model~\citep{ursino2022dice} and TITE-PK~\citep{gunhan2020bayesian}, we propose a Time-to-event model for Acute and Cumulative Toxicities using PharmacoKinetics information (TACT-PK). An adverse event (AE) may result either from intense exposure early in the treatment schedule or from cumulative exposure to repeated administrations. The acute component represents the risk associated with early exposure and is assumed to decrease over successive treatment intervals, whereas the cumulative component captures the increasing risk associated with accumulated exposure.
TACT-PK is formulated as a discrete time-to-event model in which follow-up is divided into dosing intervals. It models the conditional probability of experiencing a first toxicity event during each interval, given that no toxicity occurred previously. Aligning the intervals with treatment administrations allows the model to capture interval-specific toxicity risk, accommodate different dosing schedules, and incorporate PK exposure measures directly.

Let $\mathcal{G}$ be the vector of interval boundaries induced by the longest dosing schedule $\mathbf{s}_L$ and the toxicity follow-up duration $\bar{T}_T$, such that $\mathcal{G}=\left(\tau_{L,1},\ldots,\tau_{L,U_L},\bar{T}_T\right)=\left(\bar{\tau}_1,\ldots,\bar{\tau}_G,\bar{\tau}_{G+1}\right)$, where $G$ is the number of dosing intervals and $\bar{\tau}_1=0$. For $g=1,\ldots,G$, dosing interval $g$ is defined as $\mathcal{I}_g=\left(\bar{\tau}_g,\bar{\tau}_{g+1}\right]$ and has length $\Delta_g=\bar{\tau}_{g+1}-\bar{\tau}_g$.
For regimen $m$, $x_{m,g}=\operatorname{AUC}_g(r_m)=\int_{\bar{\tau}_g}^{\bar{\tau}_{g+1}} f^{\mathrm{PK}}\left(t,r_m,\widehat{\boldsymbol{\mu}}^{\mathrm{PK}}\right),dt$ denotes the population-predicted AUC over interval $g$. Similarly, $x_{(i),g}$ denotes the individual predicted AUC for patient $i$ over interval $g$, obtained from the individual PK parameters and the assigned regimen $r_{(i)}$. Let $\mathbf{x}_m=(x_{m,1},\ldots,x_{m,G})$ denote the vector of predicted population AUCs over all dosing intervals for regimen $m$, and let $\mathbf{x}_{(i)}=(x_{(i),1},\ldots,x_{(i),G})$ denote the corresponding vector of predicted individual AUCs for patient $i$.
For $g\geq2$, the cumulative AUC up to interval $g$ for regimen $m$ is defined as $S_{m,g}=\sum_{\gamma=2}^{g}x_{m,\gamma}$. Similarly, $S_{(i),g}=\sum_{\gamma=2}^{g}x_{(i),\gamma}$ denotes the cumulative individual AUC for patient $i$. The first-interval AUC is considered separately to characterize acute toxicity, whereas exposure accumulated from the second interval onward is used to characterize cumulative toxicity.
Let $T_{T,i}$ denote the toxicity event time for patient $i$, and let $C_{T,i}$ denote the corresponding censoring time. The observed follow-up time is $O_{T,i}=\min(T_{T,i},C_{T,i})$, and the toxicity event indicator is $\delta_{T,i}=\mathbf{1}(T_{T,i}\leq C_{T,i})$. After $n$ patients have been enrolled, the observed toxicity data are denoted by $\mathcal{D}_T(n)=\left\{\left(O_{T,i},\delta_{T,i},\mathbf{x}_{(i)}\right):i=1,\ldots,n\right\}$.
For patient $i$ during interval $g$, the piecewise-constant log-hazard function, with parameter vector $\boldsymbol{\theta}_T=(\alpha_0,\alpha_1,\alpha_2)$, is defined as
\begin{equation}
\begin{aligned}
\log h\left(t,\mathbf{x}_{(i)}\right) &= \log h_g\left(\mathbf{x}_{(i)}\right),\quad t\in\mathcal{I}_g \\
&=\left[\alpha_0+\alpha_1\log\left(\frac{x_{(i),1}}{x_{*,1}}\right)\right]\mathbf{1}(g=1)+\left[\alpha_2+\log\left(\frac{2S_{(i),g}}{S_{(i),g}+S_{*,G}}\right)\right]\mathbf{1}(g\geq2),
\end{aligned}
\label{loghazard_function}
\end{equation}
where $x_{*,1}$ is the population-predicted AUC during the first dosing interval for a prespecified reference regimen, and $S_{,G}=\sum_{g=2}^{G}x_{*,g}$ is its cumulative population-predicted AUC from the second through the final interval. The first component of the log-hazard function captures the acute toxicity risk during the first dosing interval through the log-transformed exposure normalized with respect to the reference regimen, with $\alpha_0$ representing the log hazard under the reference exposure. The second component captures cumulative toxicity risk through the logarithm of the normalized cumulative AUC accrued over the $g-1$ dosing intervals from the second interval through interval $g$. The coefficient associated with this normalized cumulative-exposure function is fixed to one to improve model parsimony and facilitate parameter estimation.

Because $h_g\left(\mathbf{x}_{(i)}\right)$ is a hazard rate per unit time, multiplication by the interval length $\Delta_g$ yields the cumulative hazard over interval $g$. The conditional probability that patient $i$ experiences a first toxicity event during interval $g$, given that no toxicity occurred before the beginning of the interval, is denoted by $p_g\left(\mathbf{x}_{(i)}\right)=\mathbb{P}\left(T_{T,i}\leq\bar{\tau}_{g+1}\vert T_{T,i}>\bar{\tau}_g,\mathbf{x}_{(i)}\right)$.
Under the piecewise-constant hazard assumption, this probability is related to the interval-specific hazard through the complementary log-log link, $\operatorname{cloglog}(p)=\log{-\log(1-p)}$, such that
$\operatorname{cloglog}\left(p_g\left(\mathbf{x}_{(i)}\right)\right) =\log\Delta_g+\log h_g\left(\mathbf{x}_{(i)}\right)$, or equivalently, $
p_g\left(\mathbf{x}_{(i)}\right) =1-\exp\left\{-\Delta_g h_g\left(\mathbf{x}_{(i)}\right)\right\}$.
Let $H_g\left(\mathbf{x}_{(i)}\right)=\sum_{\gamma=1}^{g}\Delta_\gamma h_\gamma\left(\mathbf{x}_{(i)}\right)$ denote the cumulative hazard up to the end of interval $g$. The corresponding cumulative probability that patient $i$ experiences a toxicity event by the end of interval $g$ is $P_{T,g}\left(\mathbf{x}_{(i)}\right) =\mathbb{P}\left(T_{T,i}\leq\bar{\tau}_{g+1}\vert\mathbf{x}_{(i)}\right) =1-\exp\left\{-H_g\left(\mathbf{x}_{(i)}\right)\right\}$.
In particular, the cumulative toxicity probability over the complete toxicity follow-up period is
\begin{equation}
P_{T,G}\left(\mathbf{x}_{(i)}\right)
=\mathbb{P}\left(T_{T,i}\leq\bar{T}_T\vert\mathbf{x}_{(i)}\right)
=1-\exp\left\{-H_G\left(\mathbf{x}_{(i)}\right)\right\}
=1-\exp\left\{-\sum_{g=1}^{G}\Delta_g h_g\left(\mathbf{x}_{(i)}\right)\right\}.
\label{cumulative_probability_of_toxicity}
\end{equation}

Once a patient experiences a first toxicity, toxicity follow-up is stopped at the end of the corresponding interval and no additional doses are administered. If no toxicity is observed, the patient continues through subsequent dosing intervals according to the treatment schedule until a toxicity occurs or the end of the follow-up period is reached, at which point the observation is censored. The probability of remaining free of toxicity through interval $g$ is obtained from the cumulative hazard function as
\begin{equation}
S_g\left(\mathbf{x}_{(i)}\right)=\mathbb{P}\left(T_{T,i}>\bar{\tau}_{g+1}\vert\mathbf{x}_{(i)}\right)=\exp\left\{-H_g\left(\mathbf{x}_{(i)}\right)\right\},
    \label{toxicity_survival_function}
\end{equation}
and the probability mass associated with a toxicity event occurring during dosing interval $g$ is
$f_g\left(\mathbf{x}_{(i)}\right)=\mathbb{P}\left(T_{T,i}\in\mathcal{I}_g\vert\mathbf{x}_{(i)}\right)=p_g\left(\mathbf{x}_{(i)}\right)\exp\left\{-H_{g-1}\left(\mathbf{x}_{(i)}\right)\right\}$,
where $H_0\left(\mathbf{x}_{(i)}\right)=0$. Let $k_i$ denote the interval in which patient $i$ experiences toxicity or is censored. By combining the two previous expressions, the likelihood can be written as
\begin{equation}
L\left(\mathcal{D}_T(n)\vert\boldsymbol{\theta}_T\right)=\prod_{i=1}^{n}L\left(O_{T,i},\delta_{T,i},\mathbf{x}_{(i)}\vert\boldsymbol{\theta}_T\right)=\prod_{i=1}^{n}f_{k_i}\left(\mathbf{x}_{(i)}\right)^{\delta_{T,i}}S_{k_i}\left(\mathbf{x}_{(i)}\right)^{1-\delta_{T,i}}.
    \label{likelihood_tact_pk}
\end{equation}
The posterior distribution of $\boldsymbol{\theta}_T$ is
$\pi\left(\boldsymbol{\theta}_T\vert\mathcal{D}_T(n)\right)\propto L\left(\mathcal{D}_T(n)\vert\boldsymbol{\theta}_T\right)\pi\left(\boldsymbol{\theta}_T\right)$, where the likelihood is given in Equation~\ref{likelihood_tact_pk} and the prior distribution $\pi\left(\boldsymbol{\theta}_T\right)$ is described later.

After enrolling a total of $n$ patients, the objective is to estimate the acute and cumulative toxicity probabilities for each regimen while accounting for inter-individual exposure variability in the PK model and posterior uncertainty in the parameters of the Bayesian toxicity model.
Therefore, we average the posterior toxicity probabilities over the predictive population distribution of the per-interval AUC vector, $\pi\left(\mathbf{X}_m\vert\widehat{\boldsymbol{\mu}}^{\mathrm{PK}},\widehat{\boldsymbol{\Omega}}^{\mathrm{PK}}\right)$, using Monte Carlo simulation, where $\mathbf{X}_m$ denotes the random vector of per-interval AUCs under regimen $m$. This yields the PK-marginalized posterior mean toxicity probability up to interval $g$ for regimen $m$, denoted by $\bar{P}^{\mathrm{PK}}_{T,g,m}$:
\begin{equation}
\begin{aligned}
    \bar{P}^{\mathrm{PK}}_{T,g,m}
&=
\mathbb{E}_{\pi\left(\boldsymbol{\theta}_T\vert\mathcal{D}_T(n)\right)}
\left[
\mathbb{E}_{\pi\left(\mathbf{X}_m\vert\widehat{\boldsymbol{\mu}}^{\mathrm{PK}},\widehat{\boldsymbol{\Omega}}^{\mathrm{PK}}\right)}
\left[
P_{T,g}\left(\mathbf{X}_m,\boldsymbol{\theta}_T\right)
\right]
\right] \\
&=
\int \int
P_{T,g}\left(\mathbf{X}_m,\boldsymbol{\theta}_T\right)
\pi\left(\boldsymbol{\theta}_T\vert\mathcal{D}_T(n)\right)
\pi\left(\mathbf{X}_m\vert\widehat{\boldsymbol{\mu}}^{\mathrm{PK}},\widehat{\boldsymbol{\Omega}}^{\mathrm{PK}}\right)
\,d\mathbf{X}_m\,d\boldsymbol{\theta}_T.
\end{aligned}
\label{pk_marginalized_posterior_mean}
\end{equation}
To apply the Bayesian overdosing rule for continuous safety monitoring described in Equation~\ref{safety_rule}, the PK-marginalized posterior mean probabilities of acute and cumulative toxicity are computed for each regimen as $\bar{P}^{\mathrm{PK}}_{T,1,m}$ and $\bar{P}^{\mathrm{PK}}_{T,G,m}$, respectively.

\subsection{Bayesian model-averaged time-to-event PK/PD-efficacy method}

The efficacy analysis aims to estimate the treatment effect of each experimental arm $(A_o)_{o=1,\ldots,O}$ relative to the control arm $A_0$ and to characterize the potential relationship between drug activity and the probability and timing of efficacy. Several joint-model formulations can link longitudinal PD biomarker trajectories, modelled through nonlinear mixed-effects approaches, with time-to-event outcomes in a frequentist or Bayesian framework \citep{mbogning2015joint, desmee2015nonlinear, kerioui2020bayesian}. This joint modelling approach allow uncertainty in one component to propagate through the others. Since the population PK/PD model and Bayesian toxicity model are already computationally intensive components, embedding them within a single fully joint Bayesian framework with efficacy would substantially increase model complexity and make repeated interim estimation almost unfeasible. Therefore, we use a two-step strategy in the PROP framework, similar to the previous PK--toxicity model, in which individual- and population-level predictions from the population PK/PD model are included as time-dependent covariates in a Bayesian survival model for efficacy. Moreover, Kazantzidis et al.\citep{kazantzidis2026mediation} showed that inference from joint models linking longitudinal nonlinear tumor dynamics to survival can depend strongly on the chosen association relationship, after assessing several biologically plausible ways of linking the two processes in a mediation analysis based on model selection. This finding provides further motivation for accounting for structural uncertainty in early-phase dose optimization, where the causal pathway linking PK, PD, and efficacy may be incompletely understood. In our framework, we address this uncertainty using Bayesian model averaging (BMA), combining a PK/PD--efficacy model with a direct PK--efficacy model. This allows uncertainty in the PK/PD--efficacy relationship to be propagated across competing mechanistic links rather than relying on a single prespecified model. Posterior model probabilities quantify the contribution of each model to model-specific efficacy estimates, enabling the framework to use PD information when it improves prediction compared with PK exposure alone, while relying on the PK--efficacy relationship when it does not.

In the motivating study, the experimental treatment is administered in addition to standard-of-care influenza antivirals. We assume that efficacy in the experimental arms is at least as high as in the control arm. Accordingly, the control arm is modelled through the baseline hazard. This assumption is specific to the simulation setting and should not be imposed without justification. Alternative modelling strategies for settings in which experimental treatments may be less effective than the control are mentioned in the Discussion section.

Let $\mathcal{M}_e$ be either of the two efficacy models and $\boldsymbol{\theta}_{E,e} = (\lambda_e, \nu_e, \beta_e) \in \boldsymbol{\Theta}_{E,e}$ the model-specific parameters with $e \in \{1,2\}$ denoting the model index, such that $\mathcal{M}_1$ represents the PK/PD-efficacy model and $\mathcal{M}_2$ the direct PK-efficacy model. The treatment indicator, $\gamma_i$, equals 1 if patient $i$ is randomized to an experimental arm, and 0 otherwise. For the $i$-th patient, let $T_{E,i}$ be the efficacy event time and $C_{E, i}$ the censoring time if efficacy was not declared by the end of the follow-up duration $\bar{T}_{E}$. We observe the time of event or censoring $O_{E, i} = \min(T_{E, i},C_{E, i})$ and the efficacy event indicator $\delta_{E, i} = \mathbf{1}(T_{E, i} \le C_{E, i})$. Let $\mathcal{D}_E(n) = \left\{ \left(O_{E, i}, \delta_{E,i}, \mathbf{y}_{(i)}^\text{PD}, \mathbf{\tilde{x}}_{(i)} \right) : 1 \leq i \leq n \right\}$ denote the observed time-to-event efficacy data, the predicted individual PD biomarker trajectories, and the predicted individual AUC profiles after the enrollment of $n$ patients in the platform trial. For patient $i$, the latter is defined as $\tilde{x}_{(i)}(t)=\int_0^t f^{\text{PK}}\left(u,r_{(i)},\widehat{\boldsymbol{\psi}}^{\text{PK}}_{(i)}\right)\,du$, with $0\leq t\leq\bar{T}_E$. 

For model $\mathcal{M}_1$, efficacy is assumed to be linked to PD activity through individual longitudinal predictions obtained from the estimated popPK/PD model during the first step of the two-step efficacy framework and used to inform the time-to-event endpoint in the second stage. The time-varying covariate is the predicted log fold-change from baseline of the PD biomarker for patient $i$ at time $t$, denoted by $\log\left(y_{(i)}^{\text{PD}}(t)/y_{(i)}^{\text{PD}}(0)\right)$. Based on a Bayesian survival model, the individual hazard function for patient $i$ at time $t>0$ is
\begin{equation}
h_{\mathcal{M}_1}\left(t;\mathbf{y}_{(i)}^{\text{PD}}\right)=h_0(t;\lambda_1,\nu_1)\exp\left\{\gamma_i\beta_1\log\left(\frac{y_{(i)}^{\text{PD}}(t)}{y_{(i)}^{\text{PD}}(0)}\right)\right\},
    \label{PKPD_efficacy_hazard}
\end{equation}
where $h_0(t;\lambda_1,\nu_1)$ is a Weibull baseline hazard function with parameters $(\lambda_1,\nu_1)$, such that $h_0(t;\lambda,\nu)=\lambda\nu t^{\nu-1}$, and $\beta_1$ quantifies the strength of the association between the PD biomarker and the efficacy event. If $\beta_1$ is estimated to be close to zero, there is limited evidence that the PD biomarker is a relevant predictor of efficacy. Baseline covariates were not included in the efficacy model in the simulation study but could be incorporated to allow for more complex survival modelling across patient subpopulations. The individual survival function is $S_{\mathcal{M}_1}\left(t;\mathbf{y}_{(i)}^{\text{PD}}\right)=\exp\left\{-\int_0^t h_{\mathcal{M}_1}\left(u;\mathbf{y}_{(i)}^{\text{PD}}\right)\,du\right\}$.
The likelihood of the PK/PD-efficacy time-to-event model is
\begin{equation}
L\left(\mathcal{D}_E(n)\vert\boldsymbol{\theta}_{E,1},\mathcal{M}_1\right)=\prod_{i=1}^{n}h_{\mathcal{M}_1}\left(O_{E,i};\mathbf{y}_{(i)}^{\text{PD}}\right)^{\delta_{E,i}}S_{\mathcal{M}_1}\left(O_{E,i};\mathbf{y}_{(i)}^{\text{PD}}\right).
    \label{PKPD_efficacy_likelihood}
\end{equation}
Therefore, the posterior distribution of $\boldsymbol{\theta}_{E,1}$ conditional on model $\mathcal{M}_1$ is $\pi\left(\boldsymbol{\theta}_{E,1}\vert\mathcal{D}_E(n),\mathcal{M}_1\right)\propto L\left(\mathcal{D}_E(n)\vert\boldsymbol{\theta}_{E,1},\mathcal{M}_1\right)\pi\left(\boldsymbol{\theta}_{E,1}\vert\mathcal{M}_1\right)$.

For model $\mathcal{M}_2$, efficacy is linked directly to PK exposure rather than PD activity. Using the same Bayesian survival structure as for $\mathcal{M}_1$, with cumulative AUC as a time-dependent covariate, the hazard function is defined as
\begin{equation}
h_{\mathcal{M}_2}\left(t;\mathbf{\tilde{x}}_{(i)}\right)=h_0(t;\lambda_2,\nu_2)\exp\left\{\gamma_i\beta_2\log\left(1+\frac{\tilde{x}_{(i)}(t)}{\tilde{x}_*}\right)\right\},
    \label{dosing_efficacy_hazard}
\end{equation}
where $\tilde{x}_*$ is the AUC of the reference regimen over the complete efficacy follow-up period. The transformation $\log(1+x)$ ensures that the covariate is defined at zero exposure and remains non-negative, consistently with the assumed non-negative treatment effect in the motivating setting. The corresponding AUC trajectory is $\mathbf{\tilde{x}}_{(i)}=\left\{\tilde{x}_{(i)}(t):0\leq t\leq\bar{T}_E\right\}$. Values of $\beta_2$ close to zero indicate limited evidence that PK exposure predicts efficacy. The likelihood and posterior distribution under $\mathcal{M}_2$ are obtained as in Equation~\ref{PKPD_efficacy_likelihood}, replacing the model index and efficacy predictor accordingly.

For $e\in\{1,2\}$, the posterior model probability is
\begin{equation}        
    w_e=\pi\left(\mathcal{M}_e\vert\mathcal{D}_E(n)\right)=\frac{\pi\left(\mathcal{D}_E(n)\vert\mathcal{M}_e\right)\pi(\mathcal{M}_e)}{\sum_{l=1}^2\pi\left(\mathcal{D}_E(n)\vert\mathcal{M}_l\right)\pi(\mathcal{M}_l)},
\end{equation}
where $\pi\left(\mathcal{D}_E(n)\vert\mathcal{M}_e\right)=\int_{\boldsymbol{\Theta}_{E,e}}L\left(\mathcal{D}_E(n)\vert\boldsymbol{\theta}_{E,e},\mathcal{M}_e\right)\pi\left(\boldsymbol{\theta}_{E,e}\vert\mathcal{M}_e\right)\,d\boldsymbol{\theta}_{E,e}$ is the marginal likelihood and $\pi(\mathcal{M}_e)$ is the prior model probability. For any common efficacy quantity $Q$, its BMA posterior distribution is $\pi\left(Q\vert\mathcal{D}_E(n)\right)=\sum_{e=1}^2 w_e\,\pi\left(Q\vert\mathcal{D}_E(n),\mathcal{M}_e\right)$.
A larger value of $w_1$ indicates greater support for PD activity providing predictive information beyond PK exposure, whereas a larger value of $w_2$ favors the direct PK--efficacy relationship. Similar posterior model probabilities indicate that the available data do not clearly distinguish between the two models, potentially because of limited sample size, measurement error in the PD biomarker, a weak mediating role of PD activity, or a strong relationship between PK exposure and PD activity.

Similarly to Equation~\ref{pk_marginalized_posterior_mean} for the two-step PK--toxicity model, inter-individual variability in PD activity and PK exposure is propagated when estimating the posterior probability of efficacy at time $\bar{T}_E$ under the PK/PD--efficacy and PK--efficacy models, respectively. For the control arm $A_0$, whose efficacy probability depends only on the baseline hazard, posterior samples are generated by selecting one of the two candidate models according to its posterior model probability and then sampling the corresponding efficacy probability from the model-specific posterior distribution. The posterior mean probability of efficacy is estimated by the Monte Carlo average of these samples.
For each experimental arm $A_o$, the PK/PD-marginalized posterior mean probability of efficacy, denoted by $\bar{P}^{\text{PK/PD}}_{E,o}$, is computed using BMA while additionally averaging the model-specific efficacy probabilities over the predictive population distribution of the corresponding efficacy predictor. Let $\mathbf{Z}_{e,o}$ denote the random predictor trajectory under model $\mathcal{M}_e$ for experimental arm $A_o$, with $\mathbf{Z}_{1,o}=\mathbf{y}^{\text{PD}}_o$ and $\mathbf{Z}_{2,o}=\mathbf{\tilde{x}}_o$. Then,
\begin{equation}
    \bar{P}^{\text{PK/PD}}_{E,o}
    =
    \sum_{e=1}^{2}
    w_e
    \mathbb{E}_{\pi\left(\boldsymbol{\theta}_{E,e}\vert\mathcal{D}_E(n),\mathcal{M}_e\right)}
    \left[
    \mathbb{E}_{\pi\left(\mathbf{Z}_{e,o}\vert\widehat{\boldsymbol{\mu}}_e,\widehat{\boldsymbol{\Omega}}_e\right)}
    \left[
    P_{E,o}\left(\bar{T}_E;\mathbf{Z}_{e,o},\boldsymbol{\theta}_{E,e},\mathcal{M}_e\right)
    \right]
    \right],
    \label{pkpd_marginalized_posterior_mean}
\end{equation}
where $(\widehat{\boldsymbol{\mu}}_1,\widehat{\boldsymbol{\Omega}}_1)$ denote the estimated population parameters governing the PD trajectory under $\mathcal{M}_1$, and $(\widehat{\boldsymbol{\mu}}_2,\widehat{\boldsymbol{\Omega}}_2)$ denote those governing the PK exposure trajectory under $\mathcal{M}_2$.

\subsection{Phase II adaptive platform trial design for dose optimization}

Our PROP design reproduces the main adaptive features of a platform trial in an early-phase setting. Multiple dosing regimens of a single investigational treatment are compared with a common control arm, and interim analyses (IAs) are conducted at prespecified enrolment thresholds to update posterior distributions, apply Bayesian decision rules, and adapt randomization probabilities. Experimental arms may be stopped for toxicity or futility, selected as optimal regimens, or replaced by newly introduced regimens. At interim analysis $\iota$, the total sample size is $n_\iota=\iota n_{\text{interim}}$, for $\iota=1,\ldots,I$, with analyses performed after every $n_{\text{interim}}$ newly enrolled patients until the maximum sample size $N$ is reached.

\subsubsection{Decision rules}

Let $\mathcal{R}_\iota \subseteq \mathcal{R}$ be the subset of active regimens and $\mathcal{A}_{\iota}^{\text{exp}}$ the set of active experimental arms at the moment of the $\iota$-th interim analysis due to possible early termination or addition of experimental arms during the trial.  Given the time-to-event nature of the toxicity and efficacy models, analyses can be performed without requiring complete follow-up for all enrolled patients. As a result, each patient contributes at least with partial outcome data at the time of the analysis. Let $\mathcal{D}_{T}\left(n_{\iota}\right)$ and $\mathcal{D}_{E}\left(n_{\iota}\right)$ denote respectively the collection of all  observed data for both the Bayesian PK-toxicity model and the model-averaged PK(/PD)-efficacy method after enrolling a total of $n_{\iota}$ patients. At each interim analysis, both Bayesian time-to-event toxicity and efficacy models are re-estimated using the currently available data and based on the updated posterior distributions of both models, the following decision rules defined in the master protocol are evaluated.

\par\noindent\textbf{Safety rule}
Bayesian safety monitoring is performed at each interim analysis by evaluating, for each experimental arm $A_o \in \mathcal{A}^{\mathrm{exp}}$, the posterior probabilities of acute and cumulative overdosing against a prespecified safety threshold $\xi_T$. Experimental arm $A_o$ is stopped for safety if
\begin{equation}
    \mathbb{P}\left(P_{T,1}(A_o)\geq\lambda_{\mathrm{AO}}\vert\mathcal{D}_T(n_\iota)\right)\geq\xi_T
    \quad\text{or}\quad
    \mathbb{P}\left(P_{T,G}(A_o)\geq\lambda_{\mathrm{CO}}\vert\mathcal{D}_T(n_\iota)\right)\geq\xi_T,
    \label{safety_rule}
\end{equation}
where $P_{T,1}(A_o)$ and $P_{T,G}(A_o)$ denote the acute and cumulative toxicity probabilities for arm $A_o$, respectively, as defined in Equation~\ref{cumulative_probability_of_toxicity}. The thresholds $\lambda_{\mathrm{AO}}$ and $\lambda_{\mathrm{CO}}$ define unacceptable acute and cumulative toxicity levels, respectively.

\par\noindent\textbf{Graduation rule} The objective is to determine whether any experimental arm $A_o$, with $o=1,\ldots,O$, demonstrates superior efficacy compared with the control arm $A_0$. To ensure that sufficient information is available, we require at least $N_{\text{min}}$ patients to have been enrolled in arm $A_o$. Let $ \Delta_E(A_o)=P_E(A_o)-P_E(A_0)$ denote the difference in efficacy probabilities between experimental arm $A_o$ and the control arm. Provided that at least $N_{\text{min}}$ patients have been randomized to $A_o$, the arm satisfies the superiority criterion if
\begin{equation} \mathbb{P}\left(\Delta_E(A_o)\geq\delta\vert\mathcal{D}_E(n_\iota)\right)\geq\xi_E,
\label{superiority_criterion}
\end{equation}
where $\delta$ is a clinically meaningful superiority margin and $\xi_E$ is the posterior probability threshold required to declare superiority. Thus, arm $A_o$ graduates in the set of recommended regimens if the posterior probability that its efficacy exceeds that of the control arm by at least $\delta$ is greater than or equal to $\xi_E$, providing that the regimen associated with the arm is safe.

\par\noindent\textbf{Futility rule} To avoid exposing patients to ineffective experimental treatments, an arm may be stopped for futility at an interim analysis. Using the same posterior superiority probability as in the graduation rule, experimental arm $A_o$ is stopped for futility if
\begin{equation}
\mathbb{P}\left(\Delta_E(A_o)\geq\delta\vert\mathcal{D}_E(n_\iota)\right)<\xi_F,
    \label{futility_rule}
\end{equation}
where $\xi_F < \xi_E$ is the prespecified futility threshold.

\par\noindent\textbf{Adding an experimental arm} Regimens initially included in the trial can be discontinued for excessive toxicity or futility, while accumulated PK/PD, toxicity, and efficacy data may help to identify a more promising regimen among those not initially investigated. Let $\mathcal{C}=\mathcal{R}\setminus\mathcal{R}_0$ denote the set of candidate regimens not included at trial initiation. At interim analysis $\iota$, posterior estimates from the PK--toxicity and model-averaged PK(/PD)--efficacy models are used, together with population PK/PD predictions, to estimate the toxicity and efficacy probabilities of each candidate regimen $r_m\in\mathcal{C}$. A candidate regimen is eligible for inclusion if its predicted acute and cumulative toxicity probabilities are below the thresholds $\lambda_{\text{AO}}$ and $\lambda_{\text{CO}}$, respectively, as specified in Eq.~\ref{safety_rule}, and if its predicted efficacy exceeds that of the control arm by at least the clinically meaningful margin, i.e., $\Delta_E(r_m)\geq \delta$, as required by Eq.~\ref{superiority_criterion}. If several candidate regimens satisfy these criteria, the regimen maximizing $\Delta_E(r_m)$ is selected. If no regimen is eligible, no arm is added and the rule is reassessed at the next interim analysis, unless the final analysis has been reached. Once selected, the new regimen is introduced as experimental arm $A_{O+1}$, and the experimental-arm set is updated to $\mathcal{A}^{\text{exp}}=\{A_o:o=1,\ldots,O+1\}$.
To limit the sample size in this early-phase setting, at most one new experimental arm may be added during the trial, even if several initial arms are discontinued. Although other mechanisms for adding arms could be considered, we focus on replacement following discontinuation for safety or futility.

\subsubsection{Bayesian Response-Adaptive Randomization (RAR)}

In the proposed platform design, Bayesian response-adaptive randomization (RAR) updates allocation probabilities at each interim analysis using posterior efficacy estimates from the BMA time-to-event model. The control-arm allocation is maintained at a prespecified level to preserve a concurrent comparator and support pairwise comparisons with the experimental arms \citep{hu2006theory, robertson2023response}. Among the experimental arms, allocation probabilities favor regimens with the highest posterior probability of being the most efficacious.
To avoid excessive adaptation based on limited early data, the randomization probabilities are stabilized as in Thall et al.\cite{thall2007practical}, with stronger preferential allocation allowed as information accumulates. When a new experimental arm is introduced, a temporary catch-up mechanism adapted from Yuan et al.\cite{yuan2016midas} increases its allocation probability until a sufficient sample size is reached. Further mathematical details are provided in Supplementary Materials S3.

\section{Simulation study}

Following the reporting guidance of Morris et al. \citep{morris2019using}, we describe the simulation study setting, objectives and scenarios, data-generating mechanisms and true parameter values for the PK/PD, toxicity, and efficacy outcomes, and the performance measures used to compare the platform trial designs.

\subsection{Scenarios}

Six main scenarios explore different acute toxicity, cumulative toxicity, and efficacy profiles to assess whether regimens are correctly graduated or stopped for safety or futility, including settings in which a new arm may be added after early discontinuation. For each scenario, 1,000 phase II platform trials are simulated, each comprising one control arm and three experimental arms initially selected from a set of ($M=9$) candidate regimens, denoted by $r_{m}$, $m=1,\ldots,9$. These regimens are obtained by combining fixed dose levels $D=\{d_1,d_2,d_3\}=\{100,200,300\}$ mg with schedules $S=\{\mathbf{s}_1,\mathbf{s}_2,\mathbf{s}_3\}=\{(0,24),(0,24,48,72),(0,24,48,72,96,120)\}$ hours. Although the motivating example considered treatment durations of 3--5 days, we selected 2-, 4-, 6-days regimens for the simulations to span this clinically relevant range and enable different regimens to deliver the same total dose, as for $r_6$ and $r_8$ defined in Table~\ref{regimens}. The reference regimen used for normalization in the different models is defined as $r_{*} = r_5 = (d_2, \mathbf{s}_2)$. The three regimens included at trial initiation are $r_2$, $r_6$, and $r_8$ (see Table~\ref{regimens}), assumed to have been selected following a phase I study in healthy volunteers. Of note, regimens $r_6$ and $r_8$ deliver the same total dose over the treatment period despite differing in dose level and administration schedule. Given the population PK/PD model of the scenario 1 with population parameters detailed in Table~\ref{true_scenario_parameters} and the choice of $r_{*}$, the reference exposure $x_*$ and $S_*$ used in the TACT-PK model are defined as $x_* = 27.5$ and $S_* = 130.6$.
\begin{table}[tb]
\centering
\begin{tabular}{lllllll}
\toprule
Name & Regimen & Dose level (mg) & Administration schedule (h) & Total dose (mg) \\
\midrule
\addlinespace[0.3em]
Regimen 1 & $r_1$ & 100 & 0, 24 & 200 \\
\textbf{Regimen 2} & $\boldsymbol{r_2}$ & \textbf{100} & \textbf{0, 24, 48, 72} & \textbf{400} \\
Regimen 3 & $r_3$ & 100 & 0, 24, 48, 72, 96, 120 & 600 \\
Regimen 4 & $r_4$ & 200 & 0, 24 & 400 \\
Regimen 5 & $r_5$ & 200 & 0, 24, 48, 72 & 800 \\
\textbf{Regimen 6} & $\boldsymbol{r_6}$ & \textbf{200} & \textbf{0, 24, 48, 72, 96, 120} & \textbf{1200} \\
Regimen 7 & $r_7$ & 300 & 0, 24 & 600 \\
\textbf{Regimen 8} & $\boldsymbol{r_8}$ & \textbf{300} & \textbf{0, 24, 48, 72} & \textbf{1200} \\
Regimen 9 & $r_9$ & 300 & 0, 24, 48, 72, 96, 120 & 1800 \\
\bottomrule
\end{tabular}
\caption{Candidate dosing regimens obtained by combining $K=3$ dose levels and $L=3$ dosing schedules and \textbf{in bold} the regimens used as starting experimental arms for the trial.}
\label{regimens}
\end{table}
The maximum number of patients included in the platform is fixed at $N=80$, corresponding to approximately $20$ patients per arm. When a new experimental arm is added, the maximum sample size is increased to $N=100$. Recruitment occurs at an average rate of $10$ patients per month, corresponding to $\lambda_R\simeq 0.0134$. The simulation study considers severe influenza in ICU-hospitalized patients, with toxicity defined as time to first adverse event and efficacy as time to ICU discharge. The toxicity and efficacy follow-up durations are both fixed at $6$ days, or $144$ hours. The decision parameters are fixed across scenarios as follows: $\lambda_{\text{AO}}=0.25$, $\lambda_{\text{CO}}=0.40$, $\xi_T=0.70$, $\delta=0.10$, $\xi_E=0.80$, and $\xi_F=0.20$. These values are used in the safety, graduation, and futility rules defined in Equations~\ref{safety_rule}, \ref{superiority_criterion}, and \ref{futility_rule}.

Table~\ref{scenario_settings_matrix} summarizes the settings of the six main scenarios described below and specifies, under the true scenario assumptions, which regimens should be stopped for safety or futility and which regimens should be graduated. 
Scenario 1, referred to as the reference scenario, is designed to evaluate the proportion of correct graduations for regimens $r_6$ and $r_8$ without allowing the addition of a new arm. All regimens are assumed to have low toxicity probabilities, while the superiority probabilities of $r_6$ and $r_8$ over the control are sufficiently high for both regimens to satisfy the graduation criterion. This scenario therefore evaluates the behavior of the PROP framework in a simple setting in which no regimen should trigger the safety or futility rules.
Scenario 2 is identical to Scenario 1, except that the PK model uses Michaelis--Menten elimination instead of linear elimination. This scenario investigates the impact of additional nonlinearity in the relationship linking regimen, exposure, activity, toxicity, and efficacy on the performance of dose-based and PK/PD-integrated methods.
Scenario 3 evaluates the ability of the different platform designs to accurately estimate toxicity and stop unsafe regimens. Regimen $r_6$ should be stopped for cumulative toxicity, whereas regimen $r_8$ should trigger the acute toxicity rule. Although both regimens meet the graduation criterion in terms of efficacy, neither should graduate because of its toxicity profile. Thus, no regimen should graduate if the acute and cumulative toxicity probabilities are accurately estimated.
Scenario 4 allows the addition of a new experimental arm after regimen $r_2$ meets the futility criterion. Among the candidate regimens, $r_9$ should be selected for inclusion because it satisfies the graduation criterion, as do regimens $r_6$ and $r_8$. Although $r_9$ has the highest true efficacy probability, its cumulative toxicity probability is relatively high while remaining below the toxicity threshold. Consequently, some simulated trials may graduate only $r_6$ and $r_8$.
Scenario 5 is identical to Scenario 3, except that a new experimental arm may be added following the early discontinuation of one of the initial arms. Under this scenario, $r_3$ is the correct candidate for inclusion because it is the only regimen among those not initially selected that truly satisfies the graduation criteria.
In Scenario 6, efficacy outcomes are generated using the PK--efficacy model to assess the ability of the proposed framework to identify settings in which the longitudinal PD biomarker is not a relevant efficacy predictor. Compared with Scenario 1, this scenario assumes larger differences in the true efficacy probabilities across regimens, while the remaining simulation settings are unchanged. Regimens $r_6$ and $r_8$ remain eligible for graduation, whereas $r_2$ should be stopped for futility because of its lower true probability of superiority.

\newcommand{\yesdot}{\ding{108}} 
\newcommand{\nodot}{\ding{109}}  
\begin{table}[tb]
\centering
\scriptsize
\resizebox{0.9\textwidth}{!}{
\begin{tabular}{lcccccc}
\toprule
\textbf{Scenario setting}
& \textbf{Scenario 1}
& \textbf{Scenario 2}
& \textbf{Scenario 3}
& \textbf{Scenario 4}
& \textbf{Scenario 5}
& \textbf{Scenario 6} \\
\midrule

\multicolumn{7}{l}{\textbf{PK/PD model}} \\
\hspace{1mm} Linear elimination & \yesdot & \nodot & \yesdot  & \yesdot  & \yesdot & \yesdot  \\
\hspace{1mm} Michaelis--Menten elimination & \nodot  & \yesdot  & \nodot & \nodot & \nodot  & \nodot \\


\multicolumn{7}{l}{\textbf{Efficacy data generation}} \\
\hspace{1mm} PK/PD-efficacy model & \yesdot & \yesdot & \yesdot  & \yesdot  & \yesdot & \nodot  \\
\hspace{1mm} PK-efficacy model & \nodot  & \nodot & \nodot   & \nodot   & \nodot  & \yesdot \\


\multicolumn{7}{l}{\textbf{True toxicity}} \\
\hspace{1mm} Regimen 2 & \nodot  & \nodot  & \nodot  & \nodot  & \nodot  & \nodot  \\
\hspace{1mm} Regimen 3 & \textemdash  & \textemdash  & \textemdash & \nodot  & \nodot  & \textemdash  \\
\hspace{1mm} Regimen 4 & \textemdash  & \textemdash  & \textemdash & \nodot  & \nodot  & \textemdash  \\
\hspace{1mm} Regimen 5 & \textemdash  & \textemdash  & \textemdash & \nodot  & \yesdot  & \textemdash  \\
\hspace{1mm} Regimen 6 & \nodot  & \nodot & \yesdot  & \nodot  & \yesdot & \nodot  \\
\hspace{1mm} Regimen 7 & \textemdash  & \textemdash  & \textemdash & \nodot  & \yesdot  & \textemdash  \\
\hspace{1mm} Regimen 8 & \nodot & \nodot & \yesdot  & \nodot & \yesdot & \nodot \\
\hspace{1mm} Regimen 9 & \textemdash  & \textemdash  & \textemdash & \nodot  & \yesdot  & \textemdash  \\


\multicolumn{7}{l}{\textbf{True futility}} \\
\hspace{1mm} Regimen 2 & \nodot  & \nodot  & \nodot  & \yesdot  & \nodot  & \yesdot  \\
\hspace{1mm} Regimen 3 & \textemdash  & \textemdash  & \textemdash & \nodot  & \nodot  & \textemdash  \\
\hspace{1mm} Regimen 4 & \textemdash  & \textemdash  & \textemdash & \yesdot  & \nodot  & \textemdash  \\
\hspace{1mm} Regimen 5 & \textemdash  & \textemdash  & \textemdash & \nodot  & \nodot  & \textemdash  \\
\hspace{1mm} Regimen 6 & \nodot  & \nodot & \nodot  & \nodot  & \nodot & \nodot  \\
\hspace{1mm} Regimen 7 & \textemdash  & \textemdash  & \textemdash & \nodot  & \nodot  & \textemdash  \\
\hspace{1mm} Regimen 8 & \nodot & \nodot & \nodot  & \nodot & \nodot & \nodot \\
\hspace{1mm} Regimen 9 & \textemdash  & \textemdash  & \textemdash & \nodot  & \nodot  & \textemdash  \\


\multicolumn{7}{l}{\textbf{True superiority}} \\
\hspace{1mm} Regimen 2 & \nodot  & \nodot  & \nodot  & \nodot  & \nodot  & \nodot  \\
\hspace{1mm} Regimen 3 & \textemdash  & \textemdash  & \textemdash & \nodot  & \yesdot  & \textemdash  \\
\hspace{1mm} Regimen 4 & \textemdash  & \textemdash  & \textemdash & \nodot  & \nodot  & \textemdash  \\
\hspace{1mm} Regimen 5 & \textemdash  & \textemdash  & \textemdash & \nodot  & \yesdot  & \textemdash  \\
\hspace{1mm} Regimen 6 & \yesdot  & \yesdot & \yesdot  & \yesdot  & \yesdot & \yesdot  \\
\hspace{1mm} Regimen 7 & \textemdash  & \textemdash  & \textemdash & \nodot  & \nodot  & \textemdash  \\
\hspace{1mm} Regimen 8 & \yesdot & \yesdot & \yesdot  & \yesdot & \yesdot & \yesdot \\
\hspace{1mm} Regimen 9 & \textemdash  & \textemdash  & \textemdash & \yesdot  & \yesdot  & \textemdash  \\


\multicolumn{7}{l}{\textbf{True graduation}} \\
\hspace{1mm} Regimen 2 & \nodot  & \nodot  & \nodot  & \nodot  & \nodot  & \nodot  \\
\hspace{1mm} Regimen 3 & \textemdash  & \textemdash  & \textemdash & \nodot  & \yesdot  & \textemdash  \\
\hspace{1mm} Regimen 4 & \textemdash  & \textemdash  & \textemdash & \nodot  & \nodot  & \textemdash  \\
\hspace{1mm} Regimen 5 & \textemdash  & \textemdash  & \textemdash & \nodot  & \nodot  & \textemdash  \\
\hspace{1mm} Regimen 6 & \yesdot  & \yesdot & \nodot  & \yesdot  & \nodot & \yesdot  \\
\hspace{1mm} Regimen 7 & \textemdash  & \textemdash  & \textemdash & \nodot  & \nodot  & \textemdash  \\
\hspace{1mm} Regimen 8 & \yesdot & \yesdot & \nodot  & \yesdot & \nodot & \yesdot \\
\hspace{1mm} Regimen 9 & \textemdash  & \textemdash  & \textemdash & \yesdot  & \nodot  & \textemdash  \\

\bottomrule
\end{tabular}
}
\caption{Summary of the simulation scenario settings, including the structural form of the PK/PD model and the true regimen-level decisions for safety, futility, superiority, and graduation. A regimen is graduated if it satisfies the superiority criterion and is not classified as toxic. Filled dots indicate that the setting or regimen-level decision is applied in the corresponding scenario, whereas empty dots indicate that it does not apply.}
\label{scenario_settings_matrix}
\end{table}

\subsection{Data-generating process and prior specifications}

The data-generating process follows the PROP framework and includes a population PK/PD model, a PK--toxicity model, and one of the two time-to-event efficacy models, depending on the scenario. PK/PD model misspecification was investigated in a previous study \cite{vuorinen2026comparative}; therefore, it is not further assessed here, and the present simulation study builds on those earlier findings.

PK data are generated using a one-compartment model with first-order absorption and linear elimination in all scenarios except Scenario 2 where elimination follows a Michaelis--Menten model. Oral administration was selected to strike a balance between the simplicity of intravenous administration and the complexity of modeling pulmonary deposition and absorption following nebulization, while ensuring model identifiability given the limited number of patients. For each simulated patient, true and error-contaminated concentration profiles are generated under any possible assigned regimen, allowing regimen-specific outcomes to be evaluated within the simulation framework. Longitudinal IL-6 data are simulated using an indirect-response turnover model \citep{dayneka1993comparison} in which drug exposure inhibits IL-6 production through an $I_{\max}$ function. IL-6 was selected because elevated and sustained concentrations have been associated with greater influenza severity, ICU admission, and poorer clinical outcomes \citep{lee2011cytokine,paquette2012interleukin,hagau2010clinical}. The parameters $I_{\max}$ and $\gamma$ are fixed to 1 to improve practical identifiability, simplifying the expression of the PD model. PD measurements are scheduled at 0, 3.3, 6.3, 27.2, 45.2, 66.2, 84.1, 105.1, 123, and 144 hours after the first dose and are collected independently of censoring induced by the efficacy endpoint. Some PK and PD sampling times coincide to limit the number of blood draws. The true population PK/PD parameter values are reported in Table~\ref{true_scenario_parameters}.

The true parameters of the PK--toxicity and PK(/PD)--efficacy models are selected in each scenario to generate the prespecified acute toxicity, cumulative toxicity, and efficacy profiles. The efficacy baseline hazard follows a Weibull distribution calibrated so that the control-arm efficacy probability increases gradually and reaches $50\%$ by the end of follow-up.

Prior distributions are calibrated to yield a prior effective sample size close to 1 or 2 using the crude Beta approximation of Morita et al.\cite{morita2008priorESS} as show in Supplementary Materials 4. Prior means are set to the true parameter values used in Scenario 1 and remain unchanged across scenarios. The same initial values are used for each estimation of the population PK/PD model, without carrying forward estimates from the preceding interim analysis.

\begin{table}[htbp]
\centering
\resizebox{\textwidth}{!}{
\begin{tabular}{llllll}
\toprule
Parameter & Description & True value & Distribution & SAEM initial value & Scenario \\
\midrule

\multicolumn{5}{l}{\textbf{Population PK/PD model}} \\
\addlinespace[0.3em]
$k_a$ & Absorption rate constant & 2 & Log-normal & 8 & 1-6 \\
$Cl$ & Clearance parameter & 5 & Log-normal & 25 & 1, 3-6 \\
$V$ & Apparent volume of distribution & 100 & Log-normal & 80 & 1-6 \\
$K_m$ & Michaelis--Menten constant & 2 & Log-normal & 5 & 2 \\
$V_{\max}$ & Maximum elimination rate & 10 & Log-normal & 25 & 2 \\
$k_{\mathrm{in}}$ & Biomarker production rate & 2 & Log-normal & 4 & 1-6 \\
$k_{\mathrm{out}}$ & Biomarker elimination rate & 0.2 & Log-normal & 0.8 & 1-6 \\
$IC_{50}$ & Concentration yielding 50\% of $I_{\max}$ & 1.3 & Log-normal & 3 & 1-6 \\
$\omega_{k_a}$ & Inter-individual variability on $k_a$ & 0.2 & & 1 & 1-6 \\
$\omega_{Cl}$ & Inter-individual variability on $Cl$ & 0.4 & & 1 & 1, 3-6 \\
$\omega_{V}$ & Inter-individual variability on $V$ & 0.4 & & 1 & 1-6 \\
$\omega_{K_m}$ & Inter-individual variability on $K_m$ & 0.4 & & 1 & 2 \\
$\omega_{V_{\max}}$ & Inter-individual variability on $V_{\max}$ & 0.4 & & 1 & 2 \\
$\omega_{k_{\mathrm{in}}}$ & Inter-individual variability on $k_{\mathrm{in}}$ & 0.4 & & 1 & 1-6 \\
$\omega_{k_{\mathrm{out}}}$ & Inter-individual variability on $k_{\mathrm{out}}$ & 0.3 & & 1 & 1-6 \\
$\omega_{IC_{50}}$ & Inter-individual variability on $IC_{50}$ & 0.4 & & 1 & 1-6 \\
$\boldsymbol{\sigma}^{\text{PK}}$ & Residual PK error parameters & $(0, 0.20)$ & & $(0, 1)$ & 1-6 \\
$\boldsymbol{\sigma}^{\text{PD}}$ & Residual PD error parameters & $(0, 0.20)$ & & $(0, 1)$ & 1-6 \\
\addlinespace[0.8em]

\toprule
Parameter & Description & True value & Distribution & Prior specification & Scenario \\
\midrule
\multicolumn{5}{l}{\textbf{Bayesian PK--toxicity time-to-event model}} \\
\addlinespace[0.3em]
$\alpha_{0}$ & Log acute toxicity hazard of & -2.97 & Normal & (-2.97, 1.5) & 1, 2, 6 \\
& the reference regimen & -1.82 & & & 3, 5 \\
$\alpha_{1}$ & Acute exposure-toxicity & 0 & Lognormal & (0, 1) & 1, 2, 6 \\
& association & 1.8 & & & 3, 5 \\
$\alpha_{2}$ & Log cumulative toxicity hazard & -3.11 & Normal & (-3.11, 2) & 1, 2, 6 \\
& of the reference regimen & -1.86 & & & 3, 5 \\

\addlinespace[0.8em]
\multicolumn{5}{l}{\textbf{Model-averaged Bayesian PK(/PD)--efficacy time-to-event method}} \\
\addlinespace[0.3em]
$\lambda_{e}$ & Weibull scale parameter & 190.9 & Lognormal & (5.25, 0.5) & 1-6 \\
$\nu_{e}$ & Weibull shape parameter & 1.3 & Normal & (1.3, 2) & 1-6 \\
$\beta_{1}$ & PK/PD-efficacy association parameter & 0.8 & Normal & (0.8, 1) & 1, 2, 6 \\
& & 1.1 & & & 3, 5 \\
& & 0.58 & & & 4 \\
$\beta_{2}$ & PK-efficacy association parameter & 0.6 & Normal & (0.6, 1) & 1-6 \\
$\mathbf{w}$ & Prior model weight ($\mathcal{M}_1$, $\mathcal{M}_2$) & $(1, 0)$ & & $(0.5, 0.5)$ & 1-5 \\
& & $(0, 1)$ & & & 6 \\

\bottomrule
\end{tabular}
}
\caption{Parameters of the PROP design framework, including the true values used in the data-generating process and the initial values or prior distributions used in the simulation study, according to the simulation scenario.}
\label{true_scenario_parameters}
\end{table}

\subsection{Comparison of platform designs}

PROP is compared with a dose-based platform design, hereafter referred to as DICE-doseEff, which uses DICE for toxicity \citep{ursino2022dice} and a Bayesian survival model with dosing history as a covariate for efficacy, as detailed in Supplementary Materials S2. This comparison is intended to assess the added value of incorporating PK/PD information rather than relying directly on dose. Both designs are evaluated with either Bayesian response-adaptive randomization or fixed randomization to assess the contribution of RAR, resulting in four platform designs.
A fifth comparator, called DICE-mAGILE, combines DICE for toxicity modelling with a modified version of the AGILE efficacy framework \citep{jaki2024seamless}. Compared with the original implementation, the model was modified to estimate the baseline hazard parameters and the superiority parameter for each experimental arm within a fully Bayesian Weibull framework, thereby accommodating the comparative setting of the simulation study instead of relying on the Cox proportional hazards model used by the original authors. This simpler benchmark compares each regimen separately with the control based on its posterior probability of superiority, rather than estimating regimen-specific efficacy probabilities. It is excluded from scenarios involving arm addition because it does not support the proposed arm-addition rule.

\subsection{Performance metrics}

Performance across simulation scenarios was assessed using several operating characteristics for each platform trial design. The primary performance metrics assessed the Bayesian toxicity-efficacy decision framework, including regimen-level graduation, safety stopping, and futility discontinuation rates; the proportions of scenario-exact, correct, and incorrect stopping (PSES, PCS, and PIS). A scenario-exact decision was defined as a simulated trial in which the set of regimens assigned a specific decision --- graduation, discontinuation for futility, or discontinuation for safety --- exactly matched the corresponding prespecified set of regimens for that scenario. A correct decision was defined as the assignment of a specific decision to at least one regimen belonging to the corresponding prespecified set, irrespective of whether the same decision was also assigned to additional regimens outside that set. An incorrect decision was defined as the assignment of a specific decision to at least one regimen outside the corresponding prespecified set. Consequently, the correct- and incorrect-decision indicators were not mutually exclusive. Among the three operating characteristics, PSES is the most informative measure of performance and also the most stringent, particularly in scenarios where an additional arm can be introduced, as this may result in relatively low values even for well-performing platform designs. A high PCS alone does not necessarily indicate good performance, as it can be inflated by selecting all regimens even when only one or two truly satisfy the decision criterion. In that case, PIS also increases because some regimens are selected incorrectly, whereas PSES remains low because it requires the selected set to match exactly the scenario-specific target set of regimens. PCS should therefore always be interpreted together with PIS. In scenarios 4 and 5, which permitted arm addition during the trial, we also recorded the proportion of trials in which each candidate regimen was added and the proportion in which the correct candidate was selected for addition.

For the population PK/PD model used in the two PK/PD-based designs, bias, standard error (SE), relative standard error (RSE), and coverage of the $95\%$ confidence intervals were used to assess parameter estimation. For the toxicity model, we reported the mean estimated acute and cumulative toxicity probabilities at each dosing interval, their Monte Carlo standard errors (MCSEs), and the coverage of the corresponding $95\%$ credible intervals.

For the efficacy model, we reported in Supplementary Materials S5 the mean estimated efficacy probabilities, their MCSEs, and credible interval coverage for the control and experimental arms. For the Bayesian model-averaging approach, model-selection performance was summarized by the proportion of simulated trials in which the posterior probability of the PK/PD--efficacy exceeded $0.5$. The PK/PD--efficacy model was expected to be selected more frequently in scenarios 1-5, whereas the PK--efficacy model was expected to be favored in scenario 6.

\subsection{Implementation}

All simulations were implemented using the R statistical software (version 4.2.3) \citep{r2023}. Simulation and estimation of the population PK/PD model were carried out using Monolix (version 2024R1) and Simulx (version 2024R1), respectively, through the \textit{lixoftConnectors} R package \citep{lixoftConnectors}, which provides an interface between R and the Lixoft software suite. Bayesian inference for the proposed toxicity and efficacy methods was performed using the \textit{rstan} package \citep{rstan}.

\section{Results}

\begin{figure}
    \centering
    \includegraphics[width=0.8\linewidth]{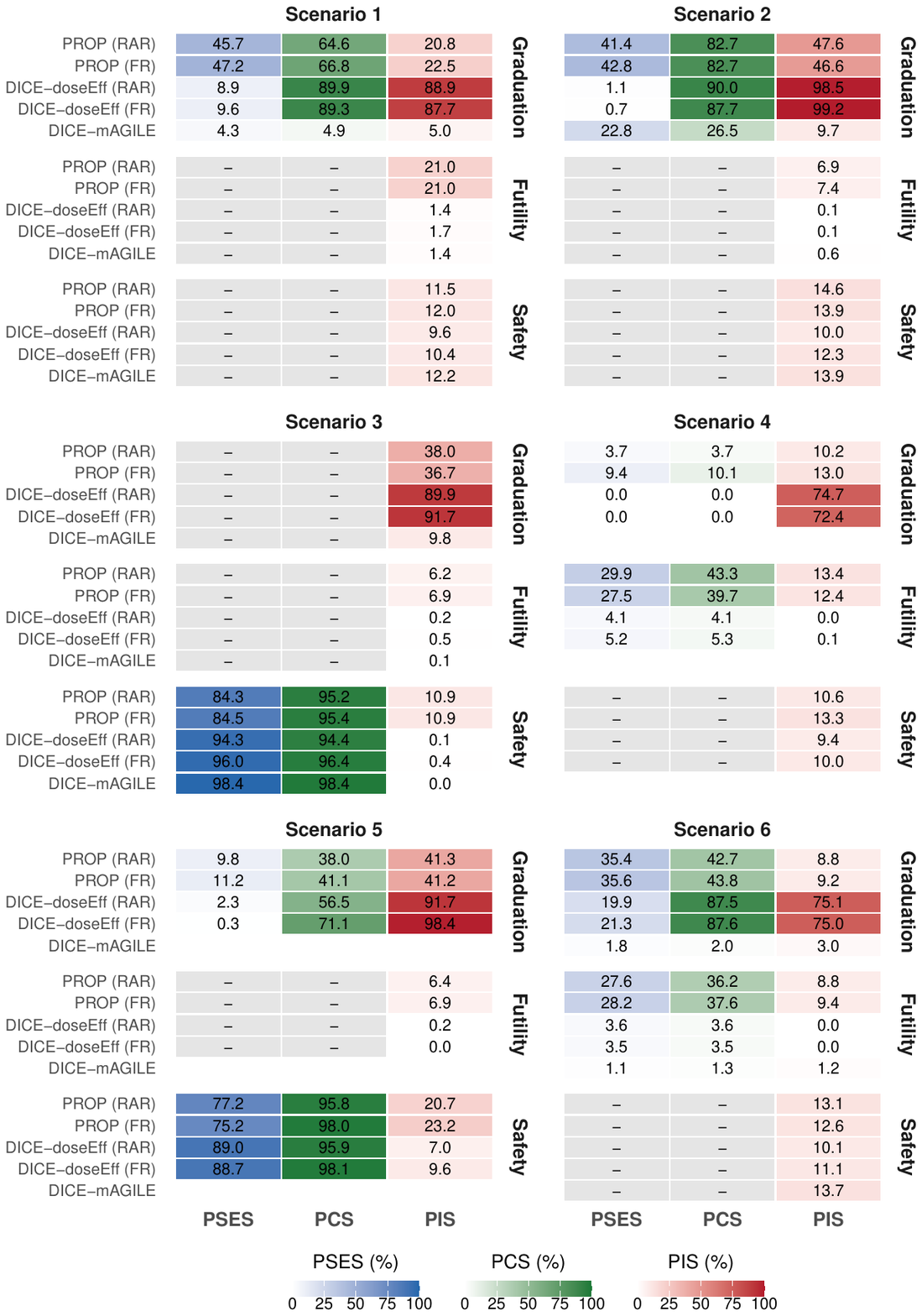}
    \caption{Decision (graduation, futility, safety) operating characteristics across platform trial designs and simulation scenarios. PSES = proportion of scenario-exact stopping; PCS = proportion of correct stopping; PIS = proportion of incorrect stopping. Color gradients indicate the magnitude of each metric. Non-applicable values are shown as grey cells containing an em dash.}
\label{decision_operating_characteristics_heatmap}
\end{figure}

\begin{figure}
    \centering
    \includegraphics[width=0.8\linewidth]{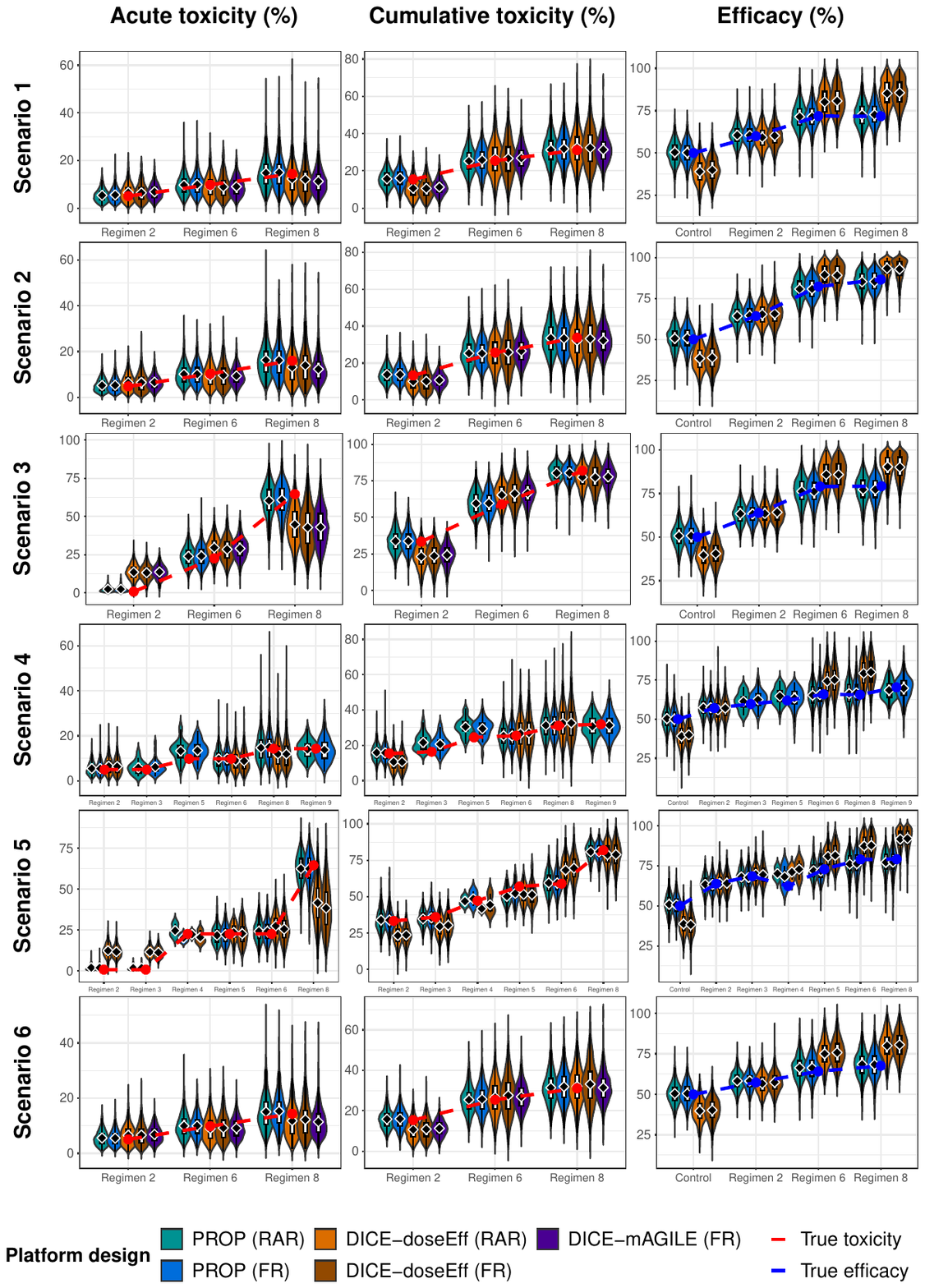}
    \caption{Estimated probabilities of acute toxicity, cumulative toxicity, and efficacy across the six simulation scenarios, with each row corresponding to one scenario and estimates shown relative to the true underlying probabilities. Violin plots grouped by platform trial design and regimen show the distribution of posterior mean estimates across simulated trials, and diamond markers indicate the overall mean.}
\label{toxicity_and_efficacy_posterior_estimates}
\end{figure}

Figure~\ref{decision_operating_characteristics_heatmap} summarizes the operating characteristics of the graduation, futility, and safety decision rules across the six simulated scenarios and platform trial designs. Regimen-level decision proportions are reported in Supplementary Materials S5. Visually, in Figure~\ref{decision_operating_characteristics_heatmap}, stronger blue (PSES) and green (PCS) tones indicate more favorable method performance, whereas darker red tones (PIS) reflect a higher proportion of incorrect decisions under the assessed trial rule and thus poorer performance of the platform design. In Figure~\ref{toxicity_and_efficacy_posterior_estimates}, the distributions of the estimated probabilities of acute toxicity, cumulative toxicity, and efficacy (from left to right) are shown using violin plots, with diamond markers indicating the overall mean for each platform design across the six simulation scenarios. Each estimate corresponds to the posterior mean of the respective probability obtained at the end of each simulated trial. The red and blue dots represent, respectively, the true mean probabilities of acute or cumulative toxicity and the true probabilities of efficacy specified for each simulation scenario, based on the true parameter values reported in Table~\ref{true_scenario_parameters}. Dashed lines of the same colors connect the associated true probabilities. Deviations of the diamond markers or violin-plot distributions from the dots or dashed lines indicate that the platform design did not accurately recover the underlying toxicity or efficacy probabilities.
In Scenarios 4 and 5, additional regimens are shown beyond those defining the initial experimental arms because of the arm-addition rule. Only regimens added at least once across the simulated trials are displayed. Since these candidate regimens were added only in a subset of simulated trials (i.e., trials in which the arm-addition criterion was met), their distributions are based on fewer estimates than those of the initial regimens and should be interpreted with greater caution, because they are subject to greater Monte Carlo variability. In Scenario 4, regimens 3, 5, and 9 were added at least once across the simulated trials under the PROP designs, whereas in Scenario 5, it occurred only for regimens 3, 4, and 5. Since some regimens shared the same dose per administration, they had identical true acute toxicity probabilities, resulting in flat segments of the true toxicity curve. The exact mean estimated probabilities of acute toxicity, cumulative toxicity, and efficacy are reported in Supplementary Materials S5.

Overall, the PROP designs provided the most balanced performance for regimen optimization, consistently achieving higher PSES for graduation than the dose-based alternatives while substantially reducing inappropriate graduation. In scenarios 1, 2, 5, and 6, DICE-doseEff frequently graduated incorrect regimens, with graduation PIS values ranging from approximately 75\% to almost 100\%, whereas the corresponding values were lower for PROP designs. DICE-mAGILE generally avoided inappropriate graduation, but often did so by rarely reaching either a graduation or a futility decision. In scenario 1, for example, the graduation PSES was approximately 46--47\% for PROP, compared with less than 10\% for the dose-based designs. Although DICE-doseEff achieved a higher PCS, close to 90\%, this was accompanied by a similarly high PIS, indicating that correct and incorrect regimens were frequently graduated within the same simulated trial. By contrast, PROP achieved a lower PCS of approximately 65--67\% but a substantially lower PIS of approximately 21--23\%, resulting in a more selective graduation behavior. A similar pattern was observed in scenario 2, where the graduation PSES remained above 40\% for PROP, while it decreased to approximately 1\% for DICE-doseEff, whose PIS approached 100\%. 

The advantage of PROP was also apparent in scenarios allowing arm addition. In scenario 4, only the PROP designs added the correct candidate regimen with substantial frequency, approximately 66--67\%, whereas DICE-doseEff never triggered arm addition. PROP also achieved substantially higher correct futility stopping for the initially evaluated ineffective regimen, with futility PCS values of approximately 40--43\%, compared with approximately 4--5\% for DICE-doseEff. In scenario 5, all eligible designs frequently added the correct candidate regimen, but PROP achieved higher scenario-exact graduation and substantially lower inappropriate graduation than DICE-doseEff. These results indicate that the PK/PD-informed framework was better able to use early discontinuation and arm addition jointly to adapt the set of regimens under evaluation.

Safety performance was more nuanced. In scenarios 3 and 5, the dose-based designs achieved higher scenario-exact safety rates and lower incorrect safety stopping than PROP. In scenario 3, for example, the safety PSES was approximately 84\% for PROP, compared with 94--96\% for DICE-doseEff and 98\% for DICE-mAGILE. Nevertheless, Figure~\ref{toxicity_and_efficacy_posterior_estimates} shows that the PROP designs generally recovered the underlying acute and cumulative toxicity probabilities more accurately, particularly for regimens at the extremes of the toxicity curve. For example, in scenario 3, PROP estimated the acute toxicity probability of the most toxic regimen at approximately 61\%, close to the true value of 64.6\%, whereas the dose-based designs estimated it at approximately 43--44\%. Thus, the less favorable safety operating characteristics observed for PROP in some scenarios did not reflect poorer estimation of regimen-specific toxicity, but rather differences in how posterior uncertainty was translated into stopping decisions.

The decision performance shown in Figure~\ref{decision_operating_characteristics_heatmap} was consistent with the estimation results in Figure~\ref{toxicity_and_efficacy_posterior_estimates}. The PROP designs closely recovered the true efficacy probabilities across experimental regimens and for the control arm. In scenario 1, for example, the true efficacy probability of regimen 8 was 71.7\%, and the PROP estimates were approximately 72\%, whereas both DICE-doseEff designs estimated it at approximately 85\%. Similarly, PROP recovered the control efficacy probability of 50\%, with estimates of 50.4\% under RAR and 50.1\% under FR, while DICE-doseEff underestimated it at approximately 38--40\%. This combination of overestimated experimental-arm efficacy and underestimated control efficacy explains the frequent inappropriate graduation observed with DICE-doseEff.

The Bayesian model-averaging framework also adapted appropriately to the assumed efficacy data-generating mechanism as shown in Supplementary Materials S5. In scenarios generated under the time-to-event PK/PD--efficacy model, the posterior probability assigned to this model was generally the largest. In scenario 1, its mean posterior probability was 71.0\% for PROP with RAR and 70.7\% with FR. However, in scenario 6, where efficacy was generated from the PK--efficacy model, the posterior probability of the PK/PD--efficacy model decreased to 21.4\% and 21.9\%, respectively, indicating that the BMA framework favored the PK--efficacy component. This flexibility contributed to the robustness of the PROP designs across different exposure--response mechanisms.

The population PK/PD model generally showed low bias and good precision for most structural and variability parameters under both RAR and FR. Greater uncertainty was observed for selected inter-individual variability parameters and, in the nonlinear-elimination scenario, for the Michaelis--Menten parameters. Nevertheless, the overall quality of the PK/PD estimation was sufficient to support accurate regimen-specific toxicity and efficacy estimation. Detailed results on regimen-level decisions, arm addition, toxicity and efficacy estimation, PK/PD parameter estimation, credible interval coverage, and model-selection performance are reported in Supplementary Materials S5.

Across modelling approaches, RAR and FR produced broadly similar operating characteristics and estimation performance. Differences between the two allocation strategies were generally small and did not alter the relative ranking of the platform trial designs. FR occasionally yielded slightly higher scenario-exact graduation, particularly in scenarios involving arm addition, whereas RAR sometimes provided marginally better futility or safety performance. However, neither allocation strategy showed a consistent advantage across scenarios, suggesting that the choice of the underlying toxicity and efficacy models had a substantially greater influence on trial performance than the choice between RAR and FR.

\section{Discussion}

We proposed a PK/PD-informed Regimen Optimization Platform (PROP) design that first estimates a population PK/PD model from longitudinal concentration and PD activity data and then incorporates the predicted individual exposure and PD response into toxicity and efficacy models. By accounting for inter-individual variability in PK and PD, this two-step framework aims to improve Bayesian decisions on regimen graduation, futility, and safety by providing an accurate characterization of the benefit--risk profile of each regimen. In our motivating setting of severe influenza requiring ICU admission, a platform design is relevant because patients are closely monitored, repeated clinical and biological measurements over time are available, and outcomes accrue over a relatively short hospitalization period. Although the therapeutic context is different from oncology, the high-risk population and the need to balance efficacy against treatment-related adverse events create similar conditions for the adaptive evaluation of multiple regimens. 

Using a simulation study, we evaluated whether PK/PD information could support adaptive regimen optimization in an early-phase platform trial. Across scenarios, the PROP design provided more reliable graduation and futility decisions than the dose-based alternatives, with higher scenario-exact graduation rates and fewer inappropriate graduations. It also accurately estimated regimen-specific acute and cumulative toxicity and arm-specific efficacy, improving characterization of the exposure--toxicity and exposure--activity--efficacy relationships. Together with its ability to add promising unexplored regimens, these features supported more informed regimen selection throughout the trial. The BMA framework further increased flexibility by assigning greater posterior probability to the efficacy model most consistent with the data-generating mechanism.

Although the dose-based designs achieved higher safety PSES and lower safety PIS in some scenarios, this did not reflect a more accurate characterization of the underlying regimen--toxicity relationship. Dose appeared to be an adequate population-level surrogate for exposure in the toxicity model, despite the PK-based data-generating mechanism. By contrast, dose alone was insufficient to characterize efficacy, resulting in poorer graduation and futility performance than PROP under both the PK--efficacy and PK/PD--efficacy mechanisms. DICE-mAGILE achieved moderate graduation performance only in scenario 2 and rarely reached graduation or futility decisions elsewhere. Its parsimonious efficacy model, based on a baseline hazard and an arm-specific hazard parameter without cumulative-dose or PK/PD summaries, required large treatment effects to establish superiority in the small-sample setting considered.

Unlike platform designs such as AGILE \citep{jaki2024seamless}, the present study did not primarily target calibration to conventional frequentist operating characteristics, such as the type-I error rate and sample size computation. Nevertheless, these properties can be evaluated and controlled through scenario-specific simulation-based calibration of the Bayesian decision thresholds. Our analyses therefore focused on the added value of PK/PD-informed modelling for adaptive dose and regimen exploration as accumulating information on the investigational treatment becomes available.

In our research context, a dosing regimen is defined as the combination of a fixed dose and an administration schedule, with all schedules nested within one another. The framework could be extended to more realistic regimens allowing dose modifications during treatment, variable interval lengths for non-nested schedules, or different routes of administration. All experimental treatments are evaluated over the first treatment cycle, reflecting the short-term clinical assessment typical of infectious diseases such as influenza. Both TACT-PK and DICE are less suited to settings involving many treatment cycles or long-term therapy. In both models, the cumulative toxicity component increases with repeated administrations and may eventually imply an excessive toxicity risk as the number of dosing intervals becomes large, even after normalization. In TACT-PK, the cumulative component is governed by a single parameter to preserve model parsimony and practical identifiability in the small-sample early-phase setting. The only hypothetical limitation is that if distinct PK profiles (e.g., one with high peaks but rapid elimination, another with a flat concentration curve but slow elimination) lead to the same AUC, the model assigns similar toxicity probabilities, even though such profiles in patients are unlikely to have the same toxicity. For the Bayesian efficacy models, if the assumption that efficacy in the experimental arms is at least as high as in the control arm cannot be justified or if the natural history of the disease is incorporated in the PK/PD model, the hazard dynamics of the control and experimental arms could instead be modelled separately. In this case, the baseline hazard would represent a latent temporal pattern shared across arms, with multiplicative arm-specific effects allowing deviations from this common pattern.

For the simulation study, we considered a limited number of experimental arms to keep the computational burden manageable. Some scenarios allowed the addition of one new arm selected from a predefined set of candidate regimens, preserving the adaptive nature of the platform while maintaining a realistic total sample size. A systematic review by Grissbach et al. \citep{griessbach2024characteristics} reported a median of one added arm (IQR 0-4), with 58.3\% of platforms adding at least one arm during their conduct, supporting the reasonableness of restricting the design to a single new arm. In settings involving a larger number of candidate dose--schedule combinations with a higher sample size, the arm-addition rule could be extended to all multiple new regimens to enter the trial.

For PK/PD modelling, a rich sampling scheme was used to achieve reliable estimation given the limited maximum number of patients included in the trial. Additionally, identical sampling times were assumed across patients, which is unlikely in practice. Under sparse sampling, heterogeneity in sampling times is preferable, as it can improve population-level parameter estimation by increasing coverage of the concentration--time and response profiles. Implementation of the PROP framework would thus require prospective evaluation of the PK/PD sampling design using Fisher-information or simulation-based optimal design methods \citep{nyberg2015methods, hooker2005simultaneous}. In practice, a population PK model is obtained via intensive phase I sampling and then updated using sparser observed data collected in the target patient population during phase II \citep{FDA2022populationPK}. As the PK model involves linear elimination, rich single-dose data in healthy volunteers during phase I could be sufficient to accurately capture the concentration--time curves induced by repeated administrations and obtain good estimates of the PK parameters. However, in our simulation study, repeated-dose PK data were assumed to be available from phase I, reflecting the dosing regimens evaluated before entry into phase II. In a real-life setting, the ten PK and PD measurements collected over six days that we simulated in our study without missingness can be difficult to obtain systematically in critically ill patients, despite some measurements being shared by both sampling processes. When a treatment-related adverse event occurs within a dosing interval, observed PK measurements remain available until the end of that interval, after which planned treatment and PK sampling are stopped. However, longitudinal PD measurements can still be obtained after treatment discontinuation to characterize the patient's biological trajectory. Overall, the choice of PK/PD settings in the simulation study can be viewed as an ideal case for the implementation of the proposed PROP framework. The maximum inhibition parameter $I_{\max}$ and the Hill coefficient were fixed to one. This choice was based on preliminary sensitivity analyses of more complex turnover $I_{\max}$ models and was made to ensure practical identifiability and adequate estimation performance. It limits the complexity of the exposure--activity and exposure--efficacy relationships that can be represented, including stronger saturation effects, delayed biomarker responses, and more heterogeneous exposure--efficacy patterns. Initial values for the PK/PD parameters were chosen to differ slightly from the true simulation values to assess whether the SAEM algorithm could reliably converge toward the true population parameters. Most of the simulated clinical trials yielded parameter estimates close to the true values.
For simplicity and computational feasibility, the simulation study did not include an infectious-disease component in the PK/PD model. Such a component could provide a more realistic representation of individual PD trajectories by accounting for the natural history and time-varying disease state. Mechanistic models of pro- and anti-inflammatory cytokine dynamics, including avian influenza, have been proposed in the literature \citep{zhang2019models}.
Finally, the compartmental model was primarily designed to compare regimens delivering the same total dose through different administration schedules. We used conventional monotonic exposure--toxicity and exposure--efficacy relationships in the simulation study, but more complex non-monotonic patterns could be evaluated as a new perspective to broaden the scope of this simulation study. An other possible extension would be to estimate the PK/PD, toxicity, and efficacy components simultaneously within a full Bayesian framework. Although such an approach would allow direct propagation of uncertainty across all model components, its repeated estimation at interim analyses would currently entail a substantial computational burden.

Bayesian model averaging was used to account for uncertainty about whether PK exposure or PD activity is the most relevant efficacy predictor. By combining the PK--efficacy and PK/PD--efficacy models, BMA avoids selecting a single pathway a priori and uses posterior model probabilities to quantify their relative support. Given the small sample size, these probabilities may remain inconclusive even when one model generated the efficacy outcomes. BMA therefore reflects predictive uncertainty rather than establishing a causal mediation mechanism. Future work could complement this framework with causal inference methods to investigate mediation and distinguish direct and indirect treatment effects along the dose--exposure--activity--efficacy pathway.

The PROP design was evaluated in a phase II dose-exploration setting following a preliminary phase I study in healthy volunteers. The framework could also be extended to a seamless phase I/II design in other therapeutic areas. In the first stage, a model-based dose-finding approach could identify regimens for further evaluation, which would then enter the randomized phase II platform stage.
Future work could also extend PROP to regimen optimization within patient subpopulations by incorporating relevant covariates into the population PK/PD model~\cite{fayette2026optimising}. This would allow heterogeneity in exposure--toxicity and exposure--activity--efficacy relationships to be characterized and Bayesian decision rules to be adapted to subgroup-specific benefit--risk profiles.

\section*{Author contributions}

Emmanuelle Comets and Moreno Ursino contributed equally to this study.

\section*{Acknowledgments}

This study was supported by a grant from Inserm and the French Ministry of Health (MESSIDORE 2022, reference number Inserm-MESSIDORE N\textsuperscript{o}94).

This work was provided with computer and storage resources by GENCI at IDRIS thanks to the grant 2025-AD010316444 on the supercomputer Jean Zay's the CSL partition.

\section*{Conflicts of interest}

The author(s) declared no potential conflicts of interest with respect to the research, authorship, and/or publication of this article.

\section*{Supporting information}

R scripts will be available on a GitHub repository, and supplementary materials are provided after the References section with additional methodological details, simulation results, and supporting analyses.

\bibliography{bibliography}

@article{gerard2022bayesian,
  title={Bayesian dose regimen assessment in early phase oncology incorporating pharmacokinetics and pharmacodynamics},
  author={Gerard, Emma and Zohar, Sarah and Thai, Hoai-Thu and Lorenzato, Christelle and Riviere, Marie-Karelle and Ursino, Moreno},
  journal={Biometrics},
  volume={78},
  number={1},
  pages={300--312},
  year={2022},
  publisher={Wiley Online Library},
  doi={https://doi.org/10.1111/biom.13433}
}

@article{ursino2022dice,
  title={DICE: A Bayesian model for early dose finding in phase I trials with multiple treatment courses},
  author={Ursino, Moreno and Biard, Lucie and Chevret, Sylvie},
  journal={Biometrical Journal},
  volume={64},
  number={8},
  pages={1486--1497},
  year={2022},
  publisher={Wiley Online Library}
}

@misc{projectoptimus,
  title={{Oncology Center of Excellence: Project Optimus}},
  author={FDA},
  year={2021},
  howpublished={\url{https://www.fda.gov/about-fda/oncology-center-excellence/project-optimus}},
  note={Accessed: 2023-05-23}
}

@manual{r2023,
    title = {{R: A Language and Environment for Statistical Computing}},
    author = {{R Core Team}},
    organization = {R Foundation for Statistical Computing},
    address = {Vienna, Austria},
    year = {2023},
    note = {\url{https://www.R-project.org/}},
}

@misc{rstan,
    title = {{RStan}: the {R} interface to {Stan}},
    author = {{Stan Development Team}},
    year = {2014},
    howpublished = {R package},
    note = {version 2.26.22, \url{https://mc-stan.org/}},
}

@article{morita2008priorESS,
author = {Morita, Satoshi and Thall, Peter F. and Müller, Peter},
title = {{Determining the Effective Sample Size of a Parametric Prior}},
journal = {Biometrics},
volume = {64},
number = {2},
pages = {595-602},
doi = {https://doi.org/10.1111/j.1541-0420.2007.00888.x},
year = {2008}
}

@article{shah2021drug,
  title={The drug-dosing conundrum in oncology—when less is more},
  author={Shah, Mirat and Rahman, Atiqur and Theoret, Marc R and Pazdur, Richard},
  journal={New England Journal of Medicine},
  volume={385},
  number={16},
  pages={1445--1447},
  year={2021},
  publisher={Mass Medical Soc}
}

@article{morris2019using,
  title={Using simulation studies to evaluate statistical methods},
  author={Morris, Tim P and White, Ian R and Crowther, Michael J},
  journal={Statistics in medicine},
  volume={38},
  number={11},
  pages={2074--2102},
  year={2019},
  publisher={Wiley Online Library}
}

@article{desmee2015nonlinear,
  title={Nonlinear mixed-effect models for prostate-specific antigen kinetics and link with survival in the context of metastatic prostate cancer: a comparison by simulation of two-stage and joint approaches},
  author={Desm{\'e}e, Sol{\`e}ne and Mentr{\'e}, France and Veyrat-Follet, Christine and Guedj, J{\'e}r{\'e}mie},
  journal={The AAPS journal},
  volume={17},
  number={3},
  pages={691--699},
  year={2015},
  publisher={Springer}
}

@article{mbogning2015joint,
  title={Joint modelling of longitudinal and repeated time-to-event data using nonlinear mixed-effects models and the stochastic approximation expectation--maximization algorithm},
  author={Mbogning, Cyprien and Bleakley, Kevin and Lavielle, Marc},
  journal={Journal of Statistical Computation and Simulation},
  volume={85},
  number={8},
  pages={1512--1528},
  year={2015},
  publisher={Taylor \& Francis}
}

@article{kerioui2020bayesian,
  title={Bayesian inference using Hamiltonian Monte-Carlo algorithm for nonlinear joint modeling in the context of cancer immunotherapy},
  author={Kerioui, Marion and Mercier, Francois and Bertrand, Julie and Tardivon, Coralie and Bruno, Ren{\'e} and Guedj, J{\'e}r{\'e}mie and Desm{\'e}e, Sol{\`e}ne},
  journal={Statistics in Medicine},
  volume={39},
  number={30},
  pages={4853--4868},
  year={2020},
  publisher={Wiley Online Library}
}

@book{hu2006theory,
  author    = {Hu, Feifang and Rosenberger, William F.},
  title     = {The Theory of Response-Adaptive Randomization in Clinical Trials},
  year      = {2006},
  publisher = {Wiley-Interscience},
  address   = {Hoboken, NJ},
  series    = {Wiley Series in Probability and Statistics},
  isbn      = {978-0-471-65396-7},
  doi       = {10.1002/047005588X}
}

@article{barker2009spy,
  title={I-SPY 2: an adaptive breast cancer trial design in the setting of neoadjuvant chemotherapy},
  author={Barker, AD and Sigman, CC and Kelloff, Gary J and Hylton, NM and Berry, Donald A and Esserman, LJs},
  journal={Clinical Pharmacology \& Therapeutics},
  volume={86},
  number={1},
  pages={97--100},
  year={2009},
  publisher={Wiley Online Library}
}

@article{thall2007practical,
  title={Practical Bayesian adaptive randomisation in clinical trials},
  author={Thall, Peter F and Wathen, J Kyle},
  journal={European Journal of Cancer},
  volume={43},
  number={5},
  pages={859--866},
  year={2007},
  publisher={Elsevier}
}

@article{robertson2023response,
  title={Response-adaptive randomization in clinical trials: from myths to practical considerations},
  author={Robertson, David S and Lee, Kim May and L{\'o}pez-Kolkovska, Boryana C and Villar, Sof{\'\i}a S},
  journal={Statistical science: a review journal of the Institute of Mathematical Statistics},
  volume={38},
  number={2},
  pages={185},
  year={2023}
}

@article{jaki2024seamless,
  title={A seamless Phase I/II platform design with a time-to-event efficacy endpoint for potential COVID-19 therapies},
  author={Jaki, Thomas and Barnett, Helen and Titman, Andrew and Mozgunov, Pavel},
  journal={Statistical Methods in Medical Research},
  volume={33},
  number={11-12},
  pages={2115--2130},
  year={2024},
  publisher={SAGE Publications Sage UK: London, England}
}

@Misc{lixoftConnectors,
    title        = {lixoftConnectors: R connectors for Lixoft Suite},
    author       = {{Lixoft}},
    year         = {2024},
    howpublished = {R package},
    note         = {Version 2024.1}
}

@article{korn2023dose,
  title={Dose optimization during drug development: whether and when to optimize},
  author={Korn, Edward L and Moscow, Jeffrey A and Freidlin, Boris},
  journal={JNCI: Journal of the National Cancer Institute},
  volume={115},
  number={5},
  pages={492--497},
  year={2023},
  publisher={Oxford University Press}
}

@article{adaptive2019adaptive,
  title={Adaptive platform trials: definition, design, conduct and reporting considerations},
  author={{The Adaptive Platform Trial Coalition}},
  journal={Nature Reviews Drug Discovery},
  volume={18},
  number={10},
  pages={797--807},
  year={2019},
  publisher={Nature Publishing Group UK London}
}

@article{wong2019estimation,
  title={Estimation of clinical trial success rates and related parameters},
  author={Wong, Chi Heem and Siah, Kien Wei and Lo, Andrew W},
  journal={Biostatistics},
  volume={20},
  number={2},
  pages={273--286},
  year={2019},
  publisher={Oxford University Press}
}

@article{sun202290,
  title={Why 90\% of clinical drug development fails and how to improve it?},
  author={Sun, Duxin and Gao, Wei and Hu, Hongxiang and Zhou, Simon},
  journal={Acta Pharmaceutica Sinica B},
  volume={12},
  number={7},
  pages={3049--3062},
  year={2022},
  publisher={Elsevier}
}

@article{alexander2018adaptive,
  title={Adaptive global innovative learning environment for glioblastoma: GBM AGILE},
  author={Alexander, Brian M and Ba, Sujuan and Berger, Mitchel S and Berry, Donald A and Cavenee, Webster K and Chang, Susan M and Cloughesy, Timothy F and Jiang, Tao and Khasraw, Mustafa and Li, Wenbin and others},
  journal={Clinical Cancer Research},
  volume={24},
  number={4},
  pages={737--743},
  year={2018},
  publisher={American Association for Cancer Research}
}

@article{griffiths2020agile,
  title={AGILE-ACCORD: A Randomized, Multicentre, Seamless, Adaptive Phase I/II Platform Study to Determine the Optimal Dose, Safety and Efficacy of Multiple Candidate Agents for the Treatment of COVID-19: A structured summary of a study protocol for a randomised platform trial},
  author={Griffiths, Gareth and Fitzgerald, Richard and Jaki, Thomas and Corkhill, Andrea and Marwood, Ellice and Reynolds, Helen and Stanton, Louise and Ewings, Sean and Condie, Susannah and Wrixon, Emma and others},
  journal={Trials},
  volume={21},
  number={1},
  pages={544},
  year={2020},
  publisher={Springer}
}

@article{mu2021bayesian,
  title={A Bayesian adaptive phase I/II platform trial design for pediatric immunotherapy trials},
  author={Mu, Rongji and Pan, Haitao and Xu, Guoying},
  journal={Statistics in Medicine},
  volume={40},
  number={2},
  pages={382--402},
  year={2021},
  publisher={Wiley Online Library}
}

@article{mu2024bayesian,
  title={A Bayesian latent-subgroup platform design for dose optimization},
  author={Mu, Rongji and Zhan, Xiaojiang and Tang, Rui and Yuan, Ying},
  journal={Biometrics},
  volume={80},
  number={3},
  pages={ujae093},
  year={2024},
  publisher={Oxford University Press}
}

@article{shi2026bayesian,
  title={A Bayesian phase I/II platform design with survival efficacy endpoint for dose optimization},
  author={Shi, Xian and Zhao, Jiangyan and Xu, Jin and Mu, Rongji},
  journal={Statistical Methods in Medical Research},
  pages={09622802261449367},
  year={2026},
  publisher={SAGE Publications Sage UK: London, England}
}

@article{mu2022bayesian,
  title={A Bayesian phase I/II platform design for co-developing drug combination therapies for multiple indications},
  author={Mu, Rongji and Xu, Jin and Tang, Rui and Kopetz, Scott and Yuan, Ying},
  journal={Statistics in Medicine},
  volume={41},
  number={2},
  pages={374--389},
  year={2022},
  publisher={Wiley Online Library}
}

@article{yuan2016midas,
  title={MIDAS: a practical Bayesian design for platform trials with molecularly targeted agents},
  author={Yuan, Ying and Guo, Beibei and Munsell, Mark and Lu, Karen and Jazaeri, Amir},
  journal={Statistics in medicine},
  volume={35},
  number={22},
  pages={3892--3906},
  year={2016},
  publisher={Wiley Online Library}
}

@article{rossoni2018phase,
  title={Phase I/II platform trial design in pediatric cancer in relapse},
  author={Rossoni, C and Geoerger, B and Paoletti, X},
  journal={Revue d'{\'E}pid{\'e}miologie et de Sant{\'e} Publique},
  volume={66},
  pages={S130},
  year={2018},
  publisher={Elsevier}
}

@article{bateman2017dian,
  title={The DIAN-TU Next Generation Alzheimer's prevention trial: adaptive design and disease progression model},
  author={Bateman, Randall J and Benzinger, Tammie L and Berry, Scott and Clifford, David B and Duggan, Cynthia and Fagan, Anne M and Fanning, Kathleen and Farlow, Martin R and Hassenstab, Jason and McDade, Eric M and others},
  journal={Alzheimer's \& Dementia},
  volume={13},
  number={1},
  pages={8--19},
  year={2017},
  publisher={Wiley Online Library}
}

@article{ritchie2016development,
  title={Development of interventions for the secondary prevention of Alzheimer's dementia: the European Prevention of Alzheimer's Dementia (EPAD) project},
  author={Ritchie, Craig W and Molinuevo, Jos{\'e} Luis and Truyen, Luc and Satlin, Andrew and Van der Geyten, Serge and Lovestone, Simon},
  journal={The Lancet Psychiatry},
  volume={3},
  number={2},
  pages={179--186},
  year={2016},
  publisher={Elsevier}
}

@article{angus2020remap,
  title={The REMAP-CAP (randomized embedded multifactorial adaptive platform for community-acquired pneumonia) study. Rationale and design},
  author={Angus, Derek C and Berry, Scott and Lewis, Roger J and Al-Beidh, Farah and Arabi, Yaseen and van Bentum-Puijk, Wilma and Bhimani, Zahra and Bonten, Marc and Broglio, Kristine and Brunkhorst, Frank and others},
  journal={Annals of the American Thoracic Society},
  volume={17},
  number={7},
  pages={879--891},
  year={2020},
  publisher={Oxford University Press}
}

@article{bongard2018antivirals,
  title={Antivirals for influenza-Like Illness? A randomised Controlled trial of Clinical and Cost effectiveness in primary CarE (ALIC4 E): the ALIC4 E protocol},
  author={Bongard, Emily and Van der Velden, Alike W and Cook, Johanna and Saville, Ben and Beutels, Philippe and Munck Aabenhus, Rune and Brugman, Curt and Chlabicz, Slawomir and Coenen, Samuel and Colliers, Annelies and others},
  journal={BMJ open},
  volume={8},
  number={7},
  pages={e021032},
  year={2018},
  publisher={British Medical Journal Publishing Group}
}

@article{kaplan2015focus4,
  title={The FOCUS4 design for biomarker stratified trials},
  author={Kaplan, Richard},
  journal={Chinese clinical oncology},
  volume={4},
  number={3},
  pages={35--35},
  year={2015},
  publisher={AME Publishing Company}
}

@article{piantadosi1996improved,
  title={Improved designs for dose escalation studies using pharmacokinetic measurements},
  author={Piantadosi, Steven and Liu, Guanghan},
  journal={Statistics in medicine},
  volume={15},
  number={15},
  pages={1605--1618},
  year={1996},
  publisher={Wiley Online Library}
}

@article{ursino2017dose,
  title={Dose-finding methods for phase I clinical trials using pharmacokinetics in small populations},
  author={Ursino, Moreno and Zohar, Sarah and Lentz, Frederike and Alberti, Corinne and Friede, Tim and Stallard, Nigel and Comets, Emmanuelle},
  journal={Biometrical Journal},
  volume={59},
  number={4},
  pages={804--825},
  year={2017},
  publisher={Wiley Online Library}
}

@article{vuorinen2026comparative,
  title={A comparative analysis of Phase I dose-finding designs incorporating pharmacokinetics information},
  author={Vuorinen, Axel and Comets, Emmanuelle and Ursino, Moreno},
  journal={The American Statistician},
  volume={80},
  number={2},
  pages={286--300},
  year={2026},
  publisher={Taylor \& Francis}
}

@article{yuan2026pharmacometrics,
  title={Pharmacometrics-Enabled DOse OPtimization (PEDOOP) for seamless phase I-II trials in oncology},
  author={Yuan, Shijie and Huang, Zhanbo and Liu, Jiaxin and Ji, Yuan},
  journal={Journal of Biopharmaceutical Statistics},
  volume={36},
  number={1},
  pages={59--78},
  year={2026},
  publisher={Taylor \& Francis}
}

@article{gunhan2020bayesian,
  title={A Bayesian time-to-event pharmacokinetic model for phase I dose-escalation trials with multiple schedules},
  author={G{\"u}nhan, Burak K{\"u}rsad and Weber, Sebastian and Friede, Tim},
  journal={Statistics in Medicine},
  volume={39},
  number={27},
  pages={3986--4000},
  year={2020},
  publisher={Wiley Online Library}
}

@article{micallef2022exposure,
  title={Exposure driven dose escalation design with overdose control: concept and first real life experience in an oncology phase I trial},
  author={Micallef, Sandrine and Sostelly, Alexandre and Zhu, Jiawen and Baverel, Paul G and Mercier, Francois},
  journal={Contemporary Clinical Trials Communications},
  volume={26},
  pages={100901},
  year={2022},
  publisher={Elsevier}
}

@article{su2022semi,
  title={A semi-mechanistic dose-finding design in oncology using pharmacokinetic/pharmacodynamic modeling},
  author={Su, Xiao and Li, Yisheng and M{\"u}ller, Peter and Hsu, Chia-Wei and Pan, Haitao and Do, Kim-Anh},
  journal={Pharmaceutical statistics},
  volume={21},
  number={6},
  pages={1149--1166},
  year={2022},
  publisher={Wiley Online Library}
}

@article{yang2024extended,
  title={An extended Bayesian semi-mechanistic dose-finding design for phase I oncology trials using pharmacokinetic and pharmacodynamic information},
  author={Yang, Chao and Li, Yisheng},
  journal={Statistics in medicine},
  volume={43},
  number={4},
  pages={689--705},
  year={2024},
  publisher={Wiley Online Library}
}

@article{fayette2026optimising,
  title={{Optimising covariate allocation at design stage using Fisher Information Matrix for Non-Linear Mixed Effects Models in pharmacometrics}},
  author={Fayette, Lucie and Brendel, Karl and Mentr{\'e}, France},
  journal={Computational Statistics \& Data Analysis},
  pages={108379},
  year={2026},
  publisher={Elsevier}
}

@article{kazantzidis2026mediation,
  title={Mediation Analysis With Bayesian Nonlinear Joint Models: Evaluation of the Treatment Causal Pathways Between Tumor Growth Kinetics and Overall Survival},
  author={Kazantzidis, Georgios and Mercier, Francois and Rondeau, Virginie},
  journal={Statistics in Medicine},
  volume={45},
  number={13-14},
  pages={e70574},
  year={2026},
  publisher={Wiley Online Library}
}

@article{iuliano2018estimates,
  title={Estimates of global seasonal influenza-associated respiratory mortality: a modelling study},
  author={Iuliano, A Danielle and Roguski, Katherine M and Chang, Howard H and Muscatello, David J and Palekar, Rakhee and Tempia, Stefano and Cohen, Cheryl and Gran, Jon Michael and Schanzer, Dena and Cowling, Benjamin J and others},
  journal={The Lancet},
  volume={391},
  number={10127},
  pages={1285--1300},
  year={2018},
  publisher={Elsevier}
}

@misc{WHO2025seasonalinfluenza,
  author  = {{World Health Organization}},
  title   = {Influenza (Seasonal)},
  year    = {2025},
  howpublished = {\url{https://www.who.int/news-room/fact-sheets/detail/influenza-(seasonal)}},
  note         = {Accessed 17 September 2026}
}

@book{WHO2024influenza,
  author    = {{World Health Organization}},
  title     = {Clinical Practice Guidelines for Influenza},
  year      = {2024},
  address   = {Geneva},
  publisher = {World Health Organization},
  isbn      = {978-92-4-009775-9},
  url       = {https://www.ncbi.nlm.nih.gov/books/NBK607904/},
  note      = {Electronic version. Print ISBN: 978-92-4-009776-6}
}

@article{jefferson2014neuraminidase,
  title={Neuraminidase inhibitors for preventing and treating influenza in healthy adults and children},
  author={Jefferson, Tom and Jones, Mark A and Doshi, Peter and Del Mar, Chris B and Hama, Rokuro and Thompson, Matthew and Spencer, Elizabeth A and Onakpoya, Igho and Mahtani, Kamal R and Nunan, David Nunan and others},
  journal={Sao Paulo Medical Journal},
  volume={132},
  number={4},
  pages={256--257},
  year={2014},
  publisher={SciELO Brasil}
}

@article{bartlett1995community,
  title={Community-acquired pneumonia},
  author={Bartlett, John G and Mundy, Linda M},
  journal={New England Journal of Medicine},
  volume={333},
  number={24},
  pages={1618--1624},
  year={1995},
  publisher={Mass Medical Soc}
}

@misc{remapcap2026oseltamivir,
  author  = {{REMAP-CAP}},
  title   = {{REMAP-CAP} results: Oseltamivir (anti-viral medication) for critically ill patients with influenza infection},
  year    = {2026},
  howpublished  = {\url{https://www.remapcap.eu/2026/06/10/remap-cap-results-oseltamivir-for-critically-ill-patients-with-influenza-infection/}},
  note = {Accessed 17 September 2026}
}

@article{may2019paracetamol,
  title={Paracetamol absorption test to detect poor enteric absorption of oseltamivir in intensive care unit patients with severe influenza: a pilot study},
  author={May, Faten and Peytavin, Gilles and Fourati, Slim and Pressiat, Claire and Carteaux, Guillaume and Razazi, Keyvan and Mekontso Dessap, Armand and de Prost, Nicolas},
  journal={Intensive Care Medicine},
  volume={45},
  number={10},
  pages={1484--1486},
  year={2019},
  publisher={Springer}
}

@article{cezard2026cis,
  title={Cis-aconitate therapy protects against influenza mortality by dual targeting of viral polymerase and ERK/AKT/NF-$\kappa$B signaling},
  author={Cezard, Adeline and Brea-Diakite, D{\'e}borah and Vasseur, Virginie and Wacquiez, Alan and Gonzalez, Loic and Le Goffic, Ronan and Da Costa, Bruno and Tinard, Ambre and Fouquenet, Delphine and Heumel, S{\'e}verine and others},
  journal={EMBO Molecular Medicine},
  volume={18},
  number={3},
  pages={1023--1053},
  year={2026},
  publisher={Springer}
}

@article{hay2014clinical,
  title={Clinical development success rates for investigational drugs},
  author={Hay, Michael and Thomas, David W and Craighead, John L and Economides, Celia and Rosenthal, Jesse},
  journal={Nature biotechnology},
  volume={32},
  number={1},
  pages={40--51},
  year={2014},
  publisher={Nature Publishing Group US New York}
}

@article{arrowsmith2011phase,
  title={Phase III and submission failures 2007--2010},
  author={Arrowsmith III, J},
  journal={Nat. Rev. Drug Discov},
  volume={10},
  number={87},
  pages={10--1038},
  year={2011}
}

@article{royston2003novel,
  title={Novel designs for multi-arm clinical trials with survival outcomes with an application in ovarian cancer},
  author={Royston, Patrick and Parmar, Mahesh KB and Qian, Wendi},
  journal={Statistics in medicine},
  volume={22},
  number={14},
  pages={2239--2256},
  year={2003},
  publisher={Wiley Online Library}
}

@article{wason2016some,
  title={Some recommendations for multi-arm multi-stage trials},
  author={Wason, James and Magirr, Dominic and Law, Martin and Jaki, Thomas},
  journal={Statistical methods in medical research},
  volume={25},
  number={2},
  pages={716--727},
  year={2016},
  publisher={SAGE Publications Sage UK: London, England}
}

@article{hagau2010clinical,
  title={Clinical aspects and cytokine response in severe H1N1 influenza A virus infection},
  author={Hagau, Natalia and Slavcovici, Adriana and Gonganau, Daniel N and Oltean, Simona and Dirzu, Dan S and Brezoszki, Erika S and Maxim, Mihaela and Ciuce, Constantin and Mlesnite, Monica and Gavrus, Rodica L and others},
  journal={Critical care},
  volume={14},
  number={6},
  pages={R203},
  year={2010},
  publisher={Springer}
}

@article{zhang2019models,
  title={Models of cytokine dynamics in the inflammatory response of viral zoonotic infectious diseases},
  author={Zhang, Wenjing and Jang, Sophia and Jonsson, Colleen B and Allen, Linda JS},
  journal={Mathematical medicine and biology: a journal of the IMA},
  volume={36},
  number={3},
  pages={269--295},
  year={2019},
  publisher={Oxford University Press}
}

@article{lee2011cytokine,
  title={Cytokine response patterns in severe pandemic 2009 H1N1 and seasonal influenza among hospitalized adults},
  author={Lee, Nelson and Wong, Chun Kwok and Chan, Paul KS and Chan, Martin CW and Wong, Rity YK and Lun, Samantha WM and Ngai, Karry LK and Lui, Grace CY and Wong, Bonnie CK and Lee, Sharon KW and others},
  journal={PloS one},
  volume={6},
  number={10},
  pages={e26050},
  year={2011},
  publisher={Public Library of Science San Francisco, USA}
}

@article{paquette2012interleukin,
  title={Interleukin-6 is a potential biomarker for severe pandemic H1N1 influenza A infection},
  author={Paquette, Stephane G and Banner, David and Zhao, Zhen and Fang, Yuan and Huang, Stephen SH and Le{\'o}n, Alberto J and Ng, Derek CK and Almansa, Raquel and Martin-Loeches, Ignacio and Ramirez, Paula and others},
  journal={PloS one},
  volume={7},
  number={6},
  pages={e38214},
  year={2012},
  publisher={Public Library of Science San Francisco, USA}
}

@article{dayneka1993comparison,
  title={Comparison of four basic models of indirect pharmacodynamic responses},
  author={Dayneka, Natalie L and Garg, Varun and Jusko, William J},
  journal={Journal of pharmacokinetics and biopharmaceutics},
  volume={21},
  number={4},
  pages={457--478},
  year={1993},
  publisher={Springer}
}

@article{griessbach2024characteristics,
  title={Characteristics, progression, and output of randomized platform trials: a systematic review},
  author={Griessbach, Alexandra and Sch{\"o}nenberger, Christof Manuel and Taji Heravi, Ala and Gloy, Viktoria and Agarwal, Arnav and Hallenberger, Tim Jonas and Schandelmaier, Stefan and Janiaud, Perrine and Amstutz, Alain and Covino, Manuela and others},
  journal={JAMA Network Open},
  volume={7},
  number={3},
  pages={e243109},
  year={2024}
}

@misc{FDA2022populationPK,
  author       = {{U.S. Food and Drug Administration}},
  title        = {Population Pharmacokinetics: Guidance for Industry},
  year         = {2022},
  howpublished = {\url{https://www.fda.gov/regulatory-information/search-fda-guidance-documents/population-pharmacokinetics}},
  note         = {Final Level 1 Guidance, Center for Drug Evaluation and Research and Center for Biologics Evaluation and Research; accessed 21 September 2026}
}

@article{nyberg2015methods,
  title={Methods and software tools for design evaluation in population pharmacokinetics--pharmacodynamics studies},
  author={Nyberg, Joakim and Bazzoli, Caroline and Ogungbenro, Kay and Aliev, Alexander and Leonov, Sergei and Duffull, Stephen and Hooker, Andrew C and Mentr{\'e}, France},
  journal={British journal of clinical pharmacology},
  volume={79},
  number={1},
  pages={6--17},
  year={2015},
  publisher={Wiley Online Library}
}

@article{hooker2005simultaneous,
  title={Simultaneous population optimal design for pharmacokinetic-pharmacodynamic experiments},
  author={Hooker, Andrew and Vicini, Paolo},
  journal={The AAPS journal},
  volume={7},
  number={4},
  pages={76},
  year={2005},
  publisher={Springer}
}

\setcounter{section}{0}
\renewcommand{\thesection}{S\arabic{section}}
\section{Mathematical framework for the Bayesian model-averaged time-to-event PK(/PD)-efficacy method}

For model $\mathcal{M}_{1}$, efficacy is assumed to be linked to PD activity with individual longitudinal predictions obtained from the estimated popPK/PD model during the first stage of the two-stage efficacy framework used to inform the time-to-event endpoint in the second stage. The link between the time-varying covariate and the efficacy hazard function is represented by the predicted log fold-change from baseline of the PD biomarker for individual $i$ at time $t$, denoted $\log \left( \frac{y_{(i)}^{\text{PD}}(t)}{y_{(i)}^{\text{PD}}(0)} \right)$. Based on a Bayesian survival model, the individual hazard function for patient $i$ at time $t>0$ is written as
\begin{equation}
    h_{\mathcal{M}_1} \left( t ; \mathbf{y}^{\text{PD}}_{(i)} \right) = h_0(t;\lambda_1,\nu_1) \exp \Biggl\{\beta_1 \log \left( \frac{y_{(i)}^{\text{PD}}(t)}{y_{(i)}^{\text{PD}}(0)} \right) \Biggl\}^{\gamma_{i}},
\end{equation}
where $h_0(t;\lambda_1,\nu_1)$ is the baseline hazard function following a Weibull distribution with parameters $(\lambda_1, \nu_1)$, such that $h_0(t;\lambda,\nu) = \lambda \nu t^{\nu-1}$, and the coefficient $\beta_1$ is quantifying the strength of the association between the PD biomarker summary measure and the efficacy event. If $\beta_1$ is estimated to be close to zero, there is limited evidence that the PD biomarker is a relevant predictor for drug efficacy. Baseline covariates were not included in the efficacy model in our simulation study, but could be readily incorporated to allow for more complex survival modeling with patient subpopulations. The individual survival function can be defined with
\begin{equation}
      S_{\mathcal{M}_1} \left( t ; \mathbf{y}^{\text{PD}}_{(i)} \right) = \mathbb{P} \left( O_{E, i} > t \mid \mathbf{y}^{\text{PD}}_{(i)} \right) =  \exp \Bigl\{ -H_{\mathcal{M}_1}\left( t ; \mathbf{y}^{\text{PD}}_{(i)} \right) \Bigl\}.
\label{PKPD_efficacy_survival}
\end{equation}
Let $\mathcal{D}_E(n) = \left\{ \left(O_{E, i}, \delta_{E}, \mathbf{y}_{(i)}^\text{PD}, \mathbf{\tilde{x}}_{(i)} \right) : 1 \leq i \leq n \right\}$ denote the observed time-to-event efficacy data, the predicted indivual PD biomarker trajectories, and the predicted individual AUC profiles collected after the enrollment of $n$ patients in the platform trial. The likelihood of the PK/PD-efficacy time-to-event model can be formulated as
\begin{align}
\begin{split}
    L \left( \mathcal{D}_E(n) \mid \boldsymbol{\theta}_{E,1}, \mathcal{M}_1 \right) &= \prod_{i=1}^{n} \pi \left( O_{E, i}, \delta_{E, i}, \mathbf{y}_{(i)}^\text{PD} \mid \boldsymbol{\theta}_{E,1}, \mathcal{M}_1 \right) \\
    &= \prod_{i=1}^{n} \left[ h_{\mathcal{M}_1} \left( T_{E, i} ; \mathbf{y}^{\text{PD}}_{(i)} \right) S_{\mathcal{M}_1} \left( T_{E, i} ; \mathbf{y}^{\text{PD}}_{(i)} \right) \right]^{\delta_{E, i}} S_{\mathcal{M}_1} \left( C_{E, i} ; \mathbf{y}^{\text{PD}}_{(i)} \right)^{1 - \delta_{E, i}}.
\label{PKPD_efficacy_likelihood_SM}
\end{split}
\end{align}
Therefore, the marginal posterior distribution of parameters $\boldsymbol{\theta}_{E}$ for $\mathcal{M}_1$ is derived from Equation \ref{PKPD_efficacy_likelihood_SM} to obtain:
\begin{equation}
    \pi \left( \boldsymbol{\theta}_{E}, \mathcal{M}_1 \mid \mathcal{D}_E(n) \right) \propto L \left( \mathcal{D}_E(n) \mid \boldsymbol{\theta}_{E}, \mathcal{M}_1 \right)  \pi(\boldsymbol{\theta}_{E} \mid \mathcal{M}_1).
\label{PKPD_efficacy_posterior}
\end{equation}

For model $\mathcal{M}_2$, efficacy is no longer linked to PD activity but instead directly to PK exposure, which captures longitudinally the cumulative dosing history over time of each regimen. Based on the same Bayesian survival model as $\mathcal{M}_1$ but using the AUC at time $t$ as the time-dependent covariate, the PK-efficacy hazard function is defined as
\begin{equation}
     h_{\mathcal{M}_2} \left( t ; \mathbf{\tilde{x}}_{(i)} \right) =  h_0(t;\lambda_2,\nu_2) \exp \left\{\beta_2 \log \left( 1 + \frac{\tilde{x}_{(i)}(t)}{\tilde{x}_*} \right) \right\}^{\gamma_{i}},
\end{equation}
where $$\tilde{x}_{(i)}(t) = \int_{0}^{t}{f^{\text{PK}} \left(u, r_{(i)}, \widehat{\boldsymbol{\psi}}^{\text{PK}}_{(i)} \right)\,du}, \quad 0 \leq t \leq \bar{T}_E,$$ denotes the predicted AUC from treatment initiation (i.e.\ $t_0 = 0$) up to time $t$ for patient $i$ receiving regimen $r_{(i)}$ and $\tilde{x}_*$ the AUC of the reference regimen over the complete follow-up period. The corresponding predicted AUC trajectory over the efficacy follow-up period is $\mathbf{\tilde{x}}_{(i)} = \left\{ \tilde{x}_{(i)}(t) : 0 \leq t \leq \bar{T}_E \right\}$. If the coefficient $\beta_2$ linking PK exposure to efficacy is estimated to be close to zero, PK exposure profiles have limited predictive value for drug efficacy. The individual survival function is expressed as
\begin{equation}
    S_{\mathcal{M}_2}(t ; \mathbf{\tilde{x}}_{(i)}) = \mathbb{P} \left( O_{E, i} > t \mid \mathbf{\tilde{x}}_{(i)} \right) =  \exp \Bigl\{ -H_{2} \left( t ; \mathbf{\tilde{x}}_{(i)} \right) \Bigl\}.
\end{equation}
The likelihood for $\mathcal{M}_2$ and the marginal posterior distribution of $\boldsymbol{\theta}_{E,2}$ can be written as in Equation \ref{PKPD_efficacy_likelihood_SM} by changing the index of the model and the efficacy predictor.

The posterior model probability of model $\mathcal{M}_e$ for $e \in \{1,2\}$, is given by
\begin{equation}
    w_e = \pi(\mathcal{M}_e \mid \mathcal{D}_{E}(n)) = \frac{\pi(\mathcal{D}_{E}(n) \mid \mathcal{M}_e) \pi(\mathcal{M}_e)}{\sum_{l=1}^{2} \pi(\mathcal{D}_{E}(n) \mid \mathcal{M}_l) \pi(\mathcal{M}_l)},
\end{equation}
where
\begin{equation*}
    \pi(\mathcal{D}_{E}(n) \mid \mathcal{M}_e) = \int_{\boldsymbol{\Theta}_{E, e}} \pi(\mathcal{D}_{E}(n) \mid \boldsymbol{\theta}_{E,e}, \mathcal{M}_e) \pi(\boldsymbol{\theta}_{E,e} \mid \mathcal{M}_e) \, d\boldsymbol{\theta}_{E,e}
\end{equation*}
is the marginal likelihood under model $\mathcal{M}_e$ and $\pi(\mathcal{M}_e)$ is the prior model probability representing the prior uncertainty regarding which model is the most likely to match the data-generating mechanism for efficacy. The BMA posterior distribution is represented as a weighted sum of the model-specific posterior distributions:
\begin{equation}
    \pi(\boldsymbol{\theta}_E \mid \mathcal{D}_{E}) = \sum_{e=1}^2 w_e \pi \left( \boldsymbol{\theta}_{E,e} \mid \mathcal{D}_{E}, \mathcal{M}_e \right) .
\end{equation}

\section{Mathematical framework for the Bayesian time-to-event efficacy dose-efficacy model}

As an alternative to the proposed model-averaged efficacy method using PK/PD information in the PROP design, a dose-based efficacy model is implemented with the cumulative dose used as covariate. The individual hazard function for patient $i$ at time $t>0$ is written as
\begin{equation}
    h \left(t, r_{(i)} \right) = h_0(t;\lambda,\nu) \exp \Biggl\{\theta \log \left( 1 + \frac{D_{(i),g}}{D_{*}} \right) \Biggl\}^{\gamma_{i}}, \quad t \in \mathcal{I}_g
\end{equation}
where $h_0(t;\lambda,\nu)$ is a Weibull baseline hazard, $D_{(i),g}$ denotes the cumulative dose for patient $i$ available at dosing interval $g$ when $t \in \mathcal{I}_g$, and $D_{*}$ is the reference cumulative dose at time $\bar{T}_{E}$. The indicator $\gamma_i$ specifies whether patient $i$ has been randomized to an experimental arm, and the coefficient $\theta$ quantifies the effect of the log-transformed cumulative dose on the hazard of an efficacy event. This efficacy model provides a direct comparison between using dosing history alone and incorporating individual exposure or activity trajectories as covariate in the Bayesian survival model with the same probability structure as the proposed efficacy model in the PROP design.

\section{Mathematical framwork of Bayesian Response-Adaptive Randomization (RAR)}

Let $\rho_{\iota}(A_o)$ be the probability that a patient is randomized to $A_o$ after the $\iota$-th interim analysis. Two constraints are enforced on the randomization probabilities to ensure $\sum_{o=0}^O \rho_{\iota}(A_o) = 1$ and $\rho_{\iota}(A_o) \in [0,1]$ for any $A_o \in \mathcal{A}_{\iota}^{\text{exp}}$. An indicator $a_{i}$ is used to register the treatment arm to which the $i$-th patient has been randomized with $a_{i} = 0$ if the $i$-th patient is randomized to the control arm $A_0$ and $a_{i} = o$ if the patient is randomized to one of the experimental arms $A_o \in \mathcal{A}^{\text{exp}}$. The randomization probability for the control arm $A_0$ is fixed to the proportion of active arms in the trial in the context of RAR to maintain a concurrent control throughout the platform trial, which is essential for platform trials spanning over a long period of time. It also ensures that a sufficient number of patients are allocated to the control arm to preserve adequate power for pairwise comparisons with experimental arms for the graduation and the safety rule \citep{hu2006theory, robertson2023response}. Therefore, the randomization probability of $A_0$ at the $\iota$-th interim analysis is defined by
\begin{equation}
    \rho_{\iota}(A_0) = \frac{1}{|\mathcal{A}_{\iota}^{\text{exp}}| +1},
\end{equation}
where $|\mathcal{A}_{\iota}^{\text{exp}}|$ is the cardinal number of the subset $\mathcal{A}_{\iota}^{\text{exp}}$. The Bayesian RAR for the experimental arms is directly driven by the BMA time-to-event efficacy model that links the PD biomarker or the PK exposure with the efficacy endpoint since the randomization adaptation is based on the posterior probability of efficacy. Unlike the graduation and futility rules, which use pairwise comparisons between each experimental arm and the control, the RAR probabilities are defined to preferentially randomize patients to the experimental arm(s) achieving the best efficacy, i.e.\ to the experimental arm(s) with the highest posterior probability of being the most efficacious among all experimental regimens. At the $\iota$-th interim analysis, $R_{\iota}(A_o)$ is defined as the posterior probability that $A_o$ is the most efficacious among all the experimental arms, such that
\begin{equation}
    R_{\iota}(A_o) = \mathbb{P}\!\left( P_{E}(A_o) = \max_{\omega \in \{1,\ldots,|\mathcal{A}_{\iota}^{\text{exp}}|\}} P_{E}(A_\omega)\; \middle|\; \mathcal{D}_{E}\!\left(n_{\iota}\right)\right).
\end{equation}
However, there might be little data available early in the platform trial to reliably estimate the efficacy due to the limited number of enrolled patients and possible long follow-up period. Using the stabilization of randomization probabilities suggested by \cite{thall2007practical}, the newly enrolled $\left(n_{\iota}+1\right)$-th patient is assigned to arm $A_o \in \mathcal{A}^{\text{exp}}$ with probability 
\begin{equation}
    \rho_{\iota}(A_o) = [1 - \rho_\iota(A_0)]\frac{R_{\iota}(A_o)^{q(n_\iota)}}{\sum_{\omega=1}^{|\mathcal{A}_{\iota}^{\text{exp}}|} R_{\iota}(A_\omega)^{q(n_\iota)}}, \quad q(n_\iota) = \frac{n_{\iota}}{2N},
\label{RAR_without_arm_addition}
\end{equation}
where $q(n_\iota)$ is a tuning parameter increasing with the number of enrolled patients to mitigate the variability of $R_{\iota}(A_o)$ in the early stage of the trial. When $q(n_\iota) = 0$, the randomization is balanced. As the number of patients enrolled in the trial increases, the randomization probabilities increasingly favor the most efficacious experimental arms, since the influence of the tuning parameter gradually decreases. If the trial reaches the maximum sample size, it eventually converges to $q(N) = \frac{1}{2}$ at the final analysis, resulting in the strongest degree of response-adaptive allocation. This approach reduces the risk of assigning patients to experimental arms with a true probability of efficacy that is inferior to that of the other arms due to high initial variability in the efficacy estimates. However, in our framework, a new experimental arm may enter the platform trial after trial initiation and equation~\ref{RAR_without_arm_addition} does not incorporate a rule to allow catch up in sample size compared to arms investigated since the beginning of the trial. Without such an adjustment, substantial sample size imbalances across arms can undermine the accuracy and stability of the estimates of $R_\iota$. When a new arm $A_{O+1}$ is added, a catch-up rule based on the one implemented in \cite{yuan2016midas} is applied: the randomization probability of $A_{O+1}$ is temporarily upweighted to reduce the sample size gap. Let $N_{\text{min}}$ be the minimum number of patients that must be enrolled in a newly added arm before the catch-up rule is lifted. For the new arm at the $\iota$-th interim analysis, the randomization probability $\rho_{\iota}(A_{O+1})$ is defined as
\begin{equation}
    \rho_{\iota}(A_{O+1}) = \max\left(\frac{R_{\iota}(A_{O+1})^{q(n_\iota)}}{\sum_{\omega=1}^{|\mathcal{A}_{\iota}^{\text{exp}}|} R_{\iota}(A_\omega)^{q(n_\iota)}}, \frac{1}{|\mathcal{A}_{\iota}^{\text{exp}}|} \right),
\label{catch_up_rule}
\end{equation}
and remains upweighted in this manner until the number of patients enrolled in the arm is greater or equal to $N_{\text{min}}$. Therefore, combining Equations \ref{RAR_without_arm_addition} and \ref{catch_up_rule}, the randomization probability of any experimental arm $A_{o} \in \mathcal{A}^{\text{exp}}_\iota \setminus \{A_{O+1}\}$, where the newly added experimental arm is excluded until it accrued $N_{\text{min}}$ patients, for the $\left(n_{\iota}+1\right)$-th patient is given by
\begin{equation}
    \rho_{\iota}(A_o) = \left[1 - \rho_\iota(A_0) - \rho_{\iota}(A_{O+1}) \right]\frac{R_{\iota}(A_o)^{q(n_\iota)}}{\sum_{\omega=1}^{|\mathcal{A}_{\iota}^{\text{exp}}|-1} R_{\iota}(A_\omega)^{q(n_\iota)}}.
\label{final_RAR_rule}
\end{equation}

\section{Sensitivity analyses for model and prior calibration}

\begin{figure}
    \centering
    \includegraphics[width=0.9\linewidth]{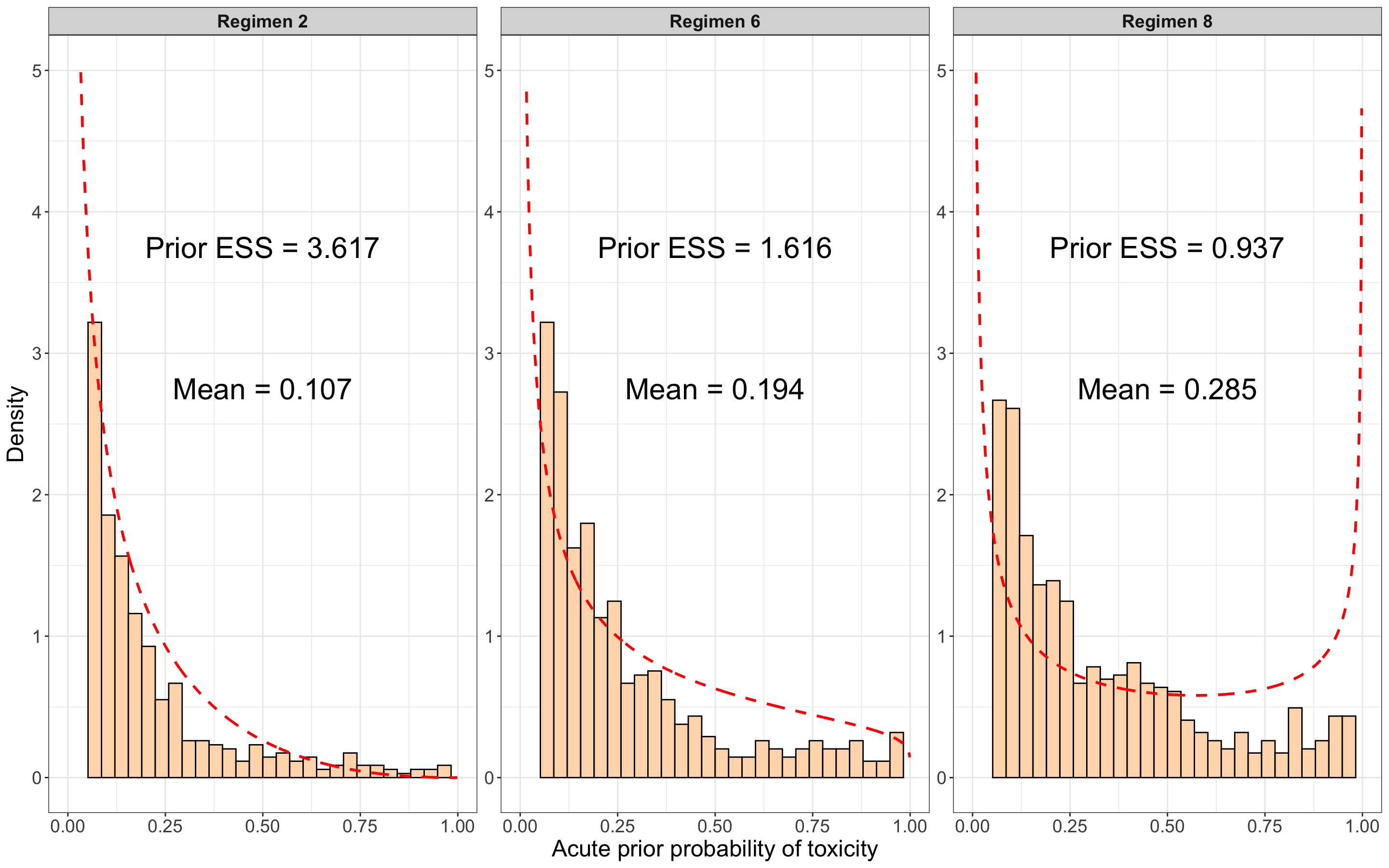}
    \includegraphics[width=0.9\linewidth]{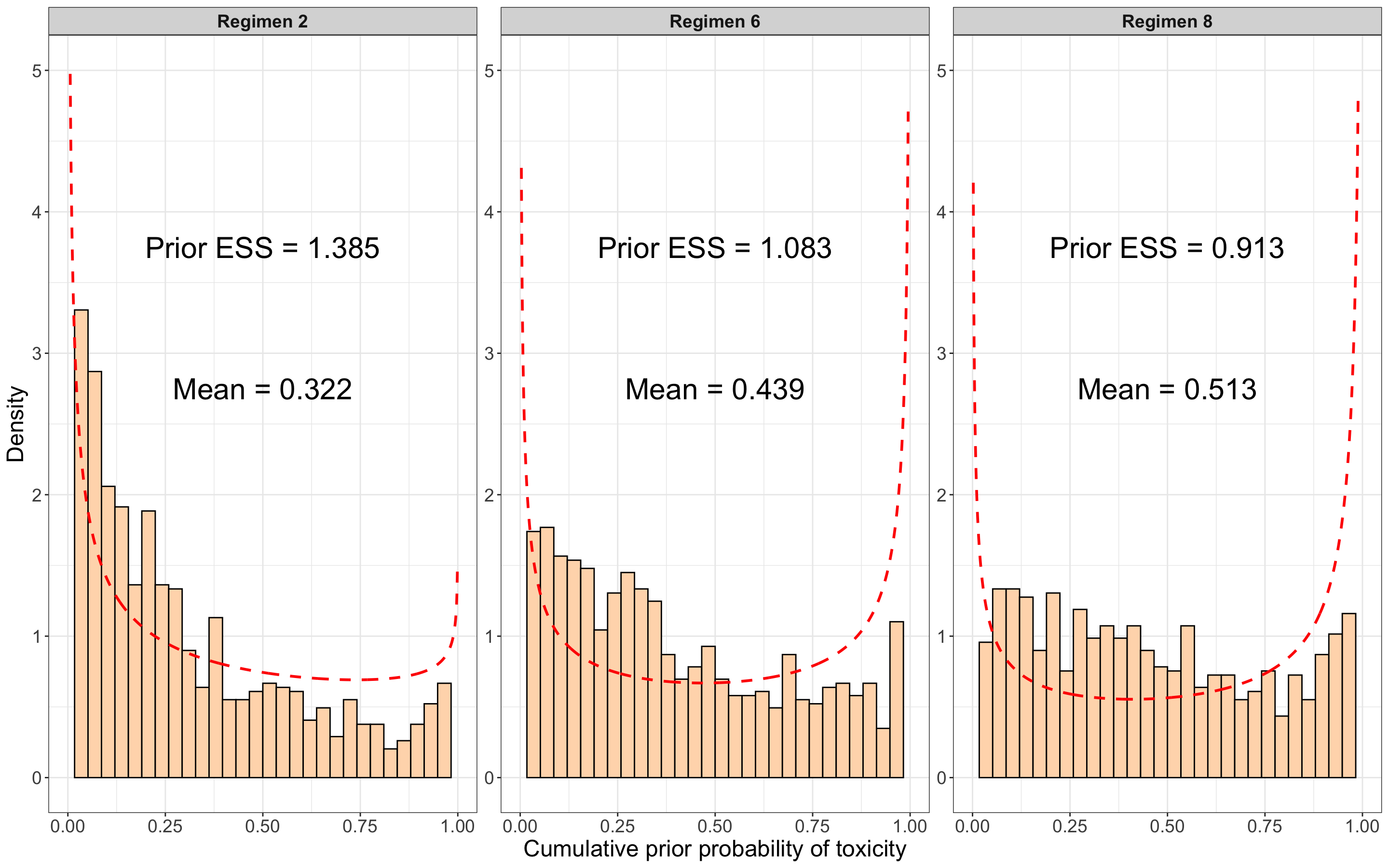}
    \caption{Histograms of prior predictive probabilities of acute and cumulative toxicity under the TACT-PK model for each starting regimen. Prior means and prior effective sample sizes (ESS) are reported for each distribution. The prior ESS is higher for acute toxicity than for cumulative toxicity, reflecting the smaller number of parameters in the acute toxicity model.}
    \label{acute_and_cum_tox_prior_predictive_p}
\end{figure}

\begin{figure}
    \centering
    \includegraphics[width=1\linewidth]{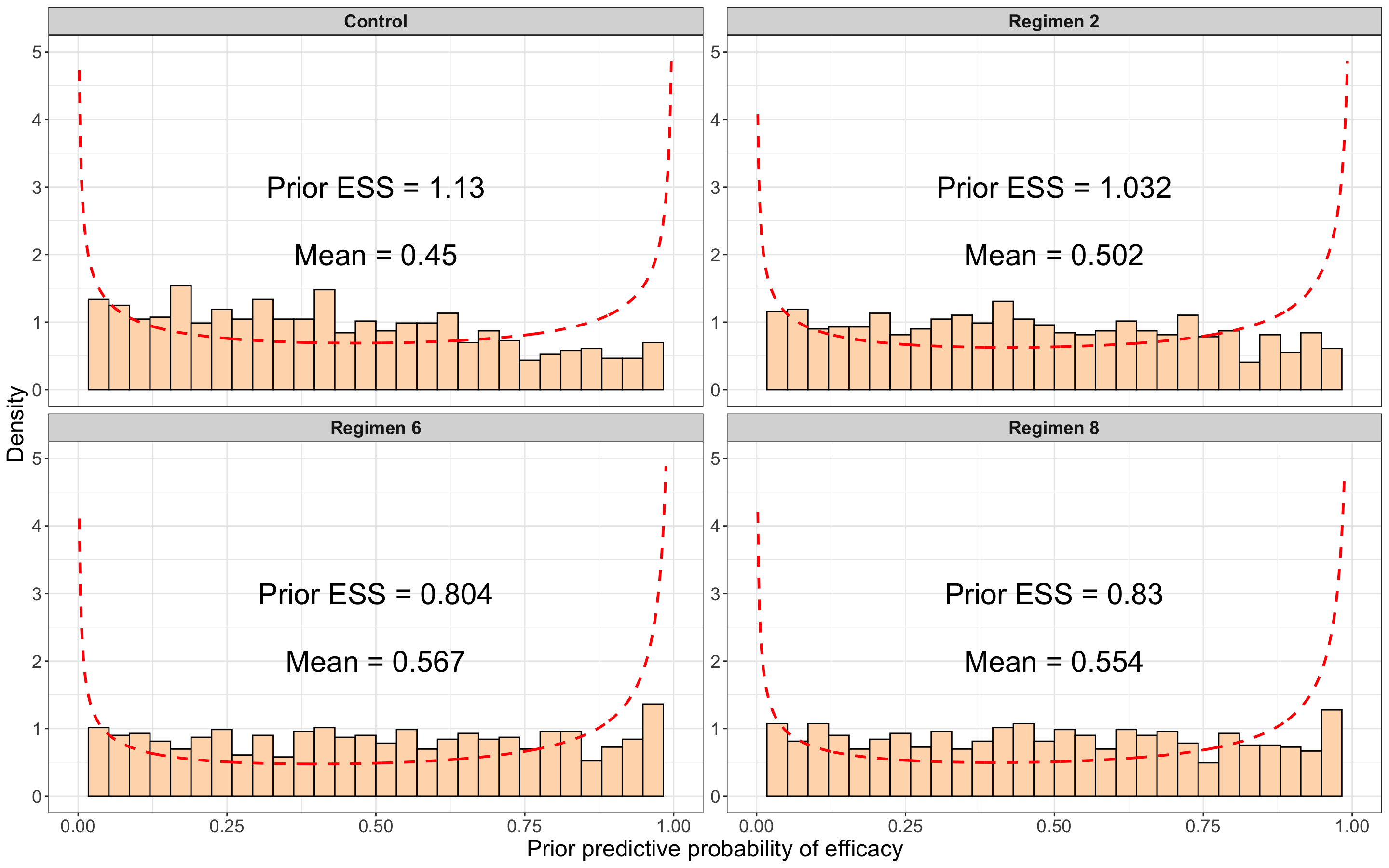}
    \caption{Histogram of prior predictive probabilities of efficacy under the the PK/PD-efficacy time-to-event model for each starting arm. Prior means and prior effective sample sizes (ESS) are reported for each distribution.}
    \label{eff_p_ess}
\end{figure}

\section{Scenario-specific simulation results}

\subsection{Scenario 1}

\begin{table}[H]
\centering
\resizebox{\textwidth}{!}{
\begin{tabular}[t]{lcccccc}
\toprule
\multicolumn{1}{l}{\textbf{Scenario 1}} & \multicolumn{3}{c}{\textbf{Graduation stop (\%)}} & \multicolumn{3}{c}{\textbf{ }} \\
\cmidrule(l{3pt}r{3pt}){2-4}
Platform design & 2 & \textbf{6} & \textbf{8} & PSES$^{1}$ & PCS$^{2}$ & PIS$^{3}$ \\
\midrule
\rowcolor{gray!10}
PROP (RAR) & 20.8 & \textbf{72.3} & \textbf{68.1} & 45.7 & 64.6 & 20.8 \\

DICE-doseEff (RAR) & 88.9 & \textbf{97.4} & \textbf{91.0} & 8.9 & 89.9 & 88.9 \\

\rowcolor{gray!10}
PROP (FR) & 22.5 & \textbf{73.2} & \textbf{71.0} & 47.2 & 66.8 & 22.5 \\

DICE-doseEff (FR) & 87.7 & \textbf{96.7} & \textbf{90.3} & 9.6 & 89.3 & 87.7 \\

\rowcolor{gray!10}
DICE-mAGILE & 5.0 & \textbf{18.1} & \textbf{18.9} & 4.3 & 4.9 & 5.0 \\

\toprule
\multicolumn{1}{l}{} & \multicolumn{3}{c}{\textbf{Futility stop (\%)}} & \multicolumn{3}{c}{} \\
\cmidrule(l{3pt}r{3pt}){2-4}
& 2 & 6 & 8 & PSES$^{14}$ & PCS$^{24}$ & PIS$^{3}$\\
\midrule
\rowcolor{gray!10}
PROP (RAR) & 20.8 & 2.6 & 1.7 & \textemdash & \textemdash & 21.0 \\

DICE-doseEff (RAR) & 1.4 & 0 & 0 & \textemdash & \textemdash & 1.4 \\

\rowcolor{gray!10}
PROP (FR) & 20.9 & 3.2 & 2.6 & \textemdash & \textemdash & 21.0 \\

DICE-doseEff (FR) & 1.7 & 0 & 0 & \textemdash & \textemdash & 1.7 \\

\rowcolor{gray!10}
DICE-mAGILE & 1.3 & 0.1 & 0.1 & \textemdash & \textemdash & 1.4 \\

\toprule
\multicolumn{1}{l}{} & \multicolumn{3}{c}{\textbf{Safety stop (\%)}} & \multicolumn{1}{c}{} \\
\cmidrule(l{3pt}r{3pt}){2-4}
& 2 & 6 & 8 & PSES$^{14}$ & PCS$^{24}$ & PIS$^{3}$\\
\midrule
\rowcolor{gray!10}
PROP (RAR) & 0 & 2.0 & 10.9 & \textemdash & \textemdash & 11.5 \\

DICE-doseEff (RAR) & 0 & 2.1 & 8.7 & \textemdash & \textemdash & 9.6 \\

\rowcolor{gray!10}
PROP (FR) & 0 & 2.4 & 11.2 & \textemdash & \textemdash & 12.0 \\

DICE-doseEff (FR) & 0 & 3.9 & 9.6 & \textemdash & \textemdash & 10.4 \\

\rowcolor{gray!10}
DICE-mAGILE & 0.1 & 4.0 & 11.6 & \textemdash & \textemdash & 12.2  \\

\bottomrule
\end{tabular}
}
\caption{Stopping proportions by dosing regimen and platform trial design in scenario 1 for graduation, futility, and safety. \textit{Note:} $^{1}$ PSES = percentage of scenario-exact stopping; $^{2}$ PCS = percentage of correct stopping; $^{3}$ PIS = percentage of incorrect stopping; $^{4}$ no true futile and toxic regimen in the scenario.}
\label{stopping_proportions_scenario1}
\end{table}

\begin{table}[H]
\centering
\resizebox{\textwidth}{!}{
\begin{tabular}{lcccccc}
\textbf{Scenario 1} & \multicolumn{6}{c}{} \\
\toprule
\multirow[c]{1}{*}{\textbf{Acute toxicity}} & \multicolumn{3}{c}{\textbf{Posterior mean (MCSE)}} & \multicolumn{3}{c}{\textbf{95\% CrI coverage}} \\
\cmidrule(lr){2-4} 
\cmidrule(lr){5-7}
\textbf{Regimen} & 2 & 6 & 8 & 2 & 6 & 8 \\
\textit{Scenario truth} & \textit{5.1} & \textit{9.8} & \textit{14.3} & \multicolumn{3}{c}{} \\
\midrule

\rowcolor{gray!10}
PROP (RAR) & 5.4 (0.097) & 9.7 (0.130) & 14.8 (0.213) & 96.9 & 94.3 & 96.7 \\

DICE-doseEff (RAR) & 6.6 (0.050) & 9.1 (0.056) & 12.3 (0.079) & 96.2 & 93.3 & 91.2 \\

\rowcolor{gray!10}
PROP (FR) & 5.5 (0.099) & 9.9 (0.135) & 14.9 (0.221) & 96.6 & 94.3 & 95.9 \\

DICE-doseEff (FR) & 6.4 (0.046) & 8.9 (0.057) & 12.0 (0.081) & 96.1 & 92.1 & 89.6 \\

\rowcolor{gray!10}
DICE-mAGILE & 6.8 (0.038) & 9.0 (0.044) & 11.2 (0.059) & 93.5 & 90.4 & 85.4 \\

\toprule
\multirow[c]{1}{*}{\textbf{Cumulative toxicity}} & \multicolumn{3}{c}{\textbf{Posterior mean (MCSE)}}
& \multicolumn{3}{c}{\textbf{95\% CrI coverage}} \\
\cmidrule(lr){2-4} 
\cmidrule(lr){5-7}
\textbf{Regimen} & 2 & 6 & 8 & 2 & 6 & 8 \\
\textit{Scenario truth} & \textit{15.4} & \textit{25.4} & \textit{31.1} & \multicolumn{3}{c}{} \\
\midrule

\rowcolor{gray!10}
PROP (RAR) & 15.7 (0.147) & 25.2 (0.201) & 31.3 (0.248) & 95.5 & 94.8 & 95.4 \\

DICE-doseEff (RAR) & 10.6 (0.057) & 26.5 (0.073) & 32.3 (0.090) & 86.1 & 92.8 & 95.4 \\

\rowcolor{gray!10}
PROP (FR) & 16.1 (0.152) & 25.8 (0.208) & 31.8 (0.257) & 94.8 & 93.6 & 95.6 \\

DICE-doseEff (FR) & 10.5 (0.056) & 26.6 (0.076) & 32.4 (0.095) & 84.1 & 92.4 & 96.4 \\

\rowcolor{gray!10}
DICE-mAGILE & 11.3 (0.047) & 26.8 (0.052) & 31.3 (0.064) & 85.3 & 94.8 & 96.7 \\

\bottomrule
\end{tabular}
}
\caption{Posterior mean toxicity probabilities with Monte-Carlo Standard Error (MCSE) and 95\% credible interval (CrI) coverage by regimen and platform trial design for scenario 1.}
\label{toxicity_estimation_scenario1}
\end{table}

\begin{table}[H]
\centering
\resizebox{\textwidth}{!}{
\begin{tabular}{lcccccccc}
\toprule
\multirow[c]{1}{*}{\textbf{Scenario 1}} 
& \multicolumn{4}{c}{\textbf{Efficacy posterior mean (MCSE)}}
& \multicolumn{4}{c}{\textbf{95\% CrI coverage}} \\
\cmidrule(lr){2-5} 
\cmidrule(lr){6-9}
Regimen & 0$^{1}$ & 2 & 6 & 8 & 0$^{1}$ & 2 & 6 & 8\\
\textit{Scenario truth} & \textit{50.0} & \textit{59.9} & \textit{71.8} & \textit{71.7} & \multicolumn{4}{c}{} \\
\midrule

\rowcolor{gray!10}
PROP (RAR) & 50.4 (0.226) & 60.5 (0.178) & 71.4 (0.217) & 72.5 (0.232) & 97.4 & 93.9 & 93.5 & 94.2 \\

DICE-doseEff (RAR) & 38.9 (0.115) & 59.2 (0.069) & 80.2 (0.073) & 85.2 (0.081) & 78.0 & 95.8 & 75.8 & 61.7 \\

\rowcolor{gray!10}
PROP (FR) & 50.1 (0.222) & 60.4 (0.179) & 71.4 (0.225) & 72.6 (0.240) & 97.0 & 95.3 & 94.4 & 95.4 \\

DICE-doseEff (FR) & 39.9 (0.114) & 60.2 (0.068) & 80.7 (0.079) & 85.6 (0.087) & 81.9 & 95.2 & 76.4 & 63.7 \\
\bottomrule
\end{tabular}
}
\caption{Posterior mean efficacy probabilities with Monte-Carlo Standard Error (MCSE) and 95\% credible interval coverage by dosing regimen and platform trial design in scenario 1. \textit{Note:} $^{1}$ Regimen 0 is the control arm.}
\label{efficacy_estimation_scenario1}
\end{table}

\begin{table}[H]
\centering
\resizebox{\textwidth}{!}{
\begin{tabular}{lrrrrrrrrrr}
\toprule
\textbf{Parameter}
& \multicolumn{5}{c}{\textbf{PROP (RAR)}}
& \multicolumn{5}{c}{\textbf{PROP (FR)}} \\
\cmidrule(lr){2-6}
\cmidrule(lr){7-11}
& Estimate & Bias & SE & RSE (\%) & Coverage (\%)
& Estimate & Bias & SE & RSE (\%) & Coverage (\%) \\
\midrule

\rowcolor{gray!10}
$k_a$        & 2.005 &  0.005 & 0.120 & 6.0 & 91.8 & 2.000 &  0.000 & 0.122 & 6.1 & 93.5 \\

$CL$         & 5.011 &  0.011 & 0.286 & 5.7 & 93.7 & 5.010 &  0.010 & 0.293 & 5.8 & 94.2 \\

\rowcolor{gray!10}
$V$          & 100.267 & 0.267 & 5.657 & 5.7 & 93.8 & 100.244 & 0.244 & 5.753 & 5.8 & 94.3 \\

$k_{\mathrm{in}}$  & 1.993 & -0.007 & 0.152 & 7.6 & 72.9 & 2.006 &  0.006 & 0.160 & 8.0 & 75.7 \\

\rowcolor{gray!10}
$k_{\mathrm{out}}$ & 0.200 &  0.000 & 0.013 & 6.4 & 69.9 & 0.200 &  0.000 & 0.013 & 6.7 & 73.1 \\

$IC_{50}$    & 1.310 &  0.010 & 0.116 & 8.8 & 90.6 & 1.306 &  0.006 & 0.119 & 9.1 & 91.5 \\

\rowcolor{gray!10}
$b_{\mathrm{PK}}$ & 0.200 &  0.000 & 0.008 & 4.2 & 94.7 & 0.200 &  0.000 & 0.009 & 4.3 & 95.0 \\

$b_{\mathrm{PD}}$ & 0.201 &  0.001 & 0.009 & 4.6 & 93.8 & 0.201 &  0.001 & 0.009 & 4.6 & 94.6 \\

\midrule

\rowcolor{gray!10}
$\omega_{k_a}$       & 0.277 & -0.023 & 0.064 & 22.8 & 92.5 & 0.281 & -0.019 & 0.065 & 22.8 & 92.4 \\

$\omega_{CL}$        & 0.394 & -0.006 & 0.042 & 10.4 & 94.1 & 0.396 & -0.004 & 0.042 & 10.7 & 94.3 \\

\rowcolor{gray!10}
$\omega_{V}$         & 0.393 & -0.007 & 0.041 & 10.5 & 92.7 & 0.392 & -0.008 & 0.042 & 10.7 & 93.8 \\

$\omega_{k_{\mathrm{in}}}$  & 0.377 & -0.023 & 0.057 & 15.1 & 80.5 & 0.375 & -0.025 & 0.059 & 15.5 & 82.6 \\

\rowcolor{gray!10}
$\omega_{k_{\mathrm{out}}}$ & 0.287 & -0.013 & 0.051 & 17.8 & 77.8 & 0.289 & -0.011 & 0.054 & 18.6 & 76.4 \\

$\omega_{IC_{50}}$   & 0.381 & -0.019 & 0.086 & 22.5 & 90.3 & 0.377 & -0.023 & 0.091 & 23.5 & 92.2 \\

\bottomrule
\end{tabular}
}
\caption{Estimation performance of the population PK/PD model parameters under the PROPs with response-adaptive randomization (RAR) and fixed randomization (FR) in scenario 1. SE denotes standard error, RSE denotes relative standard error, and coverage denotes the empirical coverage probability of the 95\% credible interval.}
\label{pkpd_tox_eff_parameter_estimation_scenario1}
\end{table}

\subsection{Scenario 2}

\begin{table}[H]
\centering
\resizebox{\textwidth}{!}{
\begin{tabular}[t]{lcccccc}
\toprule
\multicolumn{1}{l}{\textbf{Scenario 2}} & \multicolumn{3}{c}{\textbf{Graduation stop (\%)}} & \multicolumn{3}{c}{\textbf{ }} \\
\cmidrule(l{3pt}r{3pt}){2-4}
Platform design & 2 & \textbf{6} & \textbf{8} & PSES$^{1}$ & PCS$^{2}$ & PIS$^{3}$ \\
\midrule
\rowcolor{gray!10}
PROP (RAR) & 47.6 & \textbf{94.7} & \textbf{84.8} & 41.4 & 82.7 & 47.6 \\

DICE-doseEff (RAR) & 98.5 & \textbf{98.4} & \textbf{90.7} & 1.1 & 90.0 & 98.5 \\

\rowcolor{gray!10}
PROP (FR) & 46.6 & \textbf{93.3} & \textbf{85.6} & 42.8 & 82.7 & 46.6 \\

DICE-doseEff (FR) & 99.2 & \textbf{97.1} & \textbf{88.9} & 0.7 & 87.7 & 99.2 \\

\rowcolor{gray!10}
DICE-mAGILE & 9.7 & \textbf{40.8} & \textbf{53.2} & 22.8 & 26.5 & 9.7 \\

\toprule
\multicolumn{1}{l}{} & \multicolumn{3}{c}{\textbf{Futility stop (\%)}} & \multicolumn{3}{c}{} \\
\cmidrule(l{3pt}r{3pt}){2-4}
& 2 & 6 & 8 & PSES$^{14}$ & PCS$^{24}$ & PIS$^{3}$\\
\midrule
\rowcolor{gray!10}
PROP (RAR) & 6.8 & 0.7 & 0.2 & \textemdash & \textemdash & 6.9 \\

DICE-doseEff (RAR) & 0.1 & 0 & 0 & \textemdash & \textemdash & 0.1 \\

\rowcolor{gray!10}
PROP (FR) & 7.3 & 0.1 & 0.1 & \textemdash & \textemdash & 7.4 \\

DICE-doseEff (FR) & 0.1 & 0 & 0 & \textemdash & \textemdash & 0.1 \\

\rowcolor{gray!10}
DICE-mAGILE & 0.6 & 0 & 0 & \textemdash & \textemdash & 0.6 \\

\toprule
\multicolumn{1}{l}{} & \multicolumn{3}{c}{\textbf{Safety stop (\%)}} & \multicolumn{1}{c}{} \\
\cmidrule(l{3pt}r{3pt}){2-4}
& 2 & 6 & 8 & PSES$^{14}$ & PCS$^{24}$ & PIS$^{3}$\\
\midrule
\rowcolor{gray!10}
PROP (RAR) & 0 & 1.8 & 14.4 & \textemdash & \textemdash & 14.6 \\

DICE-doseEff (RAR) & 0 & 1.6 & 9.3 & \textemdash & \textemdash & 10.0 \\

\rowcolor{gray!10}
PROP (FR) & 0 & 2.4 & 13.4 & \textemdash & \textemdash & 13.9 \\

DICE-doseEff (FR) & 0 & 2.9 & 11.1 & \textemdash & \textemdash & 12.3 \\

\rowcolor{gray!10}
DICE-mAGILE & 0 & 3.0 & 13.0 & \textemdash & \textemdash & 13.9  \\

\bottomrule
\end{tabular}
}
\caption{Stopping proportions by dosing regimen and platform trial design in scenario 2 for graduation, futility, and safety. \textit{Note:} $^{1}$ PSES = percentage of scenario-exact stopping; $^{2}$ PCS = percentage of correct stopping; $^{3}$ PIS = percentage of incorrect stopping; $^{4}$ no true futile and toxic regimen in the scenario.}
\label{stopping_proportions_scenario2}
\end{table}

\begin{table}[H]
\centering
\resizebox{\textwidth}{!}{
\begin{tabular}{lcccccc}
\textbf{Scenario 2} & \multicolumn{6}{c}{} \\
\toprule
\multirow[c]{1}{*}{\textbf{Acute toxicity}} & \multicolumn{3}{c}{\textbf{Posterior mean (MCSE)}} & \multicolumn{3}{c}{\textbf{95\%-CrI coverage}} \\
\cmidrule(lr){2-4} 
\cmidrule(lr){5-7}
\textbf{Regimen} & 2 & 6 & 8 & 2 & 6 & 8 \\
\textit{Scenario truth} & \textit{4.8} & \textit{10.3} & \textit{16.0} & \multicolumn{3}{c}{} \\
\midrule

\rowcolor{gray!10}
PROP (RAR) & 5.2 (0.101) & 10.2 (0.145) & 16.3 (0.245) & 98.0 & 95.1 & 97.1 \\

DICE-doseEff (RAR) & 6.4 (0.050) & 9.3 (0.057) & 13.2 (0.081) & 97.6 & 91.6 & 87.2 \\

\rowcolor{gray!10}
PROP (FR) & 5.3 (0.102) & 10.1 (0.147) & 16.1 (0.247) & 97.0 & 95.0 & 96.9 \\

DICE-doseEff (FR) & 6.5 (0.051) & 9.6 (0.062) & 14.0 (0.091) & 97.2 & 91.5 & 90.0 \\

\rowcolor{gray!10}
DICE-mAGILE & 6.6 (0.040) & 9.4 (0.046) & 12.3 (0.065) & 94.3 & 91.0 & 85.4 \\

\toprule
\multirow[c]{1}{*}{\textbf{Cumulative toxicity}} & \multicolumn{3}{c}{\textbf{Posterior mean (MCSE)}}
& \multicolumn{3}{c}{\textbf{95\% CrI coverage}} \\
\cmidrule(lr){2-4} 
\cmidrule(lr){5-7}
\textbf{Regimen} & 2 & 6 & 8 & 2 & 6 & 8 \\
\textit{Scenario truth} & \textit{13.2} & \textit{25.6} & \textit{33.7} & \multicolumn{3}{c}{} \\
\midrule

\rowcolor{gray!10}
PROP (RAR) & 13.8 (0.146) & 25.3 (0.223) & 33.5 (0.286) & 96.6 & 95.2 & 96.4 \\

DICE-doseEff (RAR) & 10.0 (0.061) & 25.4 (0.074) & 32.2 (0.090) & 91.3 & 94.8 & 96.4 \\

\rowcolor{gray!10}
PROP (FR) & 13.8 (0.147) & 25.2 (0.226) & 33.3 (0.288) & 96.1 & 95.0 & 96.2 \\

DICE-doseEff (FR) & 10.1 (0.063) & 25.8 (0.081) & 33.2 (0.102) & 90.3 & 93.6 & 95.9 \\

\rowcolor{gray!10}
DICE-mAGILE & 10.7 (0.050) & 26.5 (0.054) & 32.2 (0.069) & 92.7 & 94.5 & 96.7 \\

\bottomrule
\end{tabular}
}
\caption{Posterior mean toxicity probabilities with Monte-Carlo Standard Error (MCSE) and 95\% credible interval (CrI) coverage by regimen and platform trial design for scenario 2.}
\label{toxicity_estimation_scenario2}
\end{table}

\begin{table}[H]
\centering
\resizebox{\textwidth}{!}{
\begin{tabular}{lcccccccc}
\toprule
\multirow[c]{1}{*}{\textbf{Scenario 2}} 
& \multicolumn{4}{c}{\textbf{Efficacy posterior mean (MCSE)}}
& \multicolumn{4}{c}{\textbf{95\% CrI coverage}} \\
\cmidrule(lr){2-5} 
\cmidrule(lr){6-9}
Regimen & 0$^{1}$ & 2 & 6 & 8 & 0$^{1}$ & 2 & 6 & 8\\
\textit{Scenario truth} & \textit{50.0} & \textit{64.2} & \textit{82.4} & \textit{86.6} & \multicolumn{4}{c}{} \\
\midrule

\rowcolor{gray!10}
PROP (RAR) & 50.6 (0.242) & 64.3 (0.187) & 80.8 (0.201) & 85.3 (0.206) & 96.9 & 94.4 & 93.0 & 94.1 \\

DICE-doseEff (RAR) & 38.3 (0.109) & 65.4 (0.066) & 89.5 (0.052) & 93.4 (0.049) & 67.7 & 93.1 & 62.5 & 55.7 \\

\rowcolor{gray!10}
PROP (FR) & 50.8 (0.234) & 64.6 (0.184) & 80.9 (0.203) & 85.4 (0.209) & 95.7 & 94.3 & 93.4 & 95.1 \\

DICE-doseEff (FR) & 38.8 (0.118) & 65.7 (0.071) & 89.2 (0.061) & 93.0 (0.059) & 76.2 & 93.1 & 68.7 & 63.8 \\
\bottomrule
\end{tabular}
}
\caption{Posterior mean efficacy probabilities with Monte-Carlo Standard Error (MCSE) and 95\% credible interval coverage by dosing regimen and platform trial design in scenario 2. \textit{Note:} $^{1}$ Regimen 0 is the control arm.}
\label{efficacy_estimation_scenario2}
\end{table}

\begin{table}[H]
\centering
\resizebox{\textwidth}{!}{
\begin{tabular}{lrrrrrrrrrr}
\toprule
\textbf{Parameter}
& \multicolumn{5}{c}{\textbf{PROP (RAR)}}
& \multicolumn{5}{c}{\textbf{PROP (FR)}} \\
\cmidrule(lr){2-6}
\cmidrule(lr){7-11}
& Estimate & Bias & SE & RSE (\%) & Coverage (\%)
& Estimate & Bias & SE & RSE (\%) & Coverage (\%) \\
\midrule

\rowcolor{gray!10}
$k_a$        & 2.002 &  0.002 & 0.127 &  6.3 & 93.7 & 2.012 &  0.012 & 0.129 &  6.4 & 94.0 \\

$V$          & 99.999 & -0.001 & 6.415 & 6.4 & 92.8 & 100.063 &  0.063 & 6.520 & 6.5 & 94.4 \\

\rowcolor{gray!10}
$K_m$        & 2.495 &  0.495 & 0.350 & 15.2 & 69.9 & 2.530 &  0.530 & 0.379 & 15.9 & 69.7 \\

$V_{\max}$   & 11.209 & 1.209 & 1.028 & 9.5 & 72.0 & 11.293 & 1.293 & 1.058 & 9.7 & 73.9 \\

\rowcolor{gray!10}
$k_{\mathrm{in}}$  & 2.016 &  0.016 & 0.226 & 11.3 & 81.0 & 2.024 &  0.024 & 0.243 & 12.1 & 80.8 \\

$k_{\mathrm{out}}$ & 0.201 &  0.001 & 0.019 & 9.4 & 79.3 & 0.202 & 0.002 & 0.021 & 10.3 & 81.2 \\

\rowcolor{gray!10}
$IC_{50}$ & 1.324 & 0.024 & 0.132 & 10.0 & 93.0 & 1.325 & 0.025 & 0.137 & 10.4 & 93.0 \\

$b_{\mathrm{PK}}$ & 0.200 &  0.000 & 0.010 & 5.1 & 95.5 & 0.200 &  0.000 & 0.011 & 5.3 & 94.1 \\

\rowcolor{gray!10}
$b_{\mathrm{PD}}$ & 0.201 &  0.001 & 0.012 & 5.8 & 95.9 & 0.200 & 0.000 & 0.012 & 6.0 & 94.7 \\

\midrule

$\omega_{k_a}$       & 0.190 & -0.010 & 0.079 & 42.3 & 90.8 & 0.189 & -0.011 & 0.082 & 44.8 & 92.0 \\

\rowcolor{gray!10}
$\omega_{V}$         & 0.392 & -0.008 & 0.046 & 11.5 & 94.0 & 0.394 & -0.006 & 0.047 & 11.8 & 94.6 \\

$\omega_{K_m}$       & 0.441 &  0.041 & 0.114 & 27.8 & 73.8 & 0.436 &  0.036 & 0.120 & 29.5 & 76.0 \\

\rowcolor{gray!10}
$\omega_{V_{\max}}$  & 0.356 & -0.044 & 0.065 & 17.9 & 87.3 & 0.353 & -0.047 & 0.067 & 18.7 & 86.4 \\

$\omega_{k_{\mathrm{in}}}$  & 0.371 & -0.029 & 0.080 & 21.5 & 87.5 & 0.367 & -0.033 & 0.084 & 23.0 & 89.1 \\

\rowcolor{gray!10}
$\omega_{k_{\mathrm{out}}}$ & 0.287 & -0.013 & 0.075 & 26.9 & 82.9 & 0.292 & -0.008 & 0.077 & 26.3 & 87.3 \\

$\omega_{IC_{50}}$  & 0.365 & -0.035 & 0.100 & 26.9 & 93.6 & 0.373 & -0.027 & 0.105 & 27.9 & 94.3 \\

\bottomrule
\end{tabular}
}
\caption{Estimation performance of the population PK/PD model parameters under the PROPs with response-adaptive randomization (RAR) and fixed randomization (FR) in scenario 2. SE denotes standard error, RSE denotes relative standard error, and coverage denotes the empirical coverage probability of the 95\% credible interval.}
\label{pkpd_tox_eff_parameter_estimation_scenario2}
\end{table}

\subsection{Scenario 3}

\begin{table}[H]
\centering
\resizebox{\textwidth}{!}{
\begin{tabular}[t]{lcccccc}
\toprule
\multicolumn{1}{l}{\textbf{Scenario 3}} & \multicolumn{3}{c}{\textbf{Graduation stop (\%)}} & \multicolumn{3}{c}{\textbf{ }} \\
\cmidrule(l{3pt}r{3pt}){2-4}
Platform design & 2 & 6 & 8 & PSES$^{14}$ & PCS$^{24}$ & PIS$^{3}$ \\
\midrule
\rowcolor{gray!10}
PROP (RAR) & 35.7 & 4.4 & 0.1 & \textemdash & \textemdash & 38.0 \\

DICE-doseEff (RAR) & 89.5 & 5.6 & 0 & \textemdash & \textemdash & 89.9 \\

\rowcolor{gray!10}
PROP (FR) & 34.7 & 4.5 & 0.1 & \textemdash & \textemdash & 36.7 \\

DICE-doseEff (FR) & 91.6 & 3.6 & 0 & \textemdash & \textemdash & 91.7 \\

\rowcolor{gray!10}
DICE-mAGILE & 9.5 & 0.3 & 0 & \textemdash & \textemdash & 9.8 \\

\toprule
\multicolumn{1}{l}{} & \multicolumn{3}{c}{\textbf{Futility stop (\%)}} & \multicolumn{3}{c}{} \\
\cmidrule(l{3pt}r{3pt}){2-4}
& 2 & 6 & 8 & PSES$^{14}$ & PCS$^{24}$ & PIS$^{3}$\\
\midrule
\rowcolor{gray!10}
PROP (RAR) & 6.2 & 0.1 & 0 & \textemdash & \textemdash & 6.2 \\

DICE-doseEff (RAR) & 0.2 & 0 & 0 & \textemdash & \textemdash & 0.2 \\

\rowcolor{gray!10}
PROP (FR) & 6.9 & 0 & 0 & \textemdash & \textemdash & 6.9 \\

DICE-doseEff (FR) & 0.5 & 0 & 0 & \textemdash & \textemdash & 0.5 \\

\rowcolor{gray!10}
DICE-mAGILE & 0.1 & 0 & 0 & \textemdash & \textemdash & 0.1 \\

\toprule
\multicolumn{1}{l}{} & \multicolumn{3}{c}{\textbf{Safety stop (\%)}} & \multicolumn{1}{c}{} \\
\cmidrule(l{3pt}r{3pt}){2-4}
& 2 & \textbf{6} & \textbf{8} & PSES$^{1}$ & PCS$^{2}$ & PIS$^{3}$\\
\midrule
\rowcolor{gray!10}
PROP (RAR) & 10.9 & \textbf{95.2} & \textbf{99.8} & 84.3 & 95.2 & 10.9 \\

DICE-doseEff (RAR) & 0.1 & \textbf{94.4} & \textbf{100} & 94.3 & 94.4 & 0.1 \\

\rowcolor{gray!10}
PROP (FR) & 10.9 & \textbf{95.4} & \textbf{99.9} & 84.5 & 95.4 & 10.9 \\

DICE-doseEff (FR) & 0.4 & \textbf{96.4} & \textbf{100} & 96.0 & 96.4 & 0.4 \\

\rowcolor{gray!10}
DICE-mAGILE & 0 & \textbf{98.9} & \textbf{99.4} & 98.4 & 98.4 & 0  \\

\bottomrule
\end{tabular}
}
\caption{Stopping proportions for graduation, futility, and safety by dosing regimen and platform trial design in scenario 3. \textit{Note:} $^{1}$ PSES = percentage of scenario-exact stopping; $^{2}$ PCS = percentage of correct stopping; $^{3}$ PIS = percentage of incorrect stopping; $^{4}$ no true graduated and futile regimen in the scenario.}
\label{stopping_proportions_scenario3}
\end{table}

\begin{table}[H]
\centering
\resizebox{\textwidth}{!}{
\begin{tabular}{lcccccc}
\textbf{Scenario 3} & \multicolumn{6}{c}{} \\
\toprule
\multirow[c]{1}{*}{\textbf{Acute toxicity}} & \multicolumn{3}{c}{\textbf{Posterior mean (MCSE)}} & \multicolumn{3}{c}{\textbf{95\% CrI coverage}} \\
\cmidrule(lr){2-4} 
\cmidrule(lr){5-7}
\textbf{Regimen} & 2 & 6 & 8 & 2 & 6 & 8 \\
\textit{Scenario truth} & \textit{0.8} & \textit{22.6} & \textit{64.6} & \multicolumn{3}{c}{} \\
\midrule

\rowcolor{gray!10}
PROP (RAR) & 2.4 (0.065) & 23.9 (0.183) & 60.3 (0.251) & 89.6 & 93.8 & 84.1 \\

DICE-doseEff (RAR) & 13.5 (0.077) & 29.6 (0.079) & 44.7 (0.119) & 6.5 & 82.3 & 52.9 \\

\rowcolor{gray!10}
PROP (FR) & 2.4 (0.066) & 24.1 (0.186) & 60.7 (0.254) & 90.4 & 92.7 & 84.1 \\

DICE-doseEff (FR) & 13.2 (0.075) & 28.4 (0.089) & 42.8 (0.137) & 6.9 & 89.6 & 54.1 \\

\rowcolor{gray!10}
DICE-mAGILE & 13.7 (0.069) & 29.0 (0.077) & 43.0 (0.121) & 3.4 & 83.7 & 46.4 \\

\toprule
\multirow[c]{1}{*}{\textbf{Cumulative toxicity}} & \multicolumn{3}{c}{\textbf{Posterior mean (MCSE)}}
& \multicolumn{3}{c}{\textbf{95\% CrI coverage}} \\
\cmidrule(lr){2-4} 
\cmidrule(lr){5-7}
\textbf{Regimen} & 2 & 6 & 8 & 2 & 6 & 8 \\
\textit{Scenario truth} & \textit{33.3} & \textit{58.9} & \textit{81.9} & \multicolumn{3}{c}{} \\
\midrule

\rowcolor{gray!10}
PROP (RAR) & 33.7 (0.218) & 59.5 (0.224) & 80.3 (0.161) & 93.8 & 91.5 & 91.7 \\

DICE-doseEff (RAR) & 23.2 (0.102) & 65.3 (0.064) & 77.5 (0.079) & 86.1 & 83.2 & 91.0 \\

\rowcolor{gray!10}
PROP (FR) & 33.7 (0.222) & 59.6 (0.228) & 80.4 (0.163) & 94.5 & 91.6 & 91.0 \\

DICE-doseEff (FR) & 23.6 (0.100) & 66.2 (0.073) & 77.3 (0.091) & 88.4 & 83.2 & 91.0 \\

\rowcolor{gray!10}
DICE-mAGILE & 24.2 (0.092) & 66.7 (0.062) & 77.7 (0.079) & 86.4 & 79.9 & 90.7 \\

\bottomrule
\end{tabular}
}
\caption{Posterior mean toxicity probabilities with Monte-Carlo Standard Error (MCSE) and 95\% credible interval (CrI) coverage by regimen and platform trial design for scenario 3.}
\label{toxicity_estimation_scenario3}
\end{table}

\begin{table}[H]
\centering
\resizebox{\textwidth}{!}{
\begin{tabular}{lcccccccc}
\toprule
\multirow[c]{1}{*}{\textbf{Scenario 3}} 
& \multicolumn{4}{c}{\textbf{Efficacy posterior mean (MCSE)}}
& \multicolumn{4}{c}{\textbf{95\% CrI coverage}} \\
\cmidrule(lr){2-5} 
\cmidrule(lr){6-9}
Regimen & 0$^{1}$ & 2 & 6 & 8 & 0$^{1}$ & 2 & 6 & 8\\
\textit{Scenario truth} & \textit{50.0} & \textit{63.8} & \textit{79.0} & \textit{79.2} & \multicolumn{4}{c}{} \\
\midrule

\rowcolor{gray!10}
PROP (RAR) & 50.7 (0.219) & 63.4 (0.176) & 76.2 (0.234) & 77.2 (0.246) & 96.4 & 96.0 & 93.7 & 95.8 \\

DICE-doseEff (RAR) & 39.6 (0.094) & 63.5 (0.051) & 85.8 (0.062) & 90.3 (0.064) & 68.5 & 92.7 & 69.8 & 52.4 \\

\rowcolor{gray!10}
PROP (FR) & 50.8 (0.216) & 63.3 (0.177) & 76.1 (0.241) & 77.1 (0.253) & 96.0 & 95.6 & 94.4 & 96.4 \\

DICE-doseEff (FR) & 40.5 (0.100) & 64.2 (0.054) & 85.8 (0.073) & 90.0 (0.075) & 77.7 & 93.7 & 74.5 & 59.2 \\
\bottomrule
\end{tabular}
}
\caption{Posterior mean efficacy probabilities with Monte-Carlo Standard Error (MCSE) and 95\% credible interval coverage by dosing regimen and platform trial design in scenario 3. \textit{Note:} $^{1}$ regimen 0 is the control arm.}
\label{efficacy_estimation_scenario3}
\end{table}

\subsection{Scenario 4}

\begin{table}[H]
\centering
\resizebox{\textwidth}{!}{
\begin{tabular}[t]{lccccc}
\toprule
\multicolumn{1}{l}{\textbf{Scenario 4}} & \multicolumn{5}{c}{\textbf{Candidate regimen addition (\%)}} \\
\cmidrule(l{3pt}r{3pt}){2-6}
Platform design & 3 & 4 & 5 & 7 & \textbf{9} \\
\midrule
\rowcolor{gray!10}
PROP (RAR) & 9.6 & 0 & 23.0 & 0 & \textbf{67.3} \\

DICE-doseEff (RAR) & 0 & 0 & 0 & 0 & \textbf{0} \\

\rowcolor{gray!10}
PROP (FR) & 9.9 & 0.3 & 23.5 & 0.3 & \textbf{65.9} \\

DICE-doseEff (FR) & 0 & 0 & 0 & 0 & \textbf{0} \\

\bottomrule
\end{tabular}
}
\caption{Proportion of simulated trials in which each candidate regimen was added as a new experimental arm, by platform trial design in scenario 4. The correct candidate regimen for addition is shown in \textbf{bold}.}
\label{arm_addition_scenario4}
\end{table}

\begin{table}[H]
\centering
\resizebox{\textwidth}{!}{
\begin{tabular}[t]{lccccccccccc}
\toprule
\multicolumn{1}{l}{\textbf{Scenario 4}} & \multicolumn{8}{c}{\textbf{Graduation stop (\%)}} & \multicolumn{3}{c}{\textbf{ }} \\
\cmidrule(l{3pt}r{3pt}){2-9}
Platform design & 2 & 3 & 4 & 5 & \textbf{6} & 7 & \textbf{8} & \textbf{9} & PSES$^{1}$ & PCS$^{2}$ & PIS$^{3}$ \\
\midrule
\rowcolor{gray!10}
PROP (RAR) & 8.0 & 0.6 & 0 & 2.5 & \textbf{51.4} & 0 & \textbf{49.4} & \textbf{6.0} & 3.7 & 3.7 & 10.2 \\

DICE-doseEff (RAR) & 74.7 & 0 & 0 & 0 & \textbf{95.5} & 0 & \textbf{89.9} & \textbf{0} & 0 & 0 & 74.7  \\

\rowcolor{gray!10}
PROP (FR) &  9.5 & 1.5 & 0 & 3.6 & \textbf{53.1} & 0 & \textbf{51.8} & \textbf{13.8} & 9.4 & 10.1 & 13.0  \\

DICE-doseEff (FR) & 72.4 & 0 & 0 & 0 & \textbf{95.6} & 0 & \textbf{89.6} & \textbf{0} & 0 & 0 & 72.4  \\

\toprule
\multicolumn{1}{l}{} & \multicolumn{8}{c}{\textbf{Futility stop (\%)}} & \multicolumn{3}{c}{} \\
\cmidrule(l{3pt}r{3pt}){2-9}
& \textbf{2} & 3 & 4 & 5 & 6 & 7 & 8 & 9 & PSES$^{1}$ & PCS$^{2}$ & PIS$^{3}$\\
\midrule
\rowcolor{gray!10}
PROP (RAR) & \textbf{43.3} & 0 & 0 & 0.1 & 13.1 & 0 & 10.7 & 0 & 29.9 & 43.3 & 13.4  \\

DICE-doseEff (RAR) & \textbf{4.1} & 0 & 0 & 0 & 0 & 0 & 0 & 0 & 4.1 & 4.1 & 0  \\

\rowcolor{gray!10}
PROP (FR) & \textbf{39.7} & 0.3 & 0 & 0.8 & 11.7 & 0 & 8.8 & 0.6 & 27.5 & 39.7 & 12.4  \\

DICE-doseEff (FR) & \textbf{5.3} & 0 & 0 & 0 & 0.1 & 0 & 0 & 0 & 5.2 & 5.3 & 0.1  \\

\toprule
\multicolumn{1}{l}{} & \multicolumn{8}{c}{\textbf{Safety stop (\%)}} & \multicolumn{1}{c}{} \\
\cmidrule(l{3pt}r{3pt}){2-9}
& 2 & 3 & 4 & 5 & 6 & 7 & 8 & 9 & PSES$^{14}$ & PCS$^{24}$ & PIS$^{3}$\\
\midrule
\rowcolor{gray!10}
PROP (RAR) & 0 & 0 & 0 & 0 & 1.7 & 0 & 10.3 & 0.1 & \textemdash & \textemdash & 10.6  \\

DICE-doseEff (RAR) & 0 & 0 & 0 & 0 & 2.3 & 0 & 8.7 & 0 & \textemdash & \textemdash & 9.4  \\

\rowcolor{gray!10}
PROP (FR) & 0.1 & 0 & 0 & 0.2 & 2.4 & 0 & 12.5 & 1.2 & \textemdash & \textemdash & 13.3  \\

DICE-doseEff (FR) & 0 & 0 & 0 & 0 & 3.1 & 0 & 9.2 & 0 & \textemdash & \textemdash & 10.0  \\

\bottomrule
\end{tabular}
}
\caption{Stopping proportions for graduation, futility, and safety by dosing regimen and platform trial design in scenario 4. \textit{Note:} $^{1}$ PSES = percentage of scenario-exact stopping; $^{2}$ PCS = percentage of correct stopping; $^{3}$ PIS = percentage of incorrect stopping; $^{4}$ no true toxic regimen in the scenario.}
\label{stopping_proportions_scenario4}
\end{table}

\begin{table}[H]
\centering
\resizebox{\textwidth}{!}{
\begin{tabular}{lcccccccccccccccc}
\textbf{Scenario 4} & \multicolumn{16}{c}{} \\
\toprule
\multirow[c]{1}{*}{\textbf{Acute toxicity}} & \multicolumn{8}{c}{\textbf{Posterior mean (MCSE)}} & \multicolumn{8}{c}{\textbf{95\% CrI coverage}} \\
\cmidrule(lr){2-9} 
\cmidrule(lr){10-17}
\textbf{Regimen} & 2 & 3 & 4 & 5 & 6 & 7 & 8 & 9 & 2 & 3 & 4 & 5 & 6 & 7 & 8 & 9 \\
\textit{Scenario truth} & \textit{5.1} & \textit{5.1} & \textit{9.8} & \textit{9.8} & \textit{9.8} & \textit{14.3} & \textit{14.3} & \textit{14.3} & \multicolumn{8}{c}{} \\
\midrule

\rowcolor{gray!10}
PROP (RAR) & 5.5 (0.098) & 5.9 (0.090) & NA & 13.3 (0.133) & 9.8 (0.125) & NA & 14.7 (0.200) & 13.9 (0.149) & 98.4 & 100 & NA & 90.7 & 95.8 & NA & 96.4 & 98.4 \\

DICE-doseEff (RAR) & 6.5 (0.045) & NA & NA & NA & 9.0 (0.054) & NA & 12.0 (0.075) & NA & 94.8 & NA & NA & NA & 91.9 & NA & 89.4 & NA \\

\rowcolor{gray!10}
PROP (FR) & 5.5 (0.097) & 6.2 (0.098) & 14.2 (0.131) & 13.5 (0.142) & 9.89 (0.127) & 13.4 (0.170) & 15.1 (0.205) & 13.8 (0.161) & 97.7 & 96.9 & 100 & 93.4 & 94.6 & 100 & 95.1 & 93.0 \\

DICE-doseEff (FR) & 6.4 (0.044) & NA & NA & NA & 8.9 (0.056) & NA & 11.9 (0.080) & NA & 96.4 & NA & NA & NA & 92.7 & NA & 91.2 & NA \\

\toprule
\multirow[c]{1}{*}{\textbf{Cumulative toxicity}} & \multicolumn{8}{c}{\textbf{Posterior mean (MCSE)}}
& \multicolumn{8}{c}{\textbf{95\% CrI coverage}} \\
\cmidrule(lr){2-9} 
\cmidrule(lr){10-17}
\textbf{Regimen} & 2 & 3 & 4 & 5 & 6 & 7 & 8 & 9 & 2 & 3 & 4 & 5 & 6 & 7 & 8 & 9 \\
\textit{Scenario truth} & \textit{15.4} & \textit{16.3} & \textit{19.8} & \textit{24.5} & \textit{25.4} & \textit{26.5} & \textit{31.1} & \textit{31.9} & \multicolumn{8}{c}{} \\
\midrule

\rowcolor{gray!10}
PROP (RAR) & 15.9 (0.144) & 21.6 (0.154) & NA & 30.4 (0.184) & 25.3 (0.193) & NA & 31.3 (0.235) & 30.8 (0.192) & 96.0 & 94.4 & NA & 88.4 & 94.5 & NA & 94.8 & 94.4 \\

DICE-doseEff (RAR) & 10.6 (0.056) & NA & NA & NA & 26.7 (0.072) & NA & 32.4 (0.086) & NA & 85.9 & NA & NA & NA & 93.7 & NA & 96.7 & NA \\

\rowcolor{gray!10}
PROP (FR) & 15.7 (0.143) & 20.6 (0.164) & 21.8 (0.149) & 29.4 (0.189) & 25.3 (0.193) & 30.7 (0.195) & 31.4 (0.238) & 31.3 (0.204) & 95.8 & 87.5 & 100 & 88.2 & 94.2 & 100 & 94.7 & 93.4 \\

DICE-doseEff (FR) & 10.6 (0.054) & NA & NA & NA & 27.0 (0.074) & NA & 32.6 (0.093) & NA & 85.4 & NA & NA & NA & 92.5 & NA & 95.2 & NA \\

\bottomrule
\end{tabular}
}
\caption{Posterior mean toxicity probabilities with Monte-Carlo Standard Error (MCSE) and 95\% credible interval (CrI) coverage by regimen and platform trial design for scenario 4.}
\label{toxicity_estimation_scenario4}
\end{table}

\begin{table}[H]
\centering
\resizebox{\textwidth}{!}{
\begin{tabular}{lccccccccc}
\toprule
\multirow[c]{1}{*}{\textbf{Scenario 4}} 
& \multicolumn{9}{c}{\textbf{Efficacy posterior mean (MCSE)}} \\
\cmidrule(lr){2-10} 
Regimen & 0$^{1}$ & 2 & 3 & 4 & 5 & 6 & 7 & 8 & 9 \\
\textit{Scenario truth} & \textit{50.0} & \textit{57.1} & \textit{59.7} & \textit{56.1} & \textit{62.0} & \textit{65.9} & \textit{58.2} & \textit{65.7} & \textit{70.4} \\
\midrule

\rowcolor{gray!10}
PROP (RAR) & 50.4 (0.225) & 57.4 (0.176) & 61.4 (0.162) & NA & 64.7 (0.185) & 65.3 (0.220) & NA & 66.4 (0.238) & 68.6 (0.217) \\

DICE-doseEff (RAR) & 39.2 (0.112) & 56.0 (0.067) & NA & NA & NA & 74.3 (0.078) & NA & 79.4 (0.090) & NA \\

\rowcolor{gray!10}
PROP (FR) & 49.9 (0.217) & 57.3 (0.172) & 62.5 (0.176) & 54.4 (0.164) & 63.5 (0.193) & 65.7 (0.221) & 57.9 (0.171) & 66.8 (0.240) & 69.8 (0.227) \\

DICE-doseEff (FR) & 39.9 (0.101) & 57.0 (0.065) & NA & NA & NA & 75.2 (0.084) & NA & 80.0 (0.098) & NA \\

\midrule

& \multicolumn{9}{c}{\textbf{95\% CrI coverage}} \\
\cmidrule(lr){2-10} 
Regimen & 0$^{1}$ & 2 & 3 & 4 & 5 & 6 & 7 & 8 & 9 \\
\midrule

\rowcolor{gray!10}
PROP (RAR) & 97.1 & 95.7 & 88.9 & NA & 97.7 & 93.1 & NA & 94.7 & 97.6 \\

DICE-doseEff (RAR) & 79.4 & 94.3 & NA & NA & NA & 79.5 & NA & 67.7 & NA \\

\rowcolor{gray!10}
PROP (FR) & 96.0 & 94.8 & 93.8 & 100 & 93.4 & 92.4 & 100 & 92.7 & 92.5 \\

DICE-doseEff (FR) & 82.4 & 94.1 & NA & NA & NA & 79.9 & NA & 68.2 & NA \\

\bottomrule
\end{tabular}
}
\caption{Posterior mean efficacy probabilities with Monte-Carlo Standard Error (MCSE) and 95\% credible interval coverage by dosing regimen and platform trial design in scenario 4. \textit{Note:} $^{1}$ regimen 0 is the control arm.}
\label{efficacy_estimation_scenario4}
\end{table}

\subsection{Scenario 5}

\begin{table}[H]
\centering
\resizebox{\textwidth}{!}{
\begin{tabular}[t]{lccccc}
\toprule
\multicolumn{1}{l}{\textbf{Scenario 5}} & \multicolumn{5}{c}{\textbf{Candidate regimen addition (\%)}} \\
\cmidrule(l{3pt}r{3pt}){2-6}
Platform design & \textbf{3} & 4 & 5 & 7 & 9 \\
\midrule
\rowcolor{gray!10}
PROP (RAR) & \textbf{87.5} & 0.5 & 12.0 & 0 & 0 \\

DICE-doseEff (RAR) & \textbf{87.5} & 0.5 & 12.0 & 0 & 0 \\

\rowcolor{gray!10}
PROP (FR) & \textbf{88.1} & 0.4 & 11.5 & 0 & 0 \\

DICE-doseEff (FR) & \textbf{88.1} & 0.4 & 11.5 & 0 & 0 \\

\bottomrule
\end{tabular}
}
\caption{Proportion of simulated trials in which each candidate regimen was added as a new experimental arm, by platform trial design in scenario 5. The correct candidate regimen for addition is shown in \textbf{bold}.}
\label{arm_addition_scenario5}
\end{table}

\begin{table}[H]
\centering
\resizebox{\textwidth}{!}{
\begin{tabular}[t]{lccccccccccc}
\toprule
\multicolumn{1}{l}{\textbf{Scenario 5}} & \multicolumn{8}{c}{\textbf{Graduation stop (\%)}} & \multicolumn{3}{c}{\textbf{ }} \\
\cmidrule(l{3pt}r{3pt}){2-9}
Platform design & 2 & \textbf{3} & 4 & 5 & 6 & 7 & 8 & 9 & PSES$^{1}$ & PCS$^{2}$ & PIS$^{3}$ \\
\midrule
\rowcolor{gray!10}
PROP (RAR) & 39.2 & \textbf{38.0} & 0 & 0.9 & 4.2 & \textemdash & 0 & \textemdash & 9.8 & 38.0 & 41.3 \\

DICE-doseEff (RAR) & 91.5 & \textbf{56.5} & 0.2 & 1.4 & 4.0 & \textemdash & 0.2 & \textemdash & 2.3 & 56.5 & 91.7  \\

\rowcolor{gray!10}
PROP (FR) & 40.2 & \textbf{41.1} & 0 & 0.8 & 2.0 & \textemdash & 0 & \textemdash & 11.2 & 41.1 & 41.2  \\

DICE-doseEff (FR) & 98.3 & \textbf{71.1} & 0.2 & 1.4 & 1.9 & \textemdash & 0 & \textemdash & 0.3 & 71.1 & 98.4  \\

\toprule
\multicolumn{1}{l}{} & \multicolumn{8}{c}{\textbf{Futility stop (\%)}} & \multicolumn{3}{c}{} \\
\cmidrule(l{3pt}r{3pt}){2-9}
& 2 & 3 & 4 & 5 & 6 & 7 & 8 & 9 & PSES$^{14}$ & PCS$^{24}$ & PIS$^{3}$\\
\midrule
\rowcolor{gray!10}
PROP (RAR) & 6.4 & 0.2 & 0 & 0 & 0 & \textemdash & 0 & \textemdash & \textemdash & \textemdash & 6.4  \\

DICE-doseEff (RAR) & 0.2 & 0 & 0 & 0 & 0 & \textemdash & 0 & \textemdash & \textemdash & \textemdash & 0.2  \\

\rowcolor{gray!10}
PROP (FR) & 6.9 & 1.2 & 0 & 0 & 0 & \textemdash & 0 & \textemdash & \textemdash & \textemdash & 6.9  \\

DICE-doseEff (FR) & 0 & 0 & 0 & 0 & 0 & \textemdash & 0 & \textemdash & \textemdash & \textemdash & 0  \\

\toprule
\multicolumn{1}{l}{} & \multicolumn{8}{c}{\textbf{Safety stop (\%)}} & \multicolumn{1}{c}{} \\
\cmidrule(l{3pt}r{3pt}){2-9}
& 2 & 3 & 4 & 5 & \textbf{6} & 7 & \textbf{8} & 9 & PSES$^{1}$ & PCS$^{2}$ & PIS$^{3}$\\
\midrule
\rowcolor{gray!10}
PROP (RAR) & 11.7 & 2.6 & 0.3 & 6.8 & \textbf{95.8} & \textemdash & \textbf{100} & \textemdash & 77.2 & 95.8 & 20.7  \\

DICE-doseEff (RAR) & 0.1 & 0.4 & 0.1 & 6.4 & \textbf{96.0} & \textemdash & \textbf{99.8} & \textemdash & 89.0 & 95.9 & 7.0  \\

\rowcolor{gray!10}
PROP (FR) & 10.1 & 7.8 & 0.3 & 8.7 & \textbf{98.0} & \textemdash & \textbf{100} & \textemdash & 75.2 & 98.0 & 23.2  \\

DICE-doseEff (FR) & 0 & 1.4 & 0.1 & 8.1 & \textbf{98.1} & \textemdash & \textbf{100} & \textemdash & 88.7 & 98.1 & 9.6  \\

\bottomrule
\end{tabular}
}
\caption{Stopping proportions for graduation, futility, and safety by dosing regimen and platform trial design in scenario 5. \textit{Note:} $^{1}$ PSES = percentage of scenario-exact stopping; $^{2}$ PCS = percentage of correct stopping; $^{3}$ PIS = percentage of incorrect stopping; $^{4}$ no true futile regimen in the scenario.}
\label{stopping_proportions_scenario5}
\end{table}

\begin{table}[H]
\centering
\resizebox{\textwidth}{!}{
\begin{tabular}{lcccccccccccccccc}
\textbf{Scenario 5} & \multicolumn{16}{c}{} \\
\toprule
\multirow[c]{1}{*}{\textbf{Acute toxicity}} & \multicolumn{8}{c}{\textbf{Posterior mean (MCSE)}} & \multicolumn{8}{c}{\textbf{95\% CrI coverage}} \\
\cmidrule(lr){2-9} 
\cmidrule(lr){10-17}
\textbf{Regimen} & 2 & 3 & 4 & 5 & 6 & 7 & 8 & 9 & 2 & 3 & 4 & 5 & 6 & 7 & 8 & 9 \\
\textit{Scenario truth} & \textit{0.8} & \textit{0.8} & \textit{22.6} & \textit{22.6} & \textit{22.6} & \textit{64.6} & \textit{64.6} & \textit{64.6} & \multicolumn{8}{c}{} \\
\midrule

\rowcolor{gray!10}
PROP (RAR) & 2.0 (0.053) & 1.8 (0.046) & 24.6 (0.135) & 21.8 (0.147) & 24.3 (0.178) & \textemdash & 62.5 (0.246) & \textemdash & 91.6 & 91.9 & 100 & 92.3 & 91.4 & \textemdash & 89.7 & \textemdash \\

DICE-doseEff (RAR) & 12.4 (0.60) & 11.6 (0.047) & 22.3 (0.070) & 23.5 (0.068) & 27.5 (0.079) & \textemdash & 41.6 (0.126) & \textemdash & 3.9 & 0.7 & 100 & 96.2 & 86.7 & \textemdash & 44.3 & \textemdash \\

\rowcolor{gray!10}
PROP (FR) & 1.9 (0.051) & 1.8 (0.047) & 20.2 (0.138) & 22.1 (0.150) & 24.2 (0.179) & \textemdash & 62.5 (0.252) & \textemdash & 93.4 & 94.0 & 100 & 89.5 & 92.0 & \textemdash & 90.5 & \textemdash \\

DICE-doseEff (FR) & 11.7 (0.056) & 11.3 (0.050) & 20.5 (0.072) & 22.8 (0.073) & 25.6 (0.086) & \textemdash & 38.4 (0.141) & \textemdash & 2.1 & 0.1 & 100 & 91.6 & 88.4 & \textemdash & 43.1 & \textemdash \\

\toprule
\multirow[c]{1}{*}{\textbf{Cumulative toxicity}} & \multicolumn{8}{c}{\textbf{Posterior mean (MCSE)}}
& \multicolumn{8}{c}{\textbf{95\% CrI coverage}} \\
\cmidrule(lr){2-9} 
\cmidrule(lr){10-17}
\textbf{Regimen} & 2 & 3 & 4 & 5 & 6 & 7 & 8 & 9 & 2 & 3 & 4 & 5 & 6 & 7 & 8 & 9 \\
\textit{Scenario truth} & \textit{33.3} & \textit{36.0} & \textit{47.2} & \textit{57.0} & \textit{58.9} & \textit{77.5} & \textit{81.9} & \textit{82.7} & \multicolumn{8}{c}{} \\
\midrule

\rowcolor{gray!10}
PROP (RAR) & 34.0 (0.197) & 34.7 (0.193) & 46.8 (0.166) & 50.2 (0.195) & 59.4 (0.208) & \textemdash & 80.8 (0.152) & \textemdash & 95.0 & 97.7 & 100 & 87.1 & 94.2 & \textemdash & 90.3 & \textemdash \\

DICE-doseEff (RAR) & 23.3 (0.081) & 29.6 (0.071) & 41.7 (0.061) & 51.6 (0.055) & 68.2 (0.066) & \textemdash & 78.8 (0.082) & \textemdash & 77.9 & 90.9 & 100 & 89.7 & 72.1 & \textemdash & 90.7 & \textemdash \\

\rowcolor{gray!10}
PROP (FR) & 33.9 (0.193) & 35.5 (0.196) & 48.6 (0.166) & 52.2 (0.198) & 59.4 (0.206) & \textemdash & 80.8 (0.154) & \textemdash & 93.1 & 94.9 & 100 & 89.5 & 93.3 & \textemdash & 91.4 & \textemdash \\

DICE-doseEff (FR) & 23.7 (0.076) & 30.3 (0.074) & 44.4 (0.064) & 51.3 (0.059) & 69.6 (0.074) & \textemdash & 78.9 (0.093) & \textemdash & 79.4 & 93.1 & 100 & 88.4 & 72.1 & \textemdash & 90.8 & \textemdash \\

\bottomrule
\end{tabular}
}
\caption{Posterior mean toxicity probabilities with Monte-Carlo Standard Error (MCSE) and 95\% credible interval (CrI) coverage by regimen and platform trial design for scenario 5.}
\label{toxicity_estimation_scenario5}
\end{table}

\begin{table}[H]
\centering
\resizebox{\textwidth}{!}{
\begin{tabular}{lccccccccc}
\toprule
\multirow[c]{1}{*}{\textbf{Scenario 5}} 
& \multicolumn{9}{c}{\textbf{Efficacy posterior mean (MCSE)}} \\
\cmidrule(lr){2-10} 
Regimen & 0$^{1}$ & 2 & 3 & 4 & 5 & 6 & 7 & 8 & 9 \\
\textit{Scenario truth} & \textit{50.0} & \textit{63.8} & \textit{68.5} & \textit{62.4} & \textit{72.9} & \textit{79.0} & \textit{66.9} & \textit{79.2} & \textit{85.4} \\
\midrule

\rowcolor{gray!10}
PROP (RAR) & 51.0 (0.208) & 63.4 (0.163) & 68.1 (0.163) & 70.2 (0.162) & 71.9 (0.179) & 76.1 (0.223) & \textemdash & 77.2 (0.235) & \textemdash \\

DICE-doseEff (RAR) & 38.8 (0.088) & 64.1 (0.047) & 69.7 (0.042) & 71.2 (0.039) & 80.9 (0.048) & 87.3 (0.058) & \textemdash & 91.7 (0.057) & \textemdash \\

\rowcolor{gray!10}
PROP (FR) & 50.7 (0.205) & 63.4 (0.161) & 67.8 (0.167) & 68.2 (0.156) & 71.8 (0.185) & 76.4 (0.220) & \textemdash & 77.4 (0.233) & \textemdash \\

DICE-doseEff (FR) & 38.3 (0.095) & 64.1 (0.050) & 69.5 (0.045) & 72.9 (0.041) & 81.5 (0.052) & 87.8 (0.063) & \textemdash & 92.0 (0.063) & \textemdash \\

\midrule

& \multicolumn{9}{c}{\textbf{95\% CrI coverage}} \\
\cmidrule(lr){2-10} 
Regimen & 0$^{1}$ & 2 & 3 & 4 & 5 & 6 & 7 & 8 & 9 \\
\midrule

\rowcolor{gray!10}
PROP (RAR) & 97.0 & 94.6 & 96.5 & 66.7 & 93.6 & 93.6 & \textemdash & 95.7 & \textemdash \\

DICE-doseEff (RAR) & 60.1 & 92.0 & 88.4 & 33.3 & 59.0 & 61.2 & \textemdash & 42.4 & \textemdash \\

\rowcolor{gray!10}
PROP (FR) & 96.4 & 94.9 & 95.2 & 66.7 & 95.8 & 93.9 & \textemdash & 95.7 & \textemdash \\

DICE-doseEff (FR) & 62.4 & 93.9 & 89.0 & 33.3 & 60.0 & 62.6 & \textemdash & 45.5 & \textemdash \\

\bottomrule
\end{tabular}
}
\caption{Posterior mean efficacy probabilities with Monte-Carlo Standard Error (MCSE) and 95\% credible interval coverage by dosing regimen and platform trial design in scenario 5. \textit{Note:} $^{1}$ regimen 0 is the control arm.}
\label{efficacy_estimation_scenario5}
\end{table}

\subsection{Scenario 6}

\begin{table}[H]
\centering
\resizebox{\textwidth}{!}{
\begin{tabular}[t]{lcccccc}
\toprule
\multicolumn{1}{l}{\textbf{Scenario 6}} & \multicolumn{3}{c}{\textbf{Graduation stop (\%)}} & \multicolumn{3}{c}{\textbf{ }} \\
\cmidrule(l{3pt}r{3pt}){2-4}
Platform design & 2 & \textbf{6} & \textbf{8} & PSES$^{1}$ & PCS$^{2}$ & PIS$^{3}$ \\
\midrule
\rowcolor{gray!10}
PROP (RAR) & 8.8 & \textbf{49.7} & \textbf{50.3} & 35.4 & 42.7 & 8.8 \\

DICE-doseEff (RAR) & 75.1 & \textbf{95.3} & \textbf{88.8} & 19.9 & 87.5 & 75.1 \\

\rowcolor{gray!10}
PROP (FR) & 9.2 & \textbf{48.0} & \textbf{51.3} & 35.6 & 43.8 & 9.2 \\

DICE-doseEff (FR) & 75.0 & \textbf{95.5} & \textbf{89.1} & 21.3 & 87.6 & 75.0 \\

\rowcolor{gray!10}
DICE-mAGILE & 3.0 & \textbf{7.6} & \textbf{13.0} & 1.8 & 2.0 & 3.0 \\

\toprule
\multicolumn{1}{l}{} & \multicolumn{3}{c}{\textbf{Futility stop (\%)}} & \multicolumn{3}{c}{} \\
\cmidrule(l{3pt}r{3pt}){2-4}
& \textbf{2} & 6 & 8 & PSES$^{1}$ & PCS$^{2}$ & PIS$^{3}$\\
\midrule
\rowcolor{gray!10}
PROP (RAR) & \textbf{36.2} & 8.3 & 5.3 & 27.6 & 36.2 & 8.8 \\

DICE-doseEff (RAR) & \textbf{3.6} & 0 & 0 & 3.6 & 3.6 & 0 \\

\rowcolor{gray!10}
PROP (FR) & \textbf{37.6} & 9.0 & 6.2 & 28.2 & 37.6 & 9.4 \\

DICE-doseEff (FR) & \textbf{3.5} & 0 & 0 & 3.5 & 3.5 & 0 \\

\rowcolor{gray!10}
DICE-mAGILE & \textbf{1.3} & 1.0 & 0.2 & 1.1 & 1.3 & 1.2 \\

\toprule
\multicolumn{1}{l}{} & \multicolumn{3}{c}{\textbf{Safety stop (\%)}} & \multicolumn{1}{c}{} \\
\cmidrule(l{3pt}r{3pt}){2-4}
& 2 & 6 & 8 & PSES$^{14}$ & PCS$^{24}$ & PIS$^{3}$\\
\midrule
\rowcolor{gray!10}
PROP (RAR) & 0 & 1.8 & 12.5 & \textemdash & \textemdash & 13.1 \\

DICE-doseEff (RAR) & 0 & 2.2 & 9.6 & \textemdash & \textemdash & 10.1 \\

\rowcolor{gray!10}
PROP (FR) & 0 & 2.4 & 12.3 & \textemdash & \textemdash & 12.6 \\

DICE-doseEff (FR) & 0 & 3.1 & 10.1 & \textemdash & \textemdash & 11.1 \\

\rowcolor{gray!10}
DICE-mAGILE & 0 & 3.7 & 12.8 & \textemdash & \textemdash & 13.7  \\

\bottomrule
\end{tabular}
}
\caption{Stopping proportions for graduation, futility, and safety by dosing regimen and platform trial design in scenario 6. \textit{Note:} $^{1}$ PSES = percentage of scenario-exact stopping; $^{2}$ PCS = percentage of correct stopping; $^{3}$ PIS = percentage of incorrect stopping; $^{4}$ no true toxic regimen in the scenario.}
\label{stopping_proportions_scenario6}
\end{table}

\begin{table}[H]
\centering
\resizebox{\textwidth}{!}{
\begin{tabular}{lcccccc}
\textbf{Scenario 6} & \multicolumn{6}{c}{} \\
\toprule
\multirow[c]{1}{*}{\textbf{Acute toxicity}} & \multicolumn{3}{c}{\textbf{Posterior mean (MCSE)}} & \multicolumn{3}{c}{\textbf{95\% CrI coverage}} \\
\cmidrule(lr){2-4} 
\cmidrule(lr){5-7}
\textbf{Regimen} & 2 & 6 & 8 & 2 & 6 & 8 \\
\textit{Scenario truth} & \textit{5.1} & \textit{9.8} & \textit{14.3} & \multicolumn{3}{c}{} \\
\midrule

\rowcolor{gray!10}
PROP (RAR) & 5.6 (0.098) & 10.0 (0.128) & 15.0 (0.204) & 97.8 & 94.8 & 96.6 \\

DICE-doseEff (RAR) & 6.5 (0.045) & 9.0 (0.053) & 11.8 (0.072) & 96.1 & 91.2 & 87.7 \\

\rowcolor{gray!10}
PROP (FR) & 5.5 (0.098) & 10.0 (0.131) & 15.2 (0.215) & 97.5 & 94.9 & 96.5 \\

DICE-doseEff (FR) & 6.6 (0.044) & 9.2 (0.057) & 12.2 (0.082) & 95.6 & 92.6 & 90.1 \\

\rowcolor{gray!10}
DICE-mAGILE & 6.8 (0.038) & 9.1 (0.044) & 11.4 (0.059) & 95.9 & 92.9 & 85.5 \\

\toprule
\multirow[c]{1}{*}{\textbf{Cumulative toxicity}} & \multicolumn{3}{c}{\textbf{Posterior mean (MCSE)}}
& \multicolumn{3}{c}{\textbf{95\% CrI coverage}} \\
\cmidrule(lr){2-4} 
\cmidrule(lr){5-7}
\textbf{Regimen} & 2 & 6 & 8 & 2 & 6 & 8 \\
\textit{Scenario truth} & \textit{15.4} & \textit{25.4} & \textit{31.1} & \multicolumn{3}{c}{} \\
\midrule

\rowcolor{gray!10}
PROP (RAR) & 15.8 (0.143) & 25.4 (0.193) & 31.4 (0.237) & 94.7 & 92.5 & 93.7 \\

DICE-doseEff (RAR) & 10.7 (0.056) & 26.6 (0.070) & 32.0 (0.083) & 83.9 & 93.6 & 96.3 \\

\rowcolor{gray!10}
PROP (FR) & 16.0 (0.147) & 25.7 (0.200) & 31.9 (0.247) & 94.6 & 93.2 & 95.3 \\

DICE-doseEff (FR) & 10.8 (0.054) & 27.6 (0.074) & 33.2 (0.095) & 84.9 & 92.3 & 96.1 \\

\rowcolor{gray!10}
DICE-mAGILE & 11.3 (0.048) & 26.9 (0.052) & 31.4 (0.065) & 86.3 & 94.3 & 96.7 \\

\bottomrule
\end{tabular}
}
\caption{Posterior mean toxicity probabilities with Monte-Carlo Standard Error (MCSE) and 95\% credible interval (CrI) coverage by regimen and platform trial design for scenario 6.}
\label{toxicity_estimation_scenario6}
\end{table}

\begin{table}[H]
\centering
\resizebox{\textwidth}{!}{
\begin{tabular}{lcccccccc}
\toprule
\multirow[c]{1}{*}{\textbf{Scenario 6}} 
& \multicolumn{4}{c}{\textbf{Efficacy posterior mean (MCSE)}}
& \multicolumn{4}{c}{\textbf{95\% CrI coverage}} \\
\cmidrule(lr){2-5} 
\cmidrule(lr){6-9}
Regimen & 0$^{1}$ & 2 & 6 & 8 & 0$^{1}$ & 2 & 6 & 8\\
\textit{Scenario truth} & \textit{50.0} & \textit{57.3} & \textit{64.2} & \textit{67.6} & \multicolumn{4}{c}{} \\
\midrule

\rowcolor{gray!10}
PROP (RAR) & 50.5 (0.227) & 58.3 (0.175) & 66.4 (0.219) & 68.5 (0.246) & 97.3 & 94.6 & 93.6 & 94.8 \\

DICE-doseEff (RAR) & 39.8 (0.109) & 56.7 (0.065) & 75.0 (0.073) & 80.0 (0.085) & 80.5 & 95.8 & 71.9 & 68.8 \\

\rowcolor{gray!10}
PROP (FR) & 50.5 (0.224) & 58.3 (0.176) & 66.3 (0.228) & 68.4 (0.257) & 97.1 & 95.4 & 94.5 & 95.6 \\

DICE-doseEff (FR) & 40.2 (0.114) & 57.4 (0.067) & 75.8 (0.084) & 80.6 (0.097) & 84.4 & 96.6 & 75.8 & 73.7 \\
\bottomrule
\end{tabular}
}
\caption{Posterior mean efficacy probabilities with Monte-Carlo Standard Error (MCSE) and 95\% credible interval coverage by dosing regimen and platform trial design in scenario 6. \textit{Note:} $^{1}$ regimen 0 is the control arm.}
\label{efficacy_estimation_scenario6}
\end{table}

\subsection{BMA Weights results}

\begin{table}[H]
\centering
\resizebox{\textwidth}{!}{
\begin{tabular}[t]{lcccccc}
\toprule
\multicolumn{1}{l}{} & \multicolumn{6}{c}{\textbf{PK/PD-efficacy model posterior selection (\%)}} \\
\cmidrule(l{3pt}r{3pt}){2-7}
Platform design & Scenario 1 & Scenario 2 & Scenario 3 & Scenario 4 & Scenario 5 & Scenario 6 \\
\midrule
\rowcolor{gray!10}
PROP (RAR) & 71.0 & 76.3 & 74.8 & 54.2 & 78.3 & 21.4 \\

PROP (FR) & 70.7 & 75.1 & 75.0 & 56.1 & 78.9 & 21.9 \\

\bottomrule
\end{tabular}
}
\caption{Proportion of simulated trials in which the time-to-event PK/PD-efficacy model was selected as the most plausible model under the BMA framework for the PROP designs by scenario.}
\label{bma_posterior_model_probabilities}
\end{table}
\end{document}